\documentclass[aps,prd,10pt,superscriptaddress,amsfonts,amssymb,amsmath,preprintnumbers,tightenlines,floats,notitlepage,letterpaper,twocolumn,nofootinbib,longbibliography]{revtex4-1}
\usepackage{graphicx}
\usepackage{aas_macros}

\usepackage[colorlinks=true,citecolor=blue,urlcolor=cyan,linkcolor=black]{hyperref}

\newcommand{\nhat}{\hat{ {n}}}

\usepackage{tabularx}
\newcolumntype{C}{>{\centering\arraybackslash}X}
\newcolumntype{R}{>{\raggedleft\arraybackslash}X}

\newcommand{\ie}{\textit{i.e.},\, }
\newcommand{\eg}{\textit{e.g.},\, }

\usepackage[usenames, dvipsnames]{color}
\usepackage[normalem]{ulem}
\usepackage{xcolor}

\newcommand{\dd}{{\rm d}}

\definecolor{colorA}{HTML}{1E90FF}
\definecolor{colorB}{HTML}{228B22}
\definecolor{colorC}{HTML}{FF7F00}
\definecolor{colorD}{HTML}{4B0082}
\definecolor{colorE}{HTML}{B22222}

\definecolor{lgreen}{HTML}{32CD32}
\definecolor{lgray}{HTML}{D3D3D3}
\definecolor{dblue}{HTML}{1E90FF}
\definecolor{dblue}{HTML}{1E90FF}
\definecolor{orange}{HTML}{FF4500}
\definecolor{indigo}{HTML}{4B0082}

\definecolor{teal}{HTML}{008080}
\definecolor{firebrick}{HTML}{B22222}
\definecolor{salmon}{HTML}{FA8072}
\definecolor{darkgreen}{HTML}{006400}

\newcommand{\perimeter}{Perimeter Institute for Theoretical Physics, 31 Caroline St N, Waterloo, ON N2L 2Y5, Canada}

\newcommand{\york}{Department of Physics and Astronomy, York University, Toronto, ON M3J 1P3, Canada}

\newcommand{\damtp}{DAMTP, Centre for Mathematical Sciences, Wilberforce Road, Cambridge CB3 0WA, UK}

\newcommand{\kavli}{Kavli Institute for Cosmology Cambridge, Madingley Road, Cambridge, CB3 0HA, UK}

\newcommand{\columbia}{Department of Physics, Columbia University, New York, NY, USA 10027}

\newcommand{\dunlap}{Dunlap Institute for Astronomy and Astrophysics, University of Toronto,
50 St.George Street, Toronto, ON M5S 3H4, Canada}

\newcommand{\waterloo}{Department of Physics and Astronomy, University of Waterloo, Waterloo, ON N2L 3G1, Canada}

\newcommand{\cca}{Center for Computational Astrophysics, Flatiron Institute, 162 5th Avenue, New York, NY 10010 USA}

\newcommand{\darkph}{\ensuremath{{ A^{\prime}}}}
\newcommand{\darksc}{\ensuremath{{\rm dsc}}}
\newcommand{\Tdsc}{\tilde{T}^{\ensuremath{{\rm dsc}}}}

\begin{document}

\title{Dark photon limits from patchy dark screening of the cosmic microwave background} 

\author{Fiona McCarthy}
\affiliation{\damtp}
\affiliation{\kavli}
\affiliation{\cca}

\author{Dalila P\^irvu}
\affiliation{\perimeter}
\affiliation{\waterloo}

\author{J. Colin Hill}
\affiliation{\columbia}

\author{Junwu Huang}
\affiliation{\perimeter}

\author{Matthew~C.~Johnson}
\affiliation{\perimeter}
\affiliation{\york}

\author{Keir K. Rogers}
\affiliation{\dunlap}

\begin{abstract}

Dark photons that kinetically mix with the Standard Model photon give rise to new spectral anisotropies (patchy dark screening) in the cosmic microwave background (CMB) due to conversion of photons to dark photons within large-scale structure. We utilize predictions for this patchy dark screening signal to provide the tightest constraints to date on the dark photon kinetic mixing parameter ($\varepsilon \lesssim 4\times 10^{-8}$ (95\% confidence level)) over the mass range $10^{-13} \,\, {\rm eV} \lesssim m_{\darkph} \lesssim 10^{-11}$ eV, almost an order of magnitude stronger than previous limits, by applying state-of-the-art component separation techniques to the cross-correlation of \textit{Planck} CMB and \textit{unWISE} galaxy survey data. 
\end{abstract}

\maketitle

\noindent
{\it Introduction} ---
The cosmic microwave background (CMB) is an exquisitely well-calibrated source of photons. It has a near-perfect blackbody frequency spectrum and small  `primary' anisotropies (anisotropies imprinted on its release), consistent with Gaussian statistics. These properties can be used to isolate `secondary' CMB anisotropies, induced by the interaction of CMB photons with large-scale structure (LSS) over cosmic history. For example, scattering from free electrons in LSS induces non-Gaussian and non-blackbody temperature and polarization anisotropies (Sunyaev-Zel'dovich, or SZ, effects~\cite{1972CoASP...4..173S,1980MNRAS.190..413S}) that can be distinguished from the primary CMB. If the photon has interactions with particles beyond the Standard Model (BSM), the associated secondary CMB anisotropies provide a powerful discovery tool to search for new physics~\cite{Pirvu:2023lch,Mondino:2024rif}.

One of the simplest extensions of the Standard Model is a light, massive vector boson~\cite{Holdom:1985ag,Okun:1982xi}, the  dark photon (DP)  $\darkph$, which can couple to the photon $\gamma$ through kinetic mixing. The DP is motivated as a low-energy consequence of string theory~\cite{Arvanitaki:2009fg,Arias:2012az,Goodsell:2009xc}, a dark matter candidate~\cite{Goodsell:2009xc,Graham:2015rva,East:2022rsi}, and a mediator of interactions with a larger dark sector (see~\cite{Knapen:2017xzo} and references within). The mass range of interest for the DP spans many orders of magnitude, generating a diverse experimental program~\cite{Pospelov:2008jkf,An:2013yua,Baryakhtar:2017ngi,Baryakhtar:2018doz,SENSEI:2020dpa,Hardy:2016kme,Lasenby:2020goo,Romanenko:2023irv,Chaudhuri:2014dla}. For a DP with mass below $\sim 10^{-7}$ eV, the signatures of interest are mainly cosmological and astrophysical~\cite{Georgi:1983sy,Mirizzi:2009iz,Caputo:2020rnx,caputo_dark_2020,Siemonsen:2022ivj,Siemonsen:2019ebd}.

As a consequence of kinetic mixing, CMB photons can convert to DPs as they propagate~\cite{Georgi:1983sy}. The strongest, {\em resonant}, conversion occurs in regions where the photon plasma mass (proportional to the square root of the number density of electrons) is equal to the mass of the DP~\cite{Mirizzi:2009iz} --- this is analogous to the Mikheyev-Smirnov-Wolfenstein (MSW) effect for neutrino oscillations in matter~\cite{Wolfenstein:1977ue,Mikheyev:1985zog}. Resonant conversion can occur due to the evolution of the cosmic mean density as the Universe expands~\cite{Mirizzi:2009iz} or due to the variation in density associated with the formation and evolution of LSS~\cite{caputo_dark_2020,Caputo:2020rnx}. Measurements of the CMB monopole spectrum~\cite{1996ApJ...473..576F} constrain the strength of the kinetic mixing~\cite{Mirizzi:2009iz,caputo_dark_2020,Caputo:2020rnx}. Ref.~\cite{Pirvu:2023lch} demonstrated that these previous constraints could be greatly improved at $\sim {\rm peV}$ DP masses by measuring the {\em patchy screening} of the CMB monopole, the frequency- and spatially-dependent reduction of intensity due to resonant conversion inside of non-linear structure. The characteristic spatial and spectral shape of the signal, as well as its correlation with tracers of LSS, can be used to separate the signal from the primary CMB and astrophysical foregrounds.\footnote{This framework can be extended to other BSM particles such as axions~(e.g.,~\cite{Mondino:2024rif,Mehta:2024pdz}).}

In this \textit{Letter}, we perform the first search for DPs via patchy dark screening of the CMB in cross-correlation with LSS, using data from the \textit{Planck} mission~\cite{2020A&A...641A...1P} cross-correlated with the \textit{unWISE} galaxy catalog~\cite{Lang_2014,Meisner_2017,Meisner_2017a,2019ApJS..240...30S,2020JCAP...05..047K} assembled from data taken by the \textit{Wide-field Infrared Survey Explorer} (\textit{WISE}) mission~\cite{2010AJ....140.1868W,2011ApJ...731...53M}. We place the strongest constraints to date on the strength of DP kinetic mixing in the mass range $10^{-13} \,\, {\rm eV}\lesssim m_{\darkph} \lesssim 10^{-11}$~eV, improving existing constraints by almost an order of magnitude. Forecasts~\cite{Pirvu:2023lch} suggest that even more dramatic improvements will be possible with data from future CMB~\cite{Ade:2018sbj,Abazajian:2019eic} and galaxy surveys~\cite{LSSTScienceCollaboration2009,DESI:2016fyo,SPHEREx:2014bgr}.

\noindent
{\it Dark Photon Model} --- A kinetically mixed DP is described by the following Lagrangian\footnote{We use units where $\hbar = k_{\rm B} = c = 1$.}~\cite{Okun:1982xi,Holdom:1985ag}:
\begin{align}
    \mathcal{L}\supset -\frac{1}{4}{F}_{\mu\nu}{F}^{\mu\nu}-\frac{1}{4}{F}^{\prime}_{\mu\nu}{F}^{\prime\mu\nu}
    -\frac{m_{\darkph}^2}{2}{\darkph}_\mu {\darkph}^\mu- \frac{\varepsilon}{2} {F}_{\mu\nu}{F}^{\prime\mu\nu},
    \label{eq:darkphoton}
\end{align}
where $m_{\darkph}$ is the mass of the DP, $\varepsilon$ is the kinetic mixing parameter, and $F_{\mu\nu}$ ($F^\prime_{\mu\nu})$ is the photon (DP) field-strength tensor. The photon obtains a medium-dependent plasma mass $m_{\gamma}(\vec{x}(t)) = \sqrt{4\pi\alpha n_e(\vec{x}(t))/m_e} = 3.7 \times 10^{-11} \, {\rm eV} \sqrt{n_e/{\rm cm}^{-3}}$, where $n_e$ is the free electron number density, and resonant conversion of CMB photons to DPs occurs at $t_{\rm res}$, when $m_{\gamma}(\vec{x}(t_{\rm res}) ) = m_{\darkph}$ along any photon trajectory $\vec{x}(t)$. The conversion probability can be computed via the Landau-Zener formula~\cite{Mirizzi:2009iz,Pirvu:2023lch}, resulting in a position- and frequency-dependent (scaling as $1/\nu$) dark screening optical depth $\tau^\darksc (\nu, \hat{n})$:
\begin{equation}\label{eq:totalprobHEP}
	\tau^\darksc (\nu, \hat{n}) = \sum_{t_{\rm res}} \frac{\pi \varepsilon \, m_{\darkph}^2}{2 \pi\nu (t_{\rm res})} \times \varepsilon\left| \frac{\dd}{\dd t} \ln m_{\gamma}^2(\hat{n},t)\right|^{-1}_{t=t_{\rm res}} \, ,
\end{equation}
where the sum is over all resonances  in the line of sight direction $\hat{n}$. 
The dark screening optical depth couples to the CMB monopole intensity $B(\nu, T_{\rm CMB})$,  where $B(\nu, T_{\rm CMB})$ is the Planck blackbody specific intensity at the CMB monopole temperature $T_{\rm CMB}$, resulting in a frequency-dependent specific intensity anisotropy $\Delta I(\nu, \hat{n}) = - \tau^\darksc (\nu, \hat{n}) {B(\nu, T_{\rm CMB})}$, which can be extracted from multi-frequency CMB maps. In CMB thermodynamic temperature units, this takes the form $\Tdsc (\nu, \hat{n}) = T_{\rm CMB} \tau^\darksc (\nu, \hat{n})\frac{1-e^{-x}}{x}$ where $x \equiv {h\nu }/{k_B T_{\rm CMB}}=2 \pi \nu/T_{\rm CMB}$ is the dimensionless photon frequency~\cite{DAmico:2015snf}, with  $T_{\rm CMB} \simeq 2.726$~K~\cite{1996ApJ...473..576F,2009ApJ...707..916F}. 

In Ref.~\cite{Pirvu:2023lch}, angular correlation functions of the dark screening map $\langle \Tdsc (\nu, \hat{n}) \Tdsc (\nu, \hat{n}^{\prime})\rangle$ and of its cross-correlation with a template $\hat{\tau}$ for the dark screening signal constructed from a tracer of LSS, $\langle \Tdsc (\nu, \hat{n}) \hat{\tau} (\hat{n}^{\prime}) \rangle$, were computed within the halo model (see,~\eg\cite{2002PhR...372....1C} for a review). In the Supplemental Material (SM),  we adapt this result to compute the cross-correlation $\langle \Tdsc (\nu, \hat{n}) \delta_g ( \hat{n}^{\prime})\rangle$, where the galaxy overdensity $\delta_g$ is modeled using the halo occupation distribution (HOD) described in Ref.~\cite{2022PhRvD.106l3517K} for the \textit{unWISE} galaxies. As in Ref.~\cite{Pirvu:2023lch}, we model the electron distribution in halos using the ``AGN feedback'' model of Ref.~\cite{Battaglia:2016xbi}. We work with the harmonic transforms of the statistics,~\ie $C_\ell^{g \Tdsc}$, where $\ell$ is the multipole moment.

\begin{figure}
\includegraphics[width=\columnwidth]{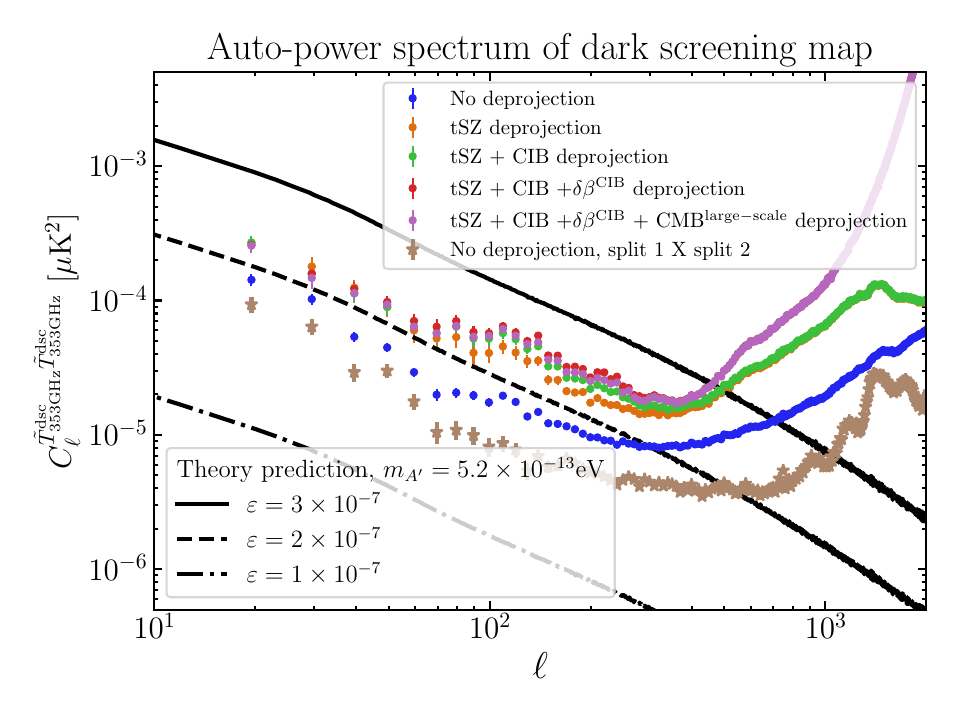}
\caption{The colored points show the measured auto-power spectrum of the NILC dark screening map for various contaminant deprojection choices.  We also show the power spectrum measured from two independent splits of the data, which does not contain correlated noise bias. We include theoretical predictions for the signal, with {$m_\darkph=5.2\times10^{-13}$ eV}, for various values of $\varepsilon$, as indicated. The data and theory signals are shown at a reference frequency of 353 GHz.} \label{fig:autopower}
\end{figure}

\noindent
{\it Data and measurement: Dark screening map} ---
The \textit{Planck} satellite~\cite{2020A&A...641A...1P} mapped the full sky in nine microwave frequency bands. In our standard measurement, we use the sky maps produced by the NPIPE (PR4) data processing~\cite{2020A&A...643A..42P} at \{30, 44, 70, 100, 143, 217, 353,  545\} GHz.  We apply several preprocessing steps to the maps, as described in Ref.~\cite{2024PhRvD.109b3528M}; in particular we mask and inpaint  regions with bright point sources, as defined by the \textit{Planck} point source masks, and a region near the Galactic center.

We use \texttt{pyilc}~\cite{2024PhRvD.109b3528M}\footnote{\url{https://github.com/jcolinhill/pyilc/}} to construct a constrained needlet internal linear combination (NILC)~\cite{2009ApJ...694..222C,2011MNRAS.410.2481R} map of the CMB spectral distortion induced by the DP conversion.  NILC is a component separation technique that allows for minimum-variance estimation of the sky map of a component with a known frequency dependence (in this case, the DP distortion) via a linear combination of the frequency maps. In CMB thermodynamic temperature units, the frequency dependence of the DP distortion that we preserve is
\begin{equation}
\label{eq.DP_SED}
    \frac{\Delta T(\nu)}{T_{\rm CMB}} \propto  \frac{1}{x} \,\left(\frac{1-e^{-x}}{x}\right).
\end{equation}
We  normalize the DP distortion in Eq.~\eqref{eq.DP_SED} at 353 GHz, and thus we denote our NILC patchy dark screening map as $\Tdsc_{353 \, {\rm GHz}}$. The angular power spectrum of the resulting map is plotted in Fig.~\ref{fig:autopower} along with the predicted dark screening signal for several values of $\varepsilon$. 

We choose constraints that allow us to directly remove, or `deproject,' contamination from residual foregrounds in the NILC, in particular the thermal Sunyaev-Zel'dovich (tSZ) effect, the cosmic infrared background (CIB), and the blackbody CMB on large scales (which contains the integrated Sachs--Wolfe~\cite{1967ApJ...147...73S} signal). These signals are correlated with the \textit{unWISE} galaxies~\cite{2023PhRvD.108l3501K,2021PhRvD.104d3518K,Ibitoye:2022oot,Yan:2023okq,Krolewski:2021znk}, and as demonstrated in the SM, bias our measurement of the cross-correlation if not removed. The tSZ effect has a well-understood spectral dependence~\cite{1970Ap&SS...7....3S} and can therefore be deprojected exactly (as can the blackbody CMB). The spectral dependence of the CIB is less well-known, but we remove it by performing a Taylor expansion around an estimate of a model for its frequency dependence following Ref.~\cite{2017MNRAS.472.1195C}; such a method was demonstrated to remove CIB contamination in an ILC setting in Refs.~\cite{2024PhRvD.109b3528M,2024PhRvD.109b3529M}.

We also construct two other versions of the undeprojected (standard ILC) map with independent noise realizations, from the independent half-ring splits of the NPIPE maps (these are co-added to produce the final NPIPE maps, such that each is twice as noisy as the final products). Their cross-power spectrum does not contain an instrumental noise term, so indicates the contribution from only foregrounds and signal.

As shown in Fig.~\ref{fig:autopower}, foreground deprojection comes at the expense of an increase in variance, since the NILC weights are constrained to explicitly remove the foregrounds rather than solely minimize the variance of the resulting map. This has the effect of weakening the sensitivity to dark screening, but increasing the robustness and interpretability of our results. We provide further details about the mapmaking procedure, including stability tests of the foreground removal, in the SM. 

\noindent
{\it  Data and measurement: LSS tracer (\textit{unWISE} galaxies)} --- We cross-correlate our dark screening map 
with galaxy catalogs extracted from \textit{unWISE}~\cite{Lang_2014,Meisner_2017,Meisner_2017a,2019ApJS..240...30S}. 
We use both the ``Blue'' and ``Green'' \textit{unWISE} samples described in Refs.~\cite{2020JCAP...05..047K,2021JCAP...12..028K}. These catalogs comprise objects from a broad range of redshifts, with the Blue sample at median redshift $\bar z\sim0.6$ and the Green sample at $\bar z\sim1.1$.  We convert the Green and Blue galaxy number density maps $n_g (\hat n)$ into overdensity maps by measuring the mean galaxy density $\bar{n}$ and calculating
\begin{equation}
\delta_g(\hat n) = \frac{n_g(\hat n) - \bar{n}}{\bar{n}} \,.
\end{equation}

These catalogs have previously been used in various CMB cross-correlation analyses, including as a tracer of matter/galaxy overdensities for CMB lensing cross-correlations~\cite{2021JCAP...12..028K,2023arXiv230905659F,2023arXiv231104213F} and of electron overdensities for kinetic SZ (kSZ) cross-correlations~\cite{2022PhRvD.106l3517K,2024arXiv240113033C,2021PhRvD.104d3518K,2024arXiv240500809B} and patchy Thomson screening cross-correlations~\cite{2024arXiv240113033C}. In this work, we similarly use these maps to trace the electron distribution.

\noindent
{\it Cross-correlation measurement} --- We use \texttt{pymaster}~\cite{2002ApJ...567....2H,2019MNRAS.484.4127A} to estimate the mask-decoupled cross-power spectrum between the NILC foreground-deprojected dark screening map and the \textit{unWISE} maps. We apply the \textit{unWISE} mask to the galaxies (retaining $51\%$ of the sky) and the union of the apodized preprocessing mask and the \textit{Planck} $70\%$ Galactic plane mask to the dark screening map (retaining $56.5\%$ of the sky). The masks are described in the SM. 

We measure power spectra in multipole bins of width $\Delta\ell=150$, starting at $\ell=5$. We do not use the largest-scale data point in our theoretical interpretation as it has a very large error bar (and is slightly unstable with respect to the choice of frequency coverage in the NILC). Our lowest multipole bin thus corresponds to $155 \leq \ell < 305$.

As described in greater detail in the SM, we find that our cross-correlation measurement is unstable to the removal of the lowest-frequency channels from the ILC, suggesting a residual extragalactic contribution from some other source in the NILC map. We expect that this is synchrotron emission from extragalactic radio sources, which we have not explicitly deprojected in the ILC.

{To assess the size of a possible radio bias, we directly measure the cross-correlation between our dark screening map and a catalog of radio-selected objects from the Very Large Array (VLA) Faint Images of the Radio
Sky at Twenty centimeters (FIRST) survey~\cite{1997ApJ...475..479W}.  The FIRST sources are detected via bright emission at 1.4~GHz, a much lower frequency than those used in our NILC map construction.  We detect a correlation coefficient of $\approx 10-20\%$ between the FIRST sources and the \textit{unWISE} samples. Additionally, the cross-correlation between our dark screening map and the FIRST catalog is positive and much higher  than the cross-correlation between the dark screening map and the \textit{unWISE} maps, indicating that there is some emission remaining in the NILC map that is more correlated with the radio sources than with \textit{unWISE}. We build a template for the residual radio-\textit{unWISE} signal $C_\ell^{g^i\,\mathrm{radio}}$ as follows:}
\begin{align}
C_\ell^{g^i\,\mathrm{radio}} = \frac{\hat {C}_\ell^{\mathrm{FIRST}\, g^i}}{\hat C_\ell^{\rm{FIRST}\,\rm{FIRST}}  } \hat C_\ell^{\rm{FIRST}\, \tilde T^{\rm dsc}_{353 \rm{GHz}}}
\end{align}
{where $\hat C_\ell^{X,\mathrm{FIRST}}$ is the measured cross-power spectrum of map $X$ with the FIRST map. This bias is small (equivalent in amplitude to a DP signal corresponding to $\varepsilon\sim5\times10^{-8}$ in our most sensitive mass range, but with opposite sign).  We describe this bias template in further detail in the SM. The bias can be measured for NILC maps constructed with different sets of frequency maps, and we find that upon subtraction of this bias the cross-correlation measurements become stable with respect to the choice of frequency coverage. {We also validate our full pipeline, including the radio bias subtraction, with a signal injection test in the SM.} Thus, in the following analysis we use the NILC map constructed from all eight \emph{Planck} channels between 30 and 545 GHz, which yields the tightest constraints on $\varepsilon$ and shows no evidence of additional bias.}

\begin{figure*}
\includegraphics[width=\columnwidth]{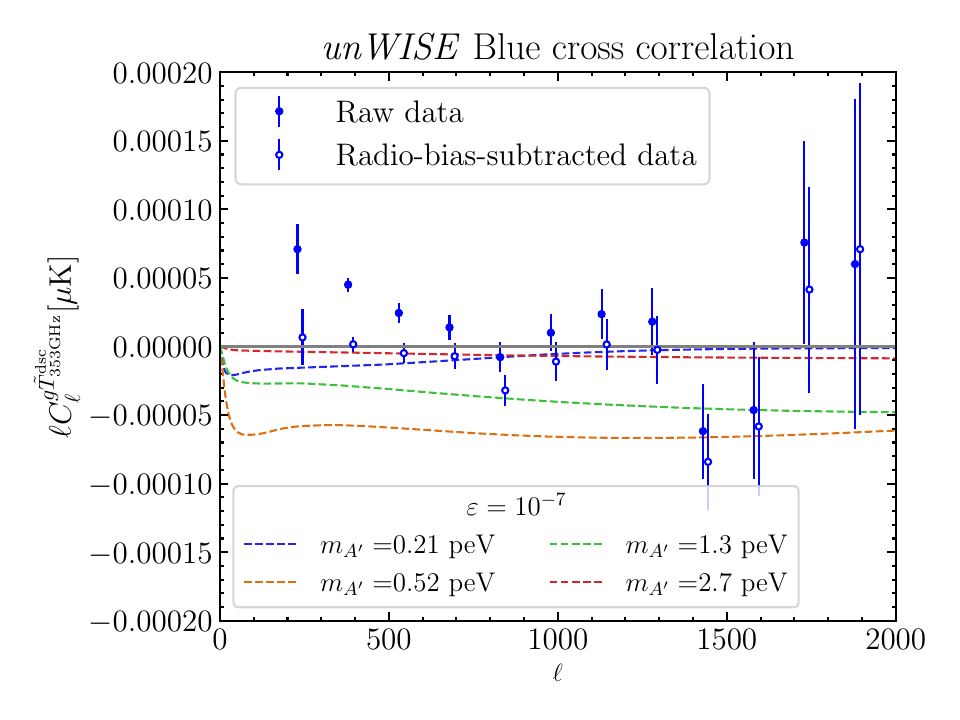}
\includegraphics[width=\columnwidth]{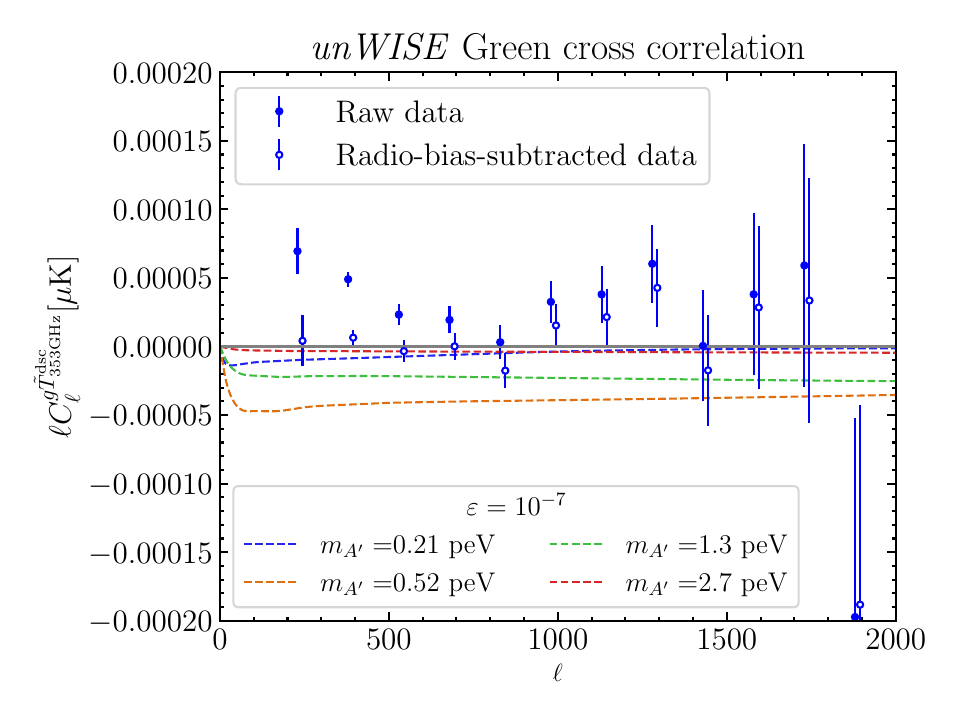}
\caption{The measured cross-correlation of our dark screening map with the \textit{unWISE} Blue (left) and Green (right) galaxy samples. In both cases we show theoretical predictions for the signal for a range of DP masses to which these measurements are sensitive (roughly $10^{-13} \, \, {\rm eV} \lesssim m_{\darkph} \lesssim 10^{-11}$~eV), and with $\varepsilon = 10^{-7}$.  {We show both the raw cross-power spectrum data and the data with the radio bias subtracted, which are offset slightly horizontally for clearer visualization.}}\label{fig:measurement}
\end{figure*}

The measured cross-power spectra, {before and after radio-bias subtraction,} are shown in Fig.~\ref{fig:measurement}. We compute the mask-deconvolved covariance matrix for the cross-power spectrum assuming that the two maps are uncorrelated Gaussian random fields with underlying power spectra given by their measured power spectra~\cite{Efstathiou:2006eb,2019MNRAS.484.4127A}. Note that, by comparison with Fig.~\ref{fig:autopower}, it is clear that the cross-correlation is a more sensitive probe of $\varepsilon$ than the auto-power spectrum, confirming the forecasts of Ref.~\cite{Pirvu:2023lch}.

The $\chi^2$ with respect to null for each of the samples, including the 12  data points, {with the radio bias subtracted,} in the $\ell$ range we use for the separate analyses, are 17.71  for Blue; 9.85  for Green; and 26.80 for the combination (including their covariance).  The associated probability-to-exceed (PTE) values are 0.12, 0.63, and 0.31, respectively, indicating consistency with a null detection.

\noindent
{\it Constraints on dark photon parameters} ---
As $C_\ell^{ g\Tdsc}$ is proportional to $\varepsilon^2$, we construct a Gaussian likelihood for $\varepsilon^2$:
\begin{eqnarray}
 && -2 \ln \mathcal L({\bf d}; \varepsilon^2,m_{\darkph}) 
 = \\
 &&\left(\mathbf{d} - \varepsilon^2 \mathbf{C}_\ell^{ g^{i}\Tdsc}(m_\darkph)\right)^T(\mathbb{C}^{ij}_{\ell\ell^\prime}){}^{-1} \left(\mathbf d - \varepsilon^2 \mathbf{C}_\ell^{ g^{j}\Tdsc}(m_\darkph)\right),   \nonumber
\end{eqnarray}
where $\mathbf{d}$ is the measured power spectrum data vector {(with the radio bias subtracted)} and $\mathbf{C}_\ell^{g^{i}\Tdsc}(m_\darkph)$ is the theoretical model for this data vector (see the SM) for DP mass $m_\darkph$, with the index ${i,j}$ denoting either the Blue or Green \textit{unWISE} sample. The covariance matrix $\mathbb{C}^{ij}_{\ell\ell^\prime}\equiv{\rm Cov}(C_\ell^{{g^{i}\Tdsc} },C_{\ell^\prime}^{{g^{j}} \Tdsc})$ is computed using \texttt{pymaster} to decouple the mask from the standard full-sky expression for {uncorrelated} Gaussian fields:
\begin{equation}\label{eq:cov}
    {\rm Cov}(C_\ell^{{g^{i} \Tdsc} },C_{\ell^\prime}^{{g^{j}\Tdsc} }) = \frac{ C_\ell^{\Tdsc\Tdsc} C_\ell^{g^{i}g^{j}} \delta_{\ell \ell{^\prime}}}{2\ell+1} \,,
\end{equation}
where $C_\ell^{\Tdsc\Tdsc}$ and $C_\ell^{{g^{i}}{g^{j}}}$ include all sources of signal and noise in the $\Tdsc$ and $g^{i}$ fields respectively.  In practice (and in particular in the absence of a model for all sources of noise and foregrounds in the $\Tdsc$ field), we use the measured auto-power spectra in place of models to evaluate Eq.~\eqref{eq:cov}.

\begin{figure}[ht!]
\includegraphics[width=\columnwidth]{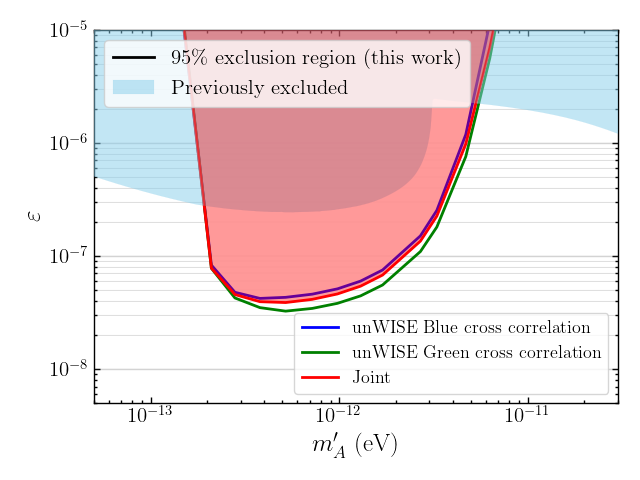}
\caption{95\% confidence upper limit constraints on $\varepsilon$ (excluded regions indicated in shaded colors). We show our constraints derived from the \textit{unWISE} Blue and Green cross-correlations (in blue and green, respectively), as well as the joint constraint in red.  We also show previous constraints on $\varepsilon$~\cite{Mirizzi:2009iz,caputo_dark_2020}, derived using CMB monopole data from \textit{COBE}/\textit{FIRAS}~\cite{1996ApJ...473..576F} to constrain the average spectral distortion induced by resonant conversion of CMB photons into DPs.}\label{fig:constraints}
\end{figure}

We scan over a range of $m_\darkph$ in the region $1.5 \times 10^{-13} \, {\rm eV} \leq m_\darkph \leq 9.7 \times 10^{-12}$ eV and compute $\mathcal L({\bf d};\varepsilon^2, m_\darkph)$. For each $m_\darkph$ value, we adopt a uniform prior on $\varepsilon$ over the domain $\varepsilon > 0$. The DP mass $m_{\darkph}$ to which our data are most sensitive is $m_{\darkph}\simeq 5\times10^{-13}$~eV. The derived 95\% confidence upper limits on $\varepsilon$ as a function of $m_\darkph$ are shown in Fig.~\ref{fig:constraints}. We find that $\varepsilon < 4 \times 10^{-8}$ for most of the covered mass range, almost an order-of-magnitude improvement compared to the previous best constraints in the literature.

For the purposes of these constraints, we hold fixed the parameters describing the electron distribution (to the AGN feedback model of Ref.~\cite{Battaglia:2016xbi}) and the galaxy HOD models. We demonstrate the validity of this approach in the SM.

\noindent
{\it Conclusion and Remarks} ---
We have improved constraints on the kinetic mixing parameter of the DP in the mass range $10^{-13} \, {\rm eV} < m_{A^\prime} < 10^{-11}$ eV by almost an order of magnitude, using a foreground-cleaned cross-correlation between LSS tracers and the anisotropic dark screening induced by resonant conversion of CMB photons into DPs. This is the first time that such a CMB-LSS cross-correlation has been used to derive a constraint on light bosons in the dark sector, and demonstrates the power of such methods to constrain BSM physics.

Improvements in the analysis can be made with higher-sensitivity CMB experiments and deeper samples of LSS tracers.  In addition, different DP masses can be probed by using LSS tracers that are sensitive to regions with different photon plasma mass, \ie different electron densities.  Theoretical calculations of such dark screening maps are underway.  This method can also be adapted to search for patchy dark screening from axions~\cite{Mondino:2024rif}; this is ongoing work. 

We make our NILC dark screening maps public at \url{https://users.flatironinstitute.org/~fmccarthy/dark_photon_screening_maps/}.

\noindent
{\bf Acknowledgements.---}

We thank Gerrit Farren, Aleksandra Kusiak, and Cristina Mondino for useful discussions, and Alex Krolewski for providing us with the \textit{unWISE} samples. This work made use of the \texttt{healpy}\footnote{\url{https://healpy.readthedocs.io/en/latest/}} library~\cite{Zonca2019} (the Python implementation of HEALPix~\cite{2005ApJ...622..759G}) as well as heavy use of \texttt{numpy}~\cite{harris2020array}. 

We thank the Scientific Computing Core staff at the Flatiron Institute for computational support. 

The Flatiron Institute is a division of the Simons Foundation. FMcC acknowledges support from the European Research Council (ERC) under the European Union's Horizon 2020 research and innovation programme (Grant agreement No.~851274). MCJ and JH are supported by the Natural Sciences and Engineering Research Council of Canada through a Discovery grant. Research at Perimeter Institute is supported in part by the Government of Canada through the Department of Innovation, Science and Economic Development Canada and by the Province of Ontario through the Ministry of Research, Innovation and Science.  JCH acknowledges support from NSF grant AST-2108536, NASA grant 80NSSC23K0463 (ADAP), NASA grant 80NSSC22K0721 (ATP), DOE HEP grant DE-SC0011941, the Sloan Foundation, and the Simons Foundation.  The Dunlap Institute is funded through an endowment established by the David Dunlap family and the University of Toronto.

{\it Note added:} An analysis appeared on arXiv~\cite{2024arXiv240505104A} during the final stage of this project, utilizing the temperature auto-correlation at 70 GHz to search for the patchy dark screening signal from dark photons. Our work improves on theirs both through the use of component separation and cross-correlation, an estimator shown to give stronger constraints than the auto-power spectrum in Ref.~\cite{Pirvu:2023lch}.  

\bibliography{references.bib}

\begin{thebibliography}{103}%
\makeatletter
\providecommand \@ifxundefined [1]{%
 \@ifx{#1\undefined}
}%
\providecommand \@ifnum [1]{%
 \ifnum #1\expandafter \@firstoftwo
 \else \expandafter \@secondoftwo
 \fi
}%
\providecommand \@ifx [1]{%
 \ifx #1\expandafter \@firstoftwo
 \else \expandafter \@secondoftwo
 \fi
}%
\providecommand \natexlab [1]{#1}%
\providecommand \enquote  [1]{``#1''}%
\providecommand \bibnamefont  [1]{#1}%
\providecommand \bibfnamefont [1]{#1}%
\providecommand \citenamefont [1]{#1}%
\providecommand \href@noop [0]{\@secondoftwo}%
\providecommand \href [0]{\begingroup \@sanitize@url \@href}%
\providecommand \@href[1]{\@@startlink{#1}\@@href}%
\providecommand \@@href[1]{\endgroup#1\@@endlink}%
\providecommand \@sanitize@url [0]{\catcode `\\12\catcode `\$12\catcode
  `\&12\catcode `\#12\catcode `\^12\catcode `\_12\catcode `\%12\relax}%
\providecommand \@@startlink[1]{}%
\providecommand \@@endlink[0]{}%
\providecommand \url  [0]{\begingroup\@sanitize@url \@url }%
\providecommand \@url [1]{\endgroup\@href {#1}{\urlprefix }}%
\providecommand \urlprefix  [0]{URL }%
\providecommand \Eprint [0]{\href }%
\providecommand \doibase [0]{http://dx.doi.org/}%
\providecommand \selectlanguage [0]{\@gobble}%
\providecommand \bibinfo  [0]{\@secondoftwo}%
\providecommand \bibfield  [0]{\@secondoftwo}%
\providecommand \translation [1]{[#1]}%
\providecommand \BibitemOpen [0]{}%
\providecommand \bibitemStop [0]{}%
\providecommand \bibitemNoStop [0]{.\EOS\space}%
\providecommand \EOS [0]{\spacefactor3000\relax}%
\providecommand \BibitemShut  [1]{\csname bibitem#1\endcsname}%
\let\auto@bib@innerbib\@empty
\bibitem [{\citenamefont {{Sunyaev}}\ and\ \citenamefont
  {{Zeldovich}}(1972)}]{1972CoASP...4..173S}%
  \BibitemOpen
  \bibfield  {author} {\bibinfo {author} {\bibfnamefont {R.~A.}\ \bibnamefont
  {{Sunyaev}}}\ and\ \bibinfo {author} {\bibfnamefont {Y.~B.}\ \bibnamefont
  {{Zeldovich}}},\ }\href@noop {} {\bibfield  {journal} {\bibinfo  {journal}
  {Comments on Astrophysics and Space Physics}\ }\textbf {\bibinfo {volume}
  {4}},\ \bibinfo {pages} {173} (\bibinfo {year} {1972})}\BibitemShut {NoStop}%
\bibitem [{\citenamefont {{Sunyaev}}\ and\ \citenamefont
  {{Zeldovich}}(1980)}]{1980MNRAS.190..413S}%
  \BibitemOpen
  \bibfield  {author} {\bibinfo {author} {\bibfnamefont {R.~A.}\ \bibnamefont
  {{Sunyaev}}}\ and\ \bibinfo {author} {\bibfnamefont {Y.~B.}\ \bibnamefont
  {{Zeldovich}}},\ }\href {\doibase 10.1093/mnras/190.3.413} {\bibfield
  {journal} {\bibinfo  {journal} {\mnras}\ }\textbf {\bibinfo {volume} {190}},\
  \bibinfo {pages} {413} (\bibinfo {year} {1980})}\BibitemShut {NoStop}%
\bibitem [{\citenamefont {Pîrvu}\ \emph {et~al.}(2024)\citenamefont {Pîrvu},
  \citenamefont {Huang},\ and\ \citenamefont {Johnson}}]{Pirvu:2023lch}%
  \BibitemOpen
  \bibfield  {author} {\bibinfo {author} {\bibfnamefont {D.}~\bibnamefont
  {Pîrvu}}, \bibinfo {author} {\bibfnamefont {J.}~\bibnamefont {Huang}}, \
  and\ \bibinfo {author} {\bibfnamefont {M.~C.}\ \bibnamefont {Johnson}},\
  }\href {\doibase 10.1088/1475-7516/2024/01/019} {\bibfield  {journal}
  {\bibinfo  {journal} {\jcap}\ }\textbf {\bibinfo {volume} {2024}},\ \bibinfo
  {pages} {019} (\bibinfo {year} {2024})}\BibitemShut {NoStop}%
\bibitem [{\citenamefont {Mondino}\ \emph {et~al.}(2024)\citenamefont
  {Mondino}, \citenamefont {P\^\i{}rvu}, \citenamefont {Huang},\ and\
  \citenamefont {Johnson}}]{Mondino:2024rif}%
  \BibitemOpen
  \bibfield  {author} {\bibinfo {author} {\bibfnamefont {C.}~\bibnamefont
  {Mondino}}, \bibinfo {author} {\bibfnamefont {D.}~\bibnamefont {P\^\i{}rvu}},
  \bibinfo {author} {\bibfnamefont {J.}~\bibnamefont {Huang}}, \ and\ \bibinfo
  {author} {\bibfnamefont {M.~C.}\ \bibnamefont {Johnson}},\ }\href@noop {}
  {\bibfield  {journal} {\bibinfo  {journal} {arXiv e-prints}\ } (\bibinfo
  {year} {2024})},\ \Eprint {http://arxiv.org/abs/2405.08059} {2405.08059}
  \BibitemShut {NoStop}%
\bibitem [{\citenamefont {Holdom}(1986)}]{Holdom:1985ag}%
  \BibitemOpen
  \bibfield  {author} {\bibinfo {author} {\bibfnamefont {B.}~\bibnamefont
  {Holdom}},\ }\href {\doibase 10.1016/0370-2693(86)91377-8} {\bibfield
  {journal} {\bibinfo  {journal} {Phys. Lett. B}\ }\textbf {\bibinfo {volume}
  {166}},\ \bibinfo {pages} {196} (\bibinfo {year} {1986})}\BibitemShut
  {NoStop}%
\bibitem [{\citenamefont {Okun}(1982)}]{Okun:1982xi}%
  \BibitemOpen
  \bibfield  {author} {\bibinfo {author} {\bibfnamefont {L.~B.}\ \bibnamefont
  {Okun}},\ }\href@noop {} {\bibfield  {journal} {\bibinfo  {journal} {Sov.
  Phys. JETP}\ }\textbf {\bibinfo {volume} {56}},\ \bibinfo {pages} {502}
  (\bibinfo {year} {1982})}\BibitemShut {NoStop}%
\bibitem [{\citenamefont {Arvanitaki}\ \emph {et~al.}(2010)\citenamefont
  {Arvanitaki}, \citenamefont {Dimopoulos}, \citenamefont {Dubovsky},
  \citenamefont {Kaloper},\ and\ \citenamefont
  {March-Russell}}]{Arvanitaki:2009fg}%
  \BibitemOpen
  \bibfield  {author} {\bibinfo {author} {\bibfnamefont {A.}~\bibnamefont
  {Arvanitaki}}, \bibinfo {author} {\bibfnamefont {S.}~\bibnamefont
  {Dimopoulos}}, \bibinfo {author} {\bibfnamefont {S.}~\bibnamefont
  {Dubovsky}}, \bibinfo {author} {\bibfnamefont {N.}~\bibnamefont {Kaloper}}, \
  and\ \bibinfo {author} {\bibfnamefont {J.}~\bibnamefont {March-Russell}},\
  }\href {\doibase 10.1103/PhysRevD.81.123530} {\bibfield  {journal} {\bibinfo
  {journal} {\prd}\ }\textbf {\bibinfo {volume} {D81}},\ \bibinfo {pages}
  {123530} (\bibinfo {year} {2010})}\BibitemShut {NoStop}%
\bibitem [{\citenamefont {Arias}\ \emph {et~al.}(2012)\citenamefont {Arias},
  \citenamefont {Cadamuro}, \citenamefont {Goodsell}, \citenamefont {Jaeckel},
  \citenamefont {Redondo},\ and\ \citenamefont {Ringwald}}]{Arias:2012az}%
  \BibitemOpen
  \bibfield  {author} {\bibinfo {author} {\bibfnamefont {P.}~\bibnamefont
  {Arias}}, \bibinfo {author} {\bibfnamefont {D.}~\bibnamefont {Cadamuro}},
  \bibinfo {author} {\bibfnamefont {M.}~\bibnamefont {Goodsell}}, \bibinfo
  {author} {\bibfnamefont {J.}~\bibnamefont {Jaeckel}}, \bibinfo {author}
  {\bibfnamefont {J.}~\bibnamefont {Redondo}}, \ and\ \bibinfo {author}
  {\bibfnamefont {A.}~\bibnamefont {Ringwald}},\ }\href {\doibase
  10.1088/1475-7516/2012/06/013} {\bibfield  {journal} {\bibinfo  {journal}
  {\jcap}\ }\textbf {\bibinfo {volume} {06}},\ \bibinfo {pages} {013} (\bibinfo
  {year} {2012})}\BibitemShut {NoStop}%
\bibitem [{\citenamefont {Goodsell}\ \emph {et~al.}(2009)\citenamefont
  {Goodsell}, \citenamefont {Jaeckel}, \citenamefont {Redondo},\ and\
  \citenamefont {Ringwald}}]{Goodsell:2009xc}%
  \BibitemOpen
  \bibfield  {author} {\bibinfo {author} {\bibfnamefont {M.}~\bibnamefont
  {Goodsell}}, \bibinfo {author} {\bibfnamefont {J.}~\bibnamefont {Jaeckel}},
  \bibinfo {author} {\bibfnamefont {J.}~\bibnamefont {Redondo}}, \ and\
  \bibinfo {author} {\bibfnamefont {A.}~\bibnamefont {Ringwald}},\ }\href
  {\doibase 10.1088/1126-6708/2009/11/027} {\bibfield  {journal} {\bibinfo
  {journal} {J. High Energy Phys}\ }\textbf {\bibinfo {volume} {11}},\ \bibinfo
  {pages} {027} (\bibinfo {year} {2009})}\BibitemShut {NoStop}%
\bibitem [{\citenamefont {Graham}\ \emph {et~al.}(2016)\citenamefont {Graham},
  \citenamefont {Mardon},\ and\ \citenamefont {Rajendran}}]{Graham:2015rva}%
  \BibitemOpen
  \bibfield  {author} {\bibinfo {author} {\bibfnamefont {P.~W.}\ \bibnamefont
  {Graham}}, \bibinfo {author} {\bibfnamefont {J.}~\bibnamefont {Mardon}}, \
  and\ \bibinfo {author} {\bibfnamefont {S.}~\bibnamefont {Rajendran}},\ }\href
  {\doibase 10.1103/PhysRevD.93.103520} {\bibfield  {journal} {\bibinfo
  {journal} {\prd}\ }\textbf {\bibinfo {volume} {93}},\ \bibinfo {pages}
  {103520} (\bibinfo {year} {2016})}\BibitemShut {NoStop}%
\bibitem [{\citenamefont {East}\ and\ \citenamefont
  {Huang}(2022)}]{East:2022rsi}%
  \BibitemOpen
  \bibfield  {author} {\bibinfo {author} {\bibfnamefont {W.~E.}\ \bibnamefont
  {East}}\ and\ \bibinfo {author} {\bibfnamefont {J.}~\bibnamefont {Huang}},\
  }\href {\doibase 10.1007/JHEP12(2022)089} {\bibfield  {journal} {\bibinfo
  {journal} {J. High Energy Phys}\ }\textbf {\bibinfo {volume} {12}},\ \bibinfo
  {pages} {089} (\bibinfo {year} {2022})}\BibitemShut {NoStop}%
\bibitem [{\citenamefont {Knapen}\ \emph {et~al.}(2017)\citenamefont {Knapen},
  \citenamefont {Lin},\ and\ \citenamefont {Zurek}}]{Knapen:2017xzo}%
  \BibitemOpen
  \bibfield  {author} {\bibinfo {author} {\bibfnamefont {S.}~\bibnamefont
  {Knapen}}, \bibinfo {author} {\bibfnamefont {T.}~\bibnamefont {Lin}}, \ and\
  \bibinfo {author} {\bibfnamefont {K.~M.}\ \bibnamefont {Zurek}},\ }\href
  {\doibase 10.1103/PhysRevD.96.115021} {\bibfield  {journal} {\bibinfo
  {journal} {\prd}\ }\textbf {\bibinfo {volume} {96}},\ \bibinfo {pages}
  {115021} (\bibinfo {year} {2017})}\BibitemShut {NoStop}%
\bibitem [{\citenamefont {Pospelov}\ \emph {et~al.}(2008)\citenamefont
  {Pospelov}, \citenamefont {Ritz},\ and\ \citenamefont
  {Voloshin}}]{Pospelov:2008jkf}%
  \BibitemOpen
  \bibfield  {author} {\bibinfo {author} {\bibfnamefont {M.}~\bibnamefont
  {Pospelov}}, \bibinfo {author} {\bibfnamefont {A.}~\bibnamefont {Ritz}}, \
  and\ \bibinfo {author} {\bibfnamefont {M.~B.}\ \bibnamefont {Voloshin}},\
  }\href {\doibase 10.1103/PhysRevD.78.115012} {\bibfield  {journal} {\bibinfo
  {journal} {\prd}\ }\textbf {\bibinfo {volume} {78}},\ \bibinfo {pages}
  {115012} (\bibinfo {year} {2008})}\BibitemShut {NoStop}%
\bibitem [{\citenamefont {An}\ \emph {et~al.}(2013)\citenamefont {An},
  \citenamefont {Pospelov},\ and\ \citenamefont {Pradler}}]{An:2013yua}%
  \BibitemOpen
  \bibfield  {author} {\bibinfo {author} {\bibfnamefont {H.}~\bibnamefont
  {An}}, \bibinfo {author} {\bibfnamefont {M.}~\bibnamefont {Pospelov}}, \ and\
  \bibinfo {author} {\bibfnamefont {J.}~\bibnamefont {Pradler}},\ }\href
  {\doibase 10.1103/PhysRevLett.111.041302} {\bibfield  {journal} {\bibinfo
  {journal} {\prl}\ }\textbf {\bibinfo {volume} {111}},\ \bibinfo {pages}
  {041302} (\bibinfo {year} {2013})}\BibitemShut {NoStop}%
\bibitem [{\citenamefont {Baryakhtar}\ \emph {et~al.}(2017)\citenamefont
  {Baryakhtar}, \citenamefont {Lasenby},\ and\ \citenamefont
  {Teo}}]{Baryakhtar:2017ngi}%
  \BibitemOpen
  \bibfield  {author} {\bibinfo {author} {\bibfnamefont {M.}~\bibnamefont
  {Baryakhtar}}, \bibinfo {author} {\bibfnamefont {R.}~\bibnamefont {Lasenby}},
  \ and\ \bibinfo {author} {\bibfnamefont {M.}~\bibnamefont {Teo}},\ }\href
  {\doibase 10.1103/PhysRevD.96.035019} {\bibfield  {journal} {\bibinfo
  {journal} {\prd}\ }\textbf {\bibinfo {volume} {D96}},\ \bibinfo {pages}
  {035019} (\bibinfo {year} {2017})}\BibitemShut {NoStop}%
\bibitem [{\citenamefont {Baryakhtar}\ \emph {et~al.}(2018)\citenamefont
  {Baryakhtar}, \citenamefont {Huang},\ and\ \citenamefont
  {Lasenby}}]{Baryakhtar:2018doz}%
  \BibitemOpen
  \bibfield  {author} {\bibinfo {author} {\bibfnamefont {M.}~\bibnamefont
  {Baryakhtar}}, \bibinfo {author} {\bibfnamefont {J.}~\bibnamefont {Huang}}, \
  and\ \bibinfo {author} {\bibfnamefont {R.}~\bibnamefont {Lasenby}},\ }\href
  {\doibase 10.1103/PhysRevD.98.035006} {\bibfield  {journal} {\bibinfo
  {journal} {\prd}\ }\textbf {\bibinfo {volume} {98}},\ \bibinfo {pages}
  {035006} (\bibinfo {year} {2018})}\BibitemShut {NoStop}%
\bibitem [{\citenamefont {Barak}\ \emph {et~al.}(2020)\citenamefont {Barak}
  \emph {et~al.}}]{SENSEI:2020dpa}%
  \BibitemOpen
  \bibfield  {author} {\bibinfo {author} {\bibfnamefont {L.}~\bibnamefont
  {Barak}} \emph {et~al.} (\bibinfo {collaboration} {SENSEI}),\ }\href
  {\doibase 10.1103/PhysRevLett.125.171802} {\bibfield  {journal} {\bibinfo
  {journal} {\prl}\ }\textbf {\bibinfo {volume} {125}},\ \bibinfo {pages}
  {171802} (\bibinfo {year} {2020})}\BibitemShut {NoStop}%
\bibitem [{\citenamefont {Hardy}\ and\ \citenamefont
  {Lasenby}(2017)}]{Hardy:2016kme}%
  \BibitemOpen
  \bibfield  {author} {\bibinfo {author} {\bibfnamefont {E.}~\bibnamefont
  {Hardy}}\ and\ \bibinfo {author} {\bibfnamefont {R.}~\bibnamefont
  {Lasenby}},\ }\href {\doibase 10.1007/JHEP02(2017)033} {\bibfield  {journal}
  {\bibinfo  {journal} {J. High Energy Phys}\ }\textbf {\bibinfo {volume}
  {02}},\ \bibinfo {pages} {033} (\bibinfo {year} {2017})}\BibitemShut
  {NoStop}%
\bibitem [{\citenamefont {Lasenby}\ and\ \citenamefont
  {Van~Tilburg}(2021)}]{Lasenby:2020goo}%
  \BibitemOpen
  \bibfield  {author} {\bibinfo {author} {\bibfnamefont {R.}~\bibnamefont
  {Lasenby}}\ and\ \bibinfo {author} {\bibfnamefont {K.}~\bibnamefont
  {Van~Tilburg}},\ }\href {\doibase 10.1103/PhysRevD.104.023020} {\bibfield
  {journal} {\bibinfo  {journal} {\prd}\ }\textbf {\bibinfo {volume} {104}},\
  \bibinfo {pages} {023020} (\bibinfo {year} {2021})}\BibitemShut {NoStop}%
\bibitem [{\citenamefont {Romanenko}\ \emph {et~al.}(2023)\citenamefont
  {Romanenko} \emph {et~al.}}]{Romanenko:2023irv}%
  \BibitemOpen
  \bibfield  {author} {\bibinfo {author} {\bibfnamefont {A.}~\bibnamefont
  {Romanenko}} \emph {et~al.},\ }\href {\doibase
  10.1103/PhysRevLett.130.261801} {\bibfield  {journal} {\bibinfo  {journal}
  {\prl}\ }\textbf {\bibinfo {volume} {130}},\ \bibinfo {pages} {261801}
  (\bibinfo {year} {2023})}\BibitemShut {NoStop}%
\bibitem [{\citenamefont {Chaudhuri}\ \emph {et~al.}(2015)\citenamefont
  {Chaudhuri}, \citenamefont {Graham}, \citenamefont {Irwin}, \citenamefont
  {Mardon}, \citenamefont {Rajendran},\ and\ \citenamefont
  {Zhao}}]{Chaudhuri:2014dla}%
  \BibitemOpen
  \bibfield  {author} {\bibinfo {author} {\bibfnamefont {S.}~\bibnamefont
  {Chaudhuri}}, \bibinfo {author} {\bibfnamefont {P.~W.}\ \bibnamefont
  {Graham}}, \bibinfo {author} {\bibfnamefont {K.}~\bibnamefont {Irwin}},
  \bibinfo {author} {\bibfnamefont {J.}~\bibnamefont {Mardon}}, \bibinfo
  {author} {\bibfnamefont {S.}~\bibnamefont {Rajendran}}, \ and\ \bibinfo
  {author} {\bibfnamefont {Y.}~\bibnamefont {Zhao}},\ }\href {\doibase
  10.1103/PhysRevD.92.075012} {\bibfield  {journal} {\bibinfo  {journal}
  {\prd}\ }\textbf {\bibinfo {volume} {92}},\ \bibinfo {pages} {075012}
  (\bibinfo {year} {2015})}\BibitemShut {NoStop}%
\bibitem [{\citenamefont {Georgi}\ \emph {et~al.}(1983)\citenamefont {Georgi},
  \citenamefont {Ginsparg},\ and\ \citenamefont {Glashow}}]{Georgi:1983sy}%
  \BibitemOpen
  \bibfield  {author} {\bibinfo {author} {\bibfnamefont {H.}~\bibnamefont
  {Georgi}}, \bibinfo {author} {\bibfnamefont {P.~H.}\ \bibnamefont
  {Ginsparg}}, \ and\ \bibinfo {author} {\bibfnamefont {S.~L.}\ \bibnamefont
  {Glashow}},\ }\href {\doibase 10.1038/306765a0} {\bibfield  {journal}
  {\bibinfo  {journal} {Nature}\ }\textbf {\bibinfo {volume} {306}},\ \bibinfo
  {pages} {765} (\bibinfo {year} {1983})}\BibitemShut {NoStop}%
\bibitem [{\citenamefont {Mirizzi}\ \emph {et~al.}(2009)\citenamefont
  {Mirizzi}, \citenamefont {Redondo},\ and\ \citenamefont
  {Sigl}}]{Mirizzi:2009iz}%
  \BibitemOpen
  \bibfield  {author} {\bibinfo {author} {\bibfnamefont {A.}~\bibnamefont
  {Mirizzi}}, \bibinfo {author} {\bibfnamefont {J.}~\bibnamefont {Redondo}}, \
  and\ \bibinfo {author} {\bibfnamefont {G.}~\bibnamefont {Sigl}},\ }\href
  {\doibase 10.1088/1475-7516/2009/03/026} {\bibfield  {journal} {\bibinfo
  {journal} {\jcap}\ }\textbf {\bibinfo {volume} {03}},\ \bibinfo {pages} {026}
  (\bibinfo {year} {2009})}\BibitemShut {NoStop}%
\bibitem [{\citenamefont {Caputo}\ \emph
  {et~al.}(2020{\natexlab{a}})\citenamefont {Caputo}, \citenamefont {Liu},
  \citenamefont {Mishra-Sharma},\ and\ \citenamefont
  {Ruderman}}]{Caputo:2020rnx}%
  \BibitemOpen
  \bibfield  {author} {\bibinfo {author} {\bibfnamefont {A.}~\bibnamefont
  {Caputo}}, \bibinfo {author} {\bibfnamefont {H.}~\bibnamefont {Liu}},
  \bibinfo {author} {\bibfnamefont {S.}~\bibnamefont {Mishra-Sharma}}, \ and\
  \bibinfo {author} {\bibfnamefont {J.~T.}\ \bibnamefont {Ruderman}},\ }\href
  {\doibase 10.1103/PhysRevD.102.103533} {\bibfield  {journal} {\bibinfo
  {journal} {\prd}\ }\textbf {\bibinfo {volume} {102}},\ \bibinfo {pages}
  {103533} (\bibinfo {year} {2020}{\natexlab{a}})}\BibitemShut {NoStop}%
\bibitem [{\citenamefont {Caputo}\ \emph
  {et~al.}(2020{\natexlab{b}})\citenamefont {Caputo}, \citenamefont {Liu},
  \citenamefont {Mishra-Sharma},\ and\ \citenamefont
  {Ruderman}}]{caputo_dark_2020}%
  \BibitemOpen
  \bibfield  {author} {\bibinfo {author} {\bibfnamefont {A.}~\bibnamefont
  {Caputo}}, \bibinfo {author} {\bibfnamefont {H.}~\bibnamefont {Liu}},
  \bibinfo {author} {\bibfnamefont {S.}~\bibnamefont {Mishra-Sharma}}, \ and\
  \bibinfo {author} {\bibfnamefont {J.~T.}\ \bibnamefont {Ruderman}},\ }\href
  {\doibase 10.1103/PhysRevLett.125.221303} {\bibfield  {journal} {\bibinfo
  {journal} {\prl}\ }\textbf {\bibinfo {volume} {125}},\ \bibinfo {pages}
  {221303} (\bibinfo {year} {2020}{\natexlab{b}})}\BibitemShut {NoStop}%
\bibitem [{\citenamefont {Siemonsen}\ \emph {et~al.}(2023)\citenamefont
  {Siemonsen}, \citenamefont {Mondino}, \citenamefont {Egana-Ugrinovic},
  \citenamefont {Huang}, \citenamefont {Baryakhtar},\ and\ \citenamefont
  {East}}]{Siemonsen:2022ivj}%
  \BibitemOpen
  \bibfield  {author} {\bibinfo {author} {\bibfnamefont {N.}~\bibnamefont
  {Siemonsen}}, \bibinfo {author} {\bibfnamefont {C.}~\bibnamefont {Mondino}},
  \bibinfo {author} {\bibfnamefont {D.}~\bibnamefont {Egana-Ugrinovic}},
  \bibinfo {author} {\bibfnamefont {J.}~\bibnamefont {Huang}}, \bibinfo
  {author} {\bibfnamefont {M.}~\bibnamefont {Baryakhtar}}, \ and\ \bibinfo
  {author} {\bibfnamefont {W.~E.}\ \bibnamefont {East}},\ }\href {\doibase
  10.1103/PhysRevD.107.075025} {\bibfield  {journal} {\bibinfo  {journal}
  {\prd}\ }\textbf {\bibinfo {volume} {107}},\ \bibinfo {pages} {075025}
  (\bibinfo {year} {2023})}\BibitemShut {NoStop}%
\bibitem [{\citenamefont {Siemonsen}\ and\ \citenamefont
  {East}(2020)}]{Siemonsen:2019ebd}%
  \BibitemOpen
  \bibfield  {author} {\bibinfo {author} {\bibfnamefont {N.}~\bibnamefont
  {Siemonsen}}\ and\ \bibinfo {author} {\bibfnamefont {W.~E.}\ \bibnamefont
  {East}},\ }\href {\doibase 10.1103/PhysRevD.101.024019} {\bibfield  {journal}
  {\bibinfo  {journal} {\prd}\ }\textbf {\bibinfo {volume} {101}},\ \bibinfo
  {pages} {024019} (\bibinfo {year} {2020})}\BibitemShut {NoStop}%
\bibitem [{\citenamefont {Wolfenstein}(1978)}]{Wolfenstein:1977ue}%
  \BibitemOpen
  \bibfield  {author} {\bibinfo {author} {\bibfnamefont {L.}~\bibnamefont
  {Wolfenstein}},\ }\href {\doibase 10.1103/PhysRevD.17.2369} {\bibfield
  {journal} {\bibinfo  {journal} {\prd}\ }\textbf {\bibinfo {volume} {17}},\
  \bibinfo {pages} {2369} (\bibinfo {year} {1978})}\BibitemShut {NoStop}%
\bibitem [{\citenamefont {Mikheyev}\ and\ \citenamefont
  {Smirnov}(1985)}]{Mikheyev:1985zog}%
  \BibitemOpen
  \bibfield  {author} {\bibinfo {author} {\bibfnamefont {S.~P.}\ \bibnamefont
  {Mikheyev}}\ and\ \bibinfo {author} {\bibfnamefont {A.~Y.}\ \bibnamefont
  {Smirnov}},\ }\href@noop {} {\bibfield  {journal} {\bibinfo  {journal} {Sov.
  J. Nucl. Phys.}\ }\textbf {\bibinfo {volume} {42}},\ \bibinfo {pages} {913}
  (\bibinfo {year} {1985})}\BibitemShut {NoStop}%
\bibitem [{\citenamefont {{Fixsen}}\ \emph {et~al.}(1996)\citenamefont
  {{Fixsen}}, \citenamefont {{Cheng}}, \citenamefont {{Gales}}, \citenamefont
  {{Mather}}, \citenamefont {{Shafer}},\ and\ \citenamefont
  {{Wright}}}]{1996ApJ...473..576F}%
  \BibitemOpen
  \bibfield  {author} {\bibinfo {author} {\bibfnamefont {D.~J.}\ \bibnamefont
  {{Fixsen}}}, \bibinfo {author} {\bibfnamefont {E.~S.}\ \bibnamefont
  {{Cheng}}}, \bibinfo {author} {\bibfnamefont {J.~M.}\ \bibnamefont
  {{Gales}}}, \bibinfo {author} {\bibfnamefont {J.~C.}\ \bibnamefont
  {{Mather}}}, \bibinfo {author} {\bibfnamefont {R.~A.}\ \bibnamefont
  {{Shafer}}}, \ and\ \bibinfo {author} {\bibfnamefont {E.~L.}\ \bibnamefont
  {{Wright}}},\ }\href {\doibase 10.1086/178173} {\bibfield  {journal}
  {\bibinfo  {journal} {\apj}\ }\textbf {\bibinfo {volume} {473}},\ \bibinfo
  {pages} {576} (\bibinfo {year} {1996})}\BibitemShut {NoStop}%
\bibitem [{\citenamefont {Mehta}\ and\ \citenamefont
  {Mukherjee}(2024)}]{Mehta:2024pdz}%
  \BibitemOpen
  \bibfield  {author} {\bibinfo {author} {\bibfnamefont {H.}~\bibnamefont
  {Mehta}}\ and\ \bibinfo {author} {\bibfnamefont {S.}~\bibnamefont
  {Mukherjee}},\ }\href@noop {} {\  (\bibinfo {year} {2024})},\ \Eprint
  {http://arxiv.org/abs/2405.08879} {arXiv:2405.08879 [astro-ph.CO]}
  \BibitemShut {NoStop}%
\bibitem [{\citenamefont {Aghanim}\ \emph
  {et~al.}(2020{\natexlab{a}})\citenamefont {Aghanim} \emph
  {et~al.}}]{2020A&A...641A...1P}%
  \BibitemOpen
  \bibfield  {author} {\bibinfo {author} {\bibfnamefont {N.}~\bibnamefont
  {Aghanim}} \emph {et~al.} (\bibinfo {collaboration} {Planck collaboration}),\
  }\href {\doibase 10.1051/0004-6361/201833880} {\bibfield  {journal} {\bibinfo
   {journal} {\aap}\ }\textbf {\bibinfo {volume} {641}},\ \bibinfo {pages} {A1}
  (\bibinfo {year} {2020}{\natexlab{a}})}\BibitemShut {NoStop}%
\bibitem [{\citenamefont {Lang}(2014)}]{Lang_2014}%
  \BibitemOpen
  \bibfield  {author} {\bibinfo {author} {\bibfnamefont {D.}~\bibnamefont
  {Lang}},\ }\href {\doibase 10.1088/0004-6256/147/5/108} {\bibfield  {journal}
  {\bibinfo  {journal} {\aj}\ }\textbf {\bibinfo {volume} {147}},\ \bibinfo
  {pages} {108} (\bibinfo {year} {2014})}\BibitemShut {NoStop}%
\bibitem [{\citenamefont {Meisner}\ \emph
  {et~al.}(2017{\natexlab{a}})\citenamefont {Meisner}, \citenamefont {Lang},\
  and\ \citenamefont {Schlegel}}]{Meisner_2017}%
  \BibitemOpen
  \bibfield  {author} {\bibinfo {author} {\bibfnamefont {A.~M.}\ \bibnamefont
  {Meisner}}, \bibinfo {author} {\bibfnamefont {D.}~\bibnamefont {Lang}}, \
  and\ \bibinfo {author} {\bibfnamefont {D.~J.}\ \bibnamefont {Schlegel}},\
  }\href {\doibase 10.3847/1538-3881/153/1/38} {\bibfield  {journal} {\bibinfo
  {journal} {\aj}\ }\textbf {\bibinfo {volume} {153}},\ \bibinfo {pages} {38}
  (\bibinfo {year} {2017}{\natexlab{a}})}\BibitemShut {NoStop}%
\bibitem [{\citenamefont {Meisner}\ \emph
  {et~al.}(2017{\natexlab{b}})\citenamefont {Meisner}, \citenamefont {Lang},\
  and\ \citenamefont {Schlegel}}]{Meisner_2017a}%
  \BibitemOpen
  \bibfield  {author} {\bibinfo {author} {\bibfnamefont {A.~M.}\ \bibnamefont
  {Meisner}}, \bibinfo {author} {\bibfnamefont {D.}~\bibnamefont {Lang}}, \
  and\ \bibinfo {author} {\bibfnamefont {D.~J.}\ \bibnamefont {Schlegel}},\
  }\href {\doibase 10.3847/1538-3881/aa894e} {\bibfield  {journal} {\bibinfo
  {journal} {\aj}\ }\textbf {\bibinfo {volume} {154}},\ \bibinfo {pages} {161}
  (\bibinfo {year} {2017}{\natexlab{b}})}\BibitemShut {NoStop}%
\bibitem [{\citenamefont {{Schlafly}}\ \emph {et~al.}(2019)\citenamefont
  {{Schlafly}}, \citenamefont {{Meisner}},\ and\ \citenamefont
  {{Green}}}]{2019ApJS..240...30S}%
  \BibitemOpen
  \bibfield  {author} {\bibinfo {author} {\bibfnamefont {E.~F.}\ \bibnamefont
  {{Schlafly}}}, \bibinfo {author} {\bibfnamefont {A.~M.}\ \bibnamefont
  {{Meisner}}}, \ and\ \bibinfo {author} {\bibfnamefont {G.~M.}\ \bibnamefont
  {{Green}}},\ }\href {\doibase 10.3847/1538-4365/aafbea} {\bibfield  {journal}
  {\bibinfo  {journal} {\apjs}\ }\textbf {\bibinfo {volume} {240}},\ \bibinfo
  {pages} {30} (\bibinfo {year} {2019})}\BibitemShut {NoStop}%
\bibitem [{\citenamefont {{Krolewski}}\ \emph {et~al.}(2020)\citenamefont
  {{Krolewski}}, \citenamefont {{Ferraro}}, \citenamefont {{Schlafly}},\ and\
  \citenamefont {{White}}}]{2020JCAP...05..047K}%
  \BibitemOpen
  \bibfield  {author} {\bibinfo {author} {\bibfnamefont {A.}~\bibnamefont
  {{Krolewski}}}, \bibinfo {author} {\bibfnamefont {S.}~\bibnamefont
  {{Ferraro}}}, \bibinfo {author} {\bibfnamefont {E.~F.}\ \bibnamefont
  {{Schlafly}}}, \ and\ \bibinfo {author} {\bibfnamefont {M.}~\bibnamefont
  {{White}}},\ }\href {\doibase 10.1088/1475-7516/2020/05/047} {\bibfield
  {journal} {\bibinfo  {journal} {\jcap}\ }\textbf {\bibinfo {volume} {2020}},\
  \bibinfo {pages} {047} (\bibinfo {year} {2020})}\BibitemShut {NoStop}%
\bibitem [{\citenamefont {{Wright}}\ \emph {et~al.}(2010)\citenamefont
  {{Wright}} \emph {et~al.}}]{2010AJ....140.1868W}%
  \BibitemOpen
  \bibfield  {author} {\bibinfo {author} {\bibfnamefont {E.~L.}\ \bibnamefont
  {{Wright}}} \emph {et~al.},\ }\href {\doibase 10.1088/0004-6256/140/6/1868}
  {\bibfield  {journal} {\bibinfo  {journal} {\aj}\ }\textbf {\bibinfo {volume}
  {140}},\ \bibinfo {pages} {1868} (\bibinfo {year} {2010})}\BibitemShut
  {NoStop}%
\bibitem [{\citenamefont {{Mainzer}}\ \emph {et~al.}(2011)\citenamefont
  {{Mainzer}} \emph {et~al.}}]{2011ApJ...731...53M}%
  \BibitemOpen
  \bibfield  {author} {\bibinfo {author} {\bibfnamefont {A.}~\bibnamefont
  {{Mainzer}}} \emph {et~al.},\ }\href {\doibase 10.1088/0004-637X/731/1/53}
  {\bibfield  {journal} {\bibinfo  {journal} {\apj}\ }\textbf {\bibinfo
  {volume} {731}},\ \bibinfo {pages} {53} (\bibinfo {year} {2011})}\BibitemShut
  {NoStop}%
\bibitem [{\citenamefont {Aguirre}\ \emph {et~al.}(2019)\citenamefont {Aguirre}
  \emph {et~al.}}]{Ade:2018sbj}%
  \BibitemOpen
  \bibfield  {author} {\bibinfo {author} {\bibfnamefont {J.}~\bibnamefont
  {Aguirre}} \emph {et~al.} (\bibinfo {collaboration} {Simons Observatory}),\
  }\href {\doibase 10.1088/1475-7516/2019/02/056} {\bibfield  {journal}
  {\bibinfo  {journal} {\jcap}\ }\textbf {\bibinfo {volume} {1902}},\ \bibinfo
  {pages} {056} (\bibinfo {year} {2019})}\BibitemShut {NoStop}%
\bibitem [{\citenamefont {Abazajian}\ \emph {et~al.}(2019)\citenamefont
  {Abazajian} \emph {et~al.}}]{Abazajian:2019eic}%
  \BibitemOpen
  \bibfield  {author} {\bibinfo {author} {\bibfnamefont {K.}~\bibnamefont
  {Abazajian}} \emph {et~al.},\ }\href@noop {} {\  (\bibinfo {year} {2019})},\
  \Eprint {http://arxiv.org/abs/1907.04473} {arXiv:1907.04473 [astro-ph.IM]}
  \BibitemShut {NoStop}%
\bibitem [{\citenamefont {Abell}\ \emph {et~al.}(2009)\citenamefont {Abell}
  \emph {et~al.}}]{LSSTScienceCollaboration2009}%
  \BibitemOpen
  \bibfield  {author} {\bibinfo {author} {\bibfnamefont {P.~A.}\ \bibnamefont
  {Abell}} \emph {et~al.} (\bibinfo {collaboration} {LSST collaboration}),\
  }\href@noop {} {\bibfield  {journal} {\bibinfo  {journal} {arXiv e-prints}\ }
  (\bibinfo {year} {2009})},\ \Eprint {http://arxiv.org/abs/0912.0201}
  {0912.0201} \BibitemShut {NoStop}%
\bibitem [{\citenamefont {Aghamousa}\ \emph {et~al.}(2016)\citenamefont
  {Aghamousa} \emph {et~al.}}]{DESI:2016fyo}%
  \BibitemOpen
  \bibfield  {author} {\bibinfo {author} {\bibfnamefont {A.}~\bibnamefont
  {Aghamousa}} \emph {et~al.} (\bibinfo {collaboration} {DESI collaboration}),\
  }\href@noop {} {\bibfield  {journal} {\bibinfo  {journal} {arXiv e-prints}\ }
  (\bibinfo {year} {2016})},\ \Eprint {http://arxiv.org/abs/1611.00036}
  {1611.00036} \BibitemShut {NoStop}%
\bibitem [{\citenamefont {Dor\'e}\ \emph {et~al.}(2014)\citenamefont {Dor\'e}
  \emph {et~al.}}]{SPHEREx:2014bgr}%
  \BibitemOpen
  \bibfield  {author} {\bibinfo {author} {\bibfnamefont {O.}~\bibnamefont
  {Dor\'e}} \emph {et~al.} (\bibinfo {collaboration} {SPHEREx}),\ }\href@noop
  {} {\bibfield  {journal} {\bibinfo  {journal} {arXiv e-prints}\ } (\bibinfo
  {year} {2014})},\ \Eprint {http://arxiv.org/abs/1412.4872} {1412.4872}
  \BibitemShut {NoStop}%
\bibitem [{\citenamefont {D'Amico}\ and\ \citenamefont
  {Kaloper}(2015)}]{DAmico:2015snf}%
  \BibitemOpen
  \bibfield  {author} {\bibinfo {author} {\bibfnamefont {G.}~\bibnamefont
  {D'Amico}}\ and\ \bibinfo {author} {\bibfnamefont {N.}~\bibnamefont
  {Kaloper}},\ }\href {\doibase 10.1103/PhysRevD.91.085015} {\bibfield
  {journal} {\bibinfo  {journal} {Phys. Rev. D}\ }\textbf {\bibinfo {volume}
  {91}},\ \bibinfo {pages} {085015} (\bibinfo {year} {2015})},\ \Eprint
  {http://arxiv.org/abs/1501.01642} {arXiv:1501.01642 [astro-ph.CO]}
  \BibitemShut {NoStop}%
\bibitem [{\citenamefont {{Fixsen}}(2009)}]{2009ApJ...707..916F}%
  \BibitemOpen
  \bibfield  {author} {\bibinfo {author} {\bibfnamefont {D.~J.}\ \bibnamefont
  {{Fixsen}}},\ }\href {\doibase 10.1088/0004-637X/707/2/916} {\bibfield
  {journal} {\bibinfo  {journal} {\apj}\ }\textbf {\bibinfo {volume} {707}},\
  \bibinfo {pages} {916} (\bibinfo {year} {2009})}\BibitemShut {NoStop}%
\bibitem [{\citenamefont {{Cooray}}\ and\ \citenamefont
  {{Sheth}}(2002)}]{2002PhR...372....1C}%
  \BibitemOpen
  \bibfield  {author} {\bibinfo {author} {\bibfnamefont {A.}~\bibnamefont
  {{Cooray}}}\ and\ \bibinfo {author} {\bibfnamefont {R.}~\bibnamefont
  {{Sheth}}},\ }\href {\doibase 10.1016/S0370-1573(02)00276-4} {\bibfield
  {journal} {\bibinfo  {journal} {\physrep}\ }\textbf {\bibinfo {volume}
  {372}},\ \bibinfo {pages} {1} (\bibinfo {year} {2002})}\BibitemShut {NoStop}%
\bibitem [{\citenamefont {{Kusiak}}\ \emph {et~al.}(2022)\citenamefont
  {{Kusiak}}, \citenamefont {{Bolliet}}, \citenamefont {{Krolewski}},\ and\
  \citenamefont {{Hill}}}]{2022PhRvD.106l3517K}%
  \BibitemOpen
  \bibfield  {author} {\bibinfo {author} {\bibfnamefont {A.}~\bibnamefont
  {{Kusiak}}}, \bibinfo {author} {\bibfnamefont {B.}~\bibnamefont {{Bolliet}}},
  \bibinfo {author} {\bibfnamefont {A.}~\bibnamefont {{Krolewski}}}, \ and\
  \bibinfo {author} {\bibfnamefont {J.~C.}\ \bibnamefont {{Hill}}},\ }\href
  {\doibase 10.1103/PhysRevD.106.123517} {\bibfield  {journal} {\bibinfo
  {journal} {\prd}\ }\textbf {\bibinfo {volume} {106}},\ \bibinfo {pages}
  {123517} (\bibinfo {year} {2022})}\BibitemShut {NoStop}%
\bibitem [{\citenamefont {Battaglia}(2016)}]{Battaglia:2016xbi}%
  \BibitemOpen
  \bibfield  {author} {\bibinfo {author} {\bibfnamefont {N.}~\bibnamefont
  {Battaglia}},\ }\href {\doibase 10.1088/1475-7516/2016/08/058} {\bibfield
  {journal} {\bibinfo  {journal} {\jcap}\ }\textbf {\bibinfo {volume} {2016}},\
  \bibinfo {pages} {058} (\bibinfo {year} {2016})}\BibitemShut {NoStop}%
\bibitem [{\citenamefont {Akrami}\ \emph {et~al.}(2020)\citenamefont {Akrami}
  \emph {et~al.}}]{2020A&A...643A..42P}%
  \BibitemOpen
  \bibfield  {author} {\bibinfo {author} {\bibfnamefont {Y.}~\bibnamefont
  {Akrami}} \emph {et~al.} (\bibinfo {collaboration} {Planck collaboration}),\
  }\href {\doibase 10.1051/0004-6361/202038073} {\bibfield  {journal} {\bibinfo
   {journal} {\aap}\ }\textbf {\bibinfo {volume} {643}},\ \bibinfo {pages}
  {A42} (\bibinfo {year} {2020})}\BibitemShut {NoStop}%
\bibitem [{\citenamefont {{McCarthy}}\ and\ \citenamefont
  {{Hill}}(2024{\natexlab{a}})}]{2024PhRvD.109b3528M}%
  \BibitemOpen
  \bibfield  {author} {\bibinfo {author} {\bibfnamefont {F.}~\bibnamefont
  {{McCarthy}}}\ and\ \bibinfo {author} {\bibfnamefont {J.~C.}\ \bibnamefont
  {{Hill}}},\ }\href {\doibase 10.1103/PhysRevD.109.023528} {\bibfield
  {journal} {\bibinfo  {journal} {\prd}\ }\textbf {\bibinfo {volume} {109}},\
  \bibinfo {pages} {023528} (\bibinfo {year} {2024}{\natexlab{a}})}\BibitemShut
  {NoStop}%
\bibitem [{\citenamefont {{Chen}}\ and\ \citenamefont
  {{Wright}}(2009)}]{2009ApJ...694..222C}%
  \BibitemOpen
  \bibfield  {author} {\bibinfo {author} {\bibfnamefont {X.}~\bibnamefont
  {{Chen}}}\ and\ \bibinfo {author} {\bibfnamefont {E.~L.}\ \bibnamefont
  {{Wright}}},\ }\href {\doibase 10.1088/0004-637X/694/1/222} {\bibfield
  {journal} {\bibinfo  {journal} {\apj}\ }\textbf {\bibinfo {volume} {694}},\
  \bibinfo {pages} {222} (\bibinfo {year} {2009})}\BibitemShut {NoStop}%
\bibitem [{\citenamefont {{Remazeilles}}\ \emph {et~al.}(2011)\citenamefont
  {{Remazeilles}}, \citenamefont {{Delabrouille}},\ and\ \citenamefont
  {{Cardoso}}}]{2011MNRAS.410.2481R}%
  \BibitemOpen
  \bibfield  {author} {\bibinfo {author} {\bibfnamefont {M.}~\bibnamefont
  {{Remazeilles}}}, \bibinfo {author} {\bibfnamefont {J.}~\bibnamefont
  {{Delabrouille}}}, \ and\ \bibinfo {author} {\bibfnamefont {J.-F.}\
  \bibnamefont {{Cardoso}}},\ }\href {\doibase
  10.1111/j.1365-2966.2010.17624.x} {\bibfield  {journal} {\bibinfo  {journal}
  {\mnras}\ }\textbf {\bibinfo {volume} {410}},\ \bibinfo {pages} {2481}
  (\bibinfo {year} {2011})}\BibitemShut {NoStop}%
\bibitem [{\citenamefont {{Sachs}}\ and\ \citenamefont
  {{Wolfe}}(1967)}]{1967ApJ...147...73S}%
  \BibitemOpen
  \bibfield  {author} {\bibinfo {author} {\bibfnamefont {R.~K.}\ \bibnamefont
  {{Sachs}}}\ and\ \bibinfo {author} {\bibfnamefont {A.~M.}\ \bibnamefont
  {{Wolfe}}},\ }\href {\doibase 10.1086/148982} {\bibfield  {journal} {\bibinfo
   {journal} {\apj}\ }\textbf {\bibinfo {volume} {147}},\ \bibinfo {pages} {73}
  (\bibinfo {year} {1967})}\BibitemShut {NoStop}%
\bibitem [{\citenamefont {{Kusiak}}\ \emph {et~al.}(2023)\citenamefont
  {{Kusiak}}, \citenamefont {{Surrao}},\ and\ \citenamefont
  {{Hill}}}]{2023PhRvD.108l3501K}%
  \BibitemOpen
  \bibfield  {author} {\bibinfo {author} {\bibfnamefont {A.}~\bibnamefont
  {{Kusiak}}}, \bibinfo {author} {\bibfnamefont {K.~M.}\ \bibnamefont
  {{Surrao}}}, \ and\ \bibinfo {author} {\bibfnamefont {J.~C.}\ \bibnamefont
  {{Hill}}},\ }\href {\doibase 10.1103/PhysRevD.108.123501} {\bibfield
  {journal} {\bibinfo  {journal} {\prd}\ }\textbf {\bibinfo {volume} {108}},\
  \bibinfo {pages} {123501} (\bibinfo {year} {2023})}\BibitemShut {NoStop}%
\bibitem [{\citenamefont {{Kusiak}}\ \emph {et~al.}(2021)\citenamefont
  {{Kusiak}}, \citenamefont {{Bolliet}}, \citenamefont {{Ferraro}},
  \citenamefont {{Hill}},\ and\ \citenamefont
  {{Krolewski}}}]{2021PhRvD.104d3518K}%
  \BibitemOpen
  \bibfield  {author} {\bibinfo {author} {\bibfnamefont {A.}~\bibnamefont
  {{Kusiak}}}, \bibinfo {author} {\bibfnamefont {B.}~\bibnamefont {{Bolliet}}},
  \bibinfo {author} {\bibfnamefont {S.}~\bibnamefont {{Ferraro}}}, \bibinfo
  {author} {\bibfnamefont {J.~C.}\ \bibnamefont {{Hill}}}, \ and\ \bibinfo
  {author} {\bibfnamefont {A.}~\bibnamefont {{Krolewski}}},\ }\href {\doibase
  10.1103/PhysRevD.104.043518} {\bibfield  {journal} {\bibinfo  {journal}
  {\prd}\ }\textbf {\bibinfo {volume} {104}},\ \bibinfo {pages} {043518}
  (\bibinfo {year} {2021})}\BibitemShut {NoStop}%
\bibitem [{\citenamefont {Ibitoye}\ \emph {et~al.}(2022)\citenamefont
  {Ibitoye}, \citenamefont {Tramonte}, \citenamefont {Ma},\ and\ \citenamefont
  {Dai}}]{Ibitoye:2022oot}%
  \BibitemOpen
  \bibfield  {author} {\bibinfo {author} {\bibfnamefont {A.}~\bibnamefont
  {Ibitoye}}, \bibinfo {author} {\bibfnamefont {D.}~\bibnamefont {Tramonte}},
  \bibinfo {author} {\bibfnamefont {Y.-Z.}\ \bibnamefont {Ma}}, \ and\ \bibinfo
  {author} {\bibfnamefont {W.-M.}\ \bibnamefont {Dai}},\ }\href {\doibase
  10.3847/1538-4357/ac7b8c} {\bibfield  {journal} {\bibinfo  {journal} {\apj}\
  }\textbf {\bibinfo {volume} {935}},\ \bibinfo {pages} {18} (\bibinfo {year}
  {2022})}\BibitemShut {NoStop}%
\bibitem [{\citenamefont {Yan}\ \emph {et~al.}(2024)\citenamefont {Yan},
  \citenamefont {Maniyar},\ and\ \citenamefont {van Waerbeke}}]{Yan:2023okq}%
  \BibitemOpen
  \bibfield  {author} {\bibinfo {author} {\bibfnamefont {Z.}~\bibnamefont
  {Yan}}, \bibinfo {author} {\bibfnamefont {A.~S.}\ \bibnamefont {Maniyar}}, \
  and\ \bibinfo {author} {\bibfnamefont {L.}~\bibnamefont {van Waerbeke}},\
  }\href {\doibase 10.1088/1475-7516/2024/05/058} {\bibfield  {journal}
  {\bibinfo  {journal} {\jcap}\ }\textbf {\bibinfo {volume} {05}},\ \bibinfo
  {pages} {058} (\bibinfo {year} {2024})}\BibitemShut {NoStop}%
\bibitem [{\citenamefont {Krolewski}\ and\ \citenamefont
  {Ferraro}(2022)}]{Krolewski:2021znk}%
  \BibitemOpen
  \bibfield  {author} {\bibinfo {author} {\bibfnamefont {A.}~\bibnamefont
  {Krolewski}}\ and\ \bibinfo {author} {\bibfnamefont {S.}~\bibnamefont
  {Ferraro}},\ }\href {\doibase 10.1088/1475-7516/2022/04/033} {\bibfield
  {journal} {\bibinfo  {journal} {\jcap}\ }\textbf {\bibinfo {volume} {04}},\
  \bibinfo {pages} {033} (\bibinfo {year} {2022})}\BibitemShut {NoStop}%
\bibitem [{\citenamefont {{Sunyaev}}\ and\ \citenamefont
  {{Zeldovich}}(1970)}]{1970Ap&SS...7....3S}%
  \BibitemOpen
  \bibfield  {author} {\bibinfo {author} {\bibfnamefont {R.~A.}\ \bibnamefont
  {{Sunyaev}}}\ and\ \bibinfo {author} {\bibfnamefont {Y.~B.}\ \bibnamefont
  {{Zeldovich}}},\ }\href {\doibase 10.1007/BF00653471} {\bibfield  {journal}
  {\bibinfo  {journal} {\apss}\ }\textbf {\bibinfo {volume} {7}},\ \bibinfo
  {pages} {3} (\bibinfo {year} {1970})}\BibitemShut {NoStop}%
\bibitem [{\citenamefont {{Chluba}}\ \emph {et~al.}(2017)\citenamefont
  {{Chluba}}, \citenamefont {{Hill}},\ and\ \citenamefont
  {{Abitbol}}}]{2017MNRAS.472.1195C}%
  \BibitemOpen
  \bibfield  {author} {\bibinfo {author} {\bibfnamefont {J.}~\bibnamefont
  {{Chluba}}}, \bibinfo {author} {\bibfnamefont {J.~C.}\ \bibnamefont
  {{Hill}}}, \ and\ \bibinfo {author} {\bibfnamefont {M.~H.}\ \bibnamefont
  {{Abitbol}}},\ }\href {\doibase 10.1093/mnras/stx1982} {\bibfield  {journal}
  {\bibinfo  {journal} {\mnras}\ }\textbf {\bibinfo {volume} {472}},\ \bibinfo
  {pages} {1195} (\bibinfo {year} {2017})}\BibitemShut {NoStop}%
\bibitem [{\citenamefont {{McCarthy}}\ and\ \citenamefont
  {{Hill}}(2024{\natexlab{b}})}]{2024PhRvD.109b3529M}%
  \BibitemOpen
  \bibfield  {author} {\bibinfo {author} {\bibfnamefont {F.}~\bibnamefont
  {{McCarthy}}}\ and\ \bibinfo {author} {\bibfnamefont {J.~C.}\ \bibnamefont
  {{Hill}}},\ }\href {\doibase 10.1103/PhysRevD.109.023529} {\bibfield
  {journal} {\bibinfo  {journal} {\prd}\ }\textbf {\bibinfo {volume} {109}},\
  \bibinfo {pages} {023529} (\bibinfo {year} {2024}{\natexlab{b}})}\BibitemShut
  {NoStop}%
\bibitem [{\citenamefont {{Krolewski}}\ \emph {et~al.}(2021)\citenamefont
  {{Krolewski}}, \citenamefont {{Ferraro}},\ and\ \citenamefont
  {{White}}}]{2021JCAP...12..028K}%
  \BibitemOpen
  \bibfield  {author} {\bibinfo {author} {\bibfnamefont {A.}~\bibnamefont
  {{Krolewski}}}, \bibinfo {author} {\bibfnamefont {S.}~\bibnamefont
  {{Ferraro}}}, \ and\ \bibinfo {author} {\bibfnamefont {M.}~\bibnamefont
  {{White}}},\ }\href {\doibase 10.1088/1475-7516/2021/12/028} {\bibfield
  {journal} {\bibinfo  {journal} {\jcap}\ }\textbf {\bibinfo {volume} {2021}},\
  \bibinfo {pages} {028} (\bibinfo {year} {2021})}\BibitemShut {NoStop}%
\bibitem [{\citenamefont {Farren}\ \emph {et~al.}(2024)\citenamefont {Farren}
  \emph {et~al.}}]{2023arXiv230905659F}%
  \BibitemOpen
  \bibfield  {author} {\bibinfo {author} {\bibfnamefont {G.~S.}\ \bibnamefont
  {Farren}} \emph {et~al.} (\bibinfo {collaboration} {ACT collaboration}),\
  }\href {\doibase 10.3847/1538-4357/ad31a5} {\bibfield  {journal} {\bibinfo
  {journal} {\apj}\ }\textbf {\bibinfo {volume} {966}},\ \bibinfo {pages} {157}
  (\bibinfo {year} {2024})}\BibitemShut {NoStop}%
\bibitem [{\citenamefont {Farren}\ \emph {et~al.}(2023)\citenamefont {Farren},
  \citenamefont {Sherwin}, \citenamefont {Bolliet}, \citenamefont {Namikawa},
  \citenamefont {Ferraro},\ and\ \citenamefont
  {Krolewski}}]{2023arXiv231104213F}%
  \BibitemOpen
  \bibfield  {author} {\bibinfo {author} {\bibfnamefont {G.~S.}\ \bibnamefont
  {Farren}}, \bibinfo {author} {\bibfnamefont {B.~D.}\ \bibnamefont {Sherwin}},
  \bibinfo {author} {\bibfnamefont {B.}~\bibnamefont {Bolliet}}, \bibinfo
  {author} {\bibfnamefont {T.}~\bibnamefont {Namikawa}}, \bibinfo {author}
  {\bibfnamefont {S.}~\bibnamefont {Ferraro}}, \ and\ \bibinfo {author}
  {\bibfnamefont {A.}~\bibnamefont {Krolewski}},\ }\href@noop {} {\bibfield
  {journal} {\bibinfo  {journal} {arXiv e-prints}\ } (\bibinfo {year}
  {2023})},\ \Eprint {http://arxiv.org/abs/2311.04213} {2311.04213}
  \BibitemShut {NoStop}%
\bibitem [{\citenamefont {Coulton}\ \emph
  {et~al.}(2024{\natexlab{a}})\citenamefont {Coulton} \emph
  {et~al.}}]{2024arXiv240113033C}%
  \BibitemOpen
  \bibfield  {author} {\bibinfo {author} {\bibfnamefont {W.~R.}\ \bibnamefont
  {Coulton}} \emph {et~al.} (\bibinfo {collaboration} {ACT collaboration}),\
  }\href@noop {} {\bibfield  {journal} {\bibinfo  {journal} {arXiv e-prints}\ }
  (\bibinfo {year} {2024}{\natexlab{a}})},\ \Eprint
  {http://arxiv.org/abs/2401.13033} {2401.13033} \BibitemShut {NoStop}%
\bibitem [{\citenamefont {Bloch}\ and\ \citenamefont
  {Johnson}(2024)}]{2024arXiv240500809B}%
  \BibitemOpen
  \bibfield  {author} {\bibinfo {author} {\bibfnamefont {R.}~\bibnamefont
  {Bloch}}\ and\ \bibinfo {author} {\bibfnamefont {M.~C.}\ \bibnamefont
  {Johnson}},\ }\href@noop {} {\bibfield  {journal} {\bibinfo  {journal} {arXiv
  e-prints}\ } (\bibinfo {year} {2024})},\ \Eprint
  {http://arxiv.org/abs/2405.00809} {2405.00809} \BibitemShut {NoStop}%
\bibitem [{\citenamefont {{Hivon}}\ \emph {et~al.}(2002)\citenamefont
  {{Hivon}}, \citenamefont {{G{\'o}rski}}, \citenamefont {{Netterfield}},
  \citenamefont {{Crill}}, \citenamefont {{Prunet}},\ and\ \citenamefont
  {{Hansen}}}]{2002ApJ...567....2H}%
  \BibitemOpen
  \bibfield  {author} {\bibinfo {author} {\bibfnamefont {E.}~\bibnamefont
  {{Hivon}}}, \bibinfo {author} {\bibfnamefont {K.~M.}\ \bibnamefont
  {{G{\'o}rski}}}, \bibinfo {author} {\bibfnamefont {C.~B.}\ \bibnamefont
  {{Netterfield}}}, \bibinfo {author} {\bibfnamefont {B.~P.}\ \bibnamefont
  {{Crill}}}, \bibinfo {author} {\bibfnamefont {S.}~\bibnamefont {{Prunet}}}, \
  and\ \bibinfo {author} {\bibfnamefont {F.}~\bibnamefont {{Hansen}}},\ }\href
  {\doibase 10.1086/338126} {\bibfield  {journal} {\bibinfo  {journal} {\apj}\
  }\textbf {\bibinfo {volume} {567}},\ \bibinfo {pages} {2} (\bibinfo {year}
  {2002})}\BibitemShut {NoStop}%
\bibitem [{\citenamefont {{Alonso}}\ \emph {et~al.}(2019)\citenamefont
  {{Alonso}}, \citenamefont {{Sanchez}}, \citenamefont {{Slosar}},\ and\
  \citenamefont {{LSST Dark Energy Science
  Collaboration}}}]{2019MNRAS.484.4127A}%
  \BibitemOpen
  \bibfield  {author} {\bibinfo {author} {\bibfnamefont {D.}~\bibnamefont
  {{Alonso}}}, \bibinfo {author} {\bibfnamefont {J.}~\bibnamefont {{Sanchez}}},
  \bibinfo {author} {\bibfnamefont {A.}~\bibnamefont {{Slosar}}}, \ and\
  \bibinfo {author} {\bibnamefont {{LSST Dark Energy Science Collaboration}}},\
  }\href {\doibase 10.1093/mnras/stz093} {\bibfield  {journal} {\bibinfo
  {journal} {\mnras}\ }\textbf {\bibinfo {volume} {484}},\ \bibinfo {pages}
  {4127} (\bibinfo {year} {2019})}\BibitemShut {NoStop}%
\bibitem [{\citenamefont {{White}}\ \emph {et~al.}(1997)\citenamefont
  {{White}}, \citenamefont {{Becker}}, \citenamefont {{Helfand}},\ and\
  \citenamefont {{Gregg}}}]{1997ApJ...475..479W}%
  \BibitemOpen
  \bibfield  {author} {\bibinfo {author} {\bibfnamefont {R.~L.}\ \bibnamefont
  {{White}}}, \bibinfo {author} {\bibfnamefont {R.~H.}\ \bibnamefont
  {{Becker}}}, \bibinfo {author} {\bibfnamefont {D.~J.}\ \bibnamefont
  {{Helfand}}}, \ and\ \bibinfo {author} {\bibfnamefont {M.~D.}\ \bibnamefont
  {{Gregg}}},\ }\href {\doibase 10.1086/303564} {\bibfield  {journal} {\bibinfo
   {journal} {\apj}\ }\textbf {\bibinfo {volume} {475}},\ \bibinfo {pages}
  {479} (\bibinfo {year} {1997})}\BibitemShut {NoStop}%
\bibitem [{\citenamefont {Efstathiou}(2006)}]{Efstathiou:2006eb}%
  \BibitemOpen
  \bibfield  {author} {\bibinfo {author} {\bibfnamefont {G.}~\bibnamefont
  {Efstathiou}},\ }\href {\doibase 10.1111/j.1365-2966.2006.10486.x} {\bibfield
   {journal} {\bibinfo  {journal} {\mnras}\ }\textbf {\bibinfo {volume}
  {370}},\ \bibinfo {pages} {343} (\bibinfo {year} {2006})}\BibitemShut
  {NoStop}%
\bibitem [{\citenamefont {Zonca}\ \emph {et~al.}(2019)\citenamefont {Zonca},
  \citenamefont {Singer}, \citenamefont {Lenz}, \citenamefont {Reinecke},
  \citenamefont {Rosset}, \citenamefont {Hivon},\ and\ \citenamefont
  {Gorski}}]{Zonca2019}%
  \BibitemOpen
  \bibfield  {author} {\bibinfo {author} {\bibfnamefont {A.}~\bibnamefont
  {Zonca}}, \bibinfo {author} {\bibfnamefont {L.}~\bibnamefont {Singer}},
  \bibinfo {author} {\bibfnamefont {D.}~\bibnamefont {Lenz}}, \bibinfo {author}
  {\bibfnamefont {M.}~\bibnamefont {Reinecke}}, \bibinfo {author}
  {\bibfnamefont {C.}~\bibnamefont {Rosset}}, \bibinfo {author} {\bibfnamefont
  {E.}~\bibnamefont {Hivon}}, \ and\ \bibinfo {author} {\bibfnamefont
  {K.}~\bibnamefont {Gorski}},\ }\href {\doibase 10.21105/joss.01298}
  {\bibfield  {journal} {\bibinfo  {journal} {Journal of Open Source Software}\
  }\textbf {\bibinfo {volume} {4}},\ \bibinfo {pages} {1298} (\bibinfo {year}
  {2019})}\BibitemShut {NoStop}%
\bibitem [{\citenamefont {{G{\'o}rski}}\ \emph {et~al.}(2005)\citenamefont
  {{G{\'o}rski}}, \citenamefont {{Hivon}}, \citenamefont {{Banday}},
  \citenamefont {{Wandelt}}, \citenamefont {{Hansen}}, \citenamefont
  {{Reinecke}},\ and\ \citenamefont {{Bartelmann}}}]{2005ApJ...622..759G}%
  \BibitemOpen
  \bibfield  {author} {\bibinfo {author} {\bibfnamefont {K.~M.}\ \bibnamefont
  {{G{\'o}rski}}}, \bibinfo {author} {\bibfnamefont {E.}~\bibnamefont
  {{Hivon}}}, \bibinfo {author} {\bibfnamefont {A.~J.}\ \bibnamefont
  {{Banday}}}, \bibinfo {author} {\bibfnamefont {B.~D.}\ \bibnamefont
  {{Wandelt}}}, \bibinfo {author} {\bibfnamefont {F.~K.}\ \bibnamefont
  {{Hansen}}}, \bibinfo {author} {\bibfnamefont {M.}~\bibnamefont
  {{Reinecke}}}, \ and\ \bibinfo {author} {\bibfnamefont {M.}~\bibnamefont
  {{Bartelmann}}},\ }\href {\doibase 10.1086/427976} {\bibfield  {journal}
  {\bibinfo  {journal} {\apj}\ }\textbf {\bibinfo {volume} {622}},\ \bibinfo
  {pages} {759} (\bibinfo {year} {2005})}\BibitemShut {NoStop}%
\bibitem [{\citenamefont {Harris}\ \emph {et~al.}(2020)\citenamefont {Harris}
  \emph {et~al.}}]{harris2020array}%
  \BibitemOpen
  \bibfield  {author} {\bibinfo {author} {\bibfnamefont {C.~R.}\ \bibnamefont
  {Harris}} \emph {et~al.},\ }\href {\doibase 10.1038/s41586-020-2649-2}
  {\bibfield  {journal} {\bibinfo  {journal} {Nature}\ }\textbf {\bibinfo
  {volume} {585}},\ \bibinfo {pages} {357} (\bibinfo {year}
  {2020})}\BibitemShut {NoStop}%
\bibitem [{\citenamefont {{Aramburo-Garcia}}\ \emph {et~al.}(2024)\citenamefont
  {{Aramburo-Garcia}}, \citenamefont {{Bondarenko}}, \citenamefont
  {{Boyarsky}}, \citenamefont {{Kashko}}, \citenamefont {{Pradler}},
  \citenamefont {{Sokolenko}}, \citenamefont {{Kugel}}, \citenamefont
  {{Schaller}},\ and\ \citenamefont {{Schaye}}}]{2024arXiv240505104A}%
  \BibitemOpen
  \bibfield  {author} {\bibinfo {author} {\bibfnamefont {A.}~\bibnamefont
  {{Aramburo-Garcia}}}, \bibinfo {author} {\bibfnamefont {K.}~\bibnamefont
  {{Bondarenko}}}, \bibinfo {author} {\bibfnamefont {A.}~\bibnamefont
  {{Boyarsky}}}, \bibinfo {author} {\bibfnamefont {P.}~\bibnamefont
  {{Kashko}}}, \bibinfo {author} {\bibfnamefont {J.}~\bibnamefont {{Pradler}}},
  \bibinfo {author} {\bibfnamefont {A.}~\bibnamefont {{Sokolenko}}}, \bibinfo
  {author} {\bibfnamefont {R.}~\bibnamefont {{Kugel}}}, \bibinfo {author}
  {\bibfnamefont {M.}~\bibnamefont {{Schaller}}}, \ and\ \bibinfo {author}
  {\bibfnamefont {J.}~\bibnamefont {{Schaye}}},\ }\href@noop {} {\bibfield
  {journal} {\bibinfo  {journal} {arXiv e-prints}\ } (\bibinfo {year}
  {2024})},\ \Eprint {http://arxiv.org/abs/2405.05104} {2405.05104}
  \BibitemShut {NoStop}%
\bibitem [{\citenamefont {{Peacock}}\ and\ \citenamefont
  {{Smith}}(2000)}]{2000MNRAS.318.1144P}%
  \BibitemOpen
  \bibfield  {author} {\bibinfo {author} {\bibfnamefont {J.~A.}\ \bibnamefont
  {{Peacock}}}\ and\ \bibinfo {author} {\bibfnamefont {R.~E.}\ \bibnamefont
  {{Smith}}},\ }\href {\doibase 10.1046/j.1365-8711.2000.03779.x} {\bibfield
  {journal} {\bibinfo  {journal} {\mnras}\ }\textbf {\bibinfo {volume} {318}},\
  \bibinfo {pages} {1144} (\bibinfo {year} {2000})}\BibitemShut {NoStop}%
\bibitem [{\citenamefont {{Navarro}}\ \emph {et~al.}(1996)\citenamefont
  {{Navarro}}, \citenamefont {{Frenk}},\ and\ \citenamefont
  {{White}}}]{1996ApJ462563N}%
  \BibitemOpen
  \bibfield  {author} {\bibinfo {author} {\bibfnamefont {J.~F.}\ \bibnamefont
  {{Navarro}}}, \bibinfo {author} {\bibfnamefont {C.~S.}\ \bibnamefont
  {{Frenk}}}, \ and\ \bibinfo {author} {\bibfnamefont {S.~D.~M.}\ \bibnamefont
  {{White}}},\ }\href {\doibase 10.1086/177173} {\bibfield  {journal} {\bibinfo
   {journal} {\apj}\ }\textbf {\bibinfo {volume} {462}} (\bibinfo {year}
  {1996}),\ 10.1086/177173}\BibitemShut {NoStop}%
\bibitem [{\citenamefont {{Tinker}}\ \emph {et~al.}(2008)\citenamefont
  {{Tinker}}, \citenamefont {{Kravtsov}}, \citenamefont {{Klypin}},
  \citenamefont {{Abazajian}}, \citenamefont {{Warren}}, \citenamefont
  {{Yepes}}, \citenamefont {{Gottl{\"o}ber}},\ and\ \citenamefont
  {{Holz}}}]{tinker2008}%
  \BibitemOpen
  \bibfield  {author} {\bibinfo {author} {\bibfnamefont {J.}~\bibnamefont
  {{Tinker}}}, \bibinfo {author} {\bibfnamefont {A.~V.}\ \bibnamefont
  {{Kravtsov}}}, \bibinfo {author} {\bibfnamefont {A.}~\bibnamefont
  {{Klypin}}}, \bibinfo {author} {\bibfnamefont {K.}~\bibnamefont
  {{Abazajian}}}, \bibinfo {author} {\bibfnamefont {M.}~\bibnamefont
  {{Warren}}}, \bibinfo {author} {\bibfnamefont {G.}~\bibnamefont {{Yepes}}},
  \bibinfo {author} {\bibfnamefont {S.}~\bibnamefont {{Gottl{\"o}ber}}}, \ and\
  \bibinfo {author} {\bibfnamefont {D.~E.}\ \bibnamefont {{Holz}}},\ }\href
  {\doibase 10.1086/591439} {\bibfield  {journal} {\bibinfo  {journal} {\apj}\
  }\textbf {\bibinfo {volume} {688}},\ \bibinfo {pages} {709} (\bibinfo {year}
  {2008})}\BibitemShut {NoStop}%
\bibitem [{\citenamefont {Tinker}\ \emph {et~al.}(2010)\citenamefont {Tinker},
  \citenamefont {Robertson}, \citenamefont {Kravtsov}, \citenamefont {Klypin},
  \citenamefont {Warren}, \citenamefont {Yepes},\ and\ \citenamefont
  {Gottl{\"o}ber}}]{Tinker_2010}%
  \BibitemOpen
  \bibfield  {author} {\bibinfo {author} {\bibfnamefont {J.~L.}\ \bibnamefont
  {Tinker}}, \bibinfo {author} {\bibfnamefont {B.~E.}\ \bibnamefont
  {Robertson}}, \bibinfo {author} {\bibfnamefont {A.~V.}\ \bibnamefont
  {Kravtsov}}, \bibinfo {author} {\bibfnamefont {A.}~\bibnamefont {Klypin}},
  \bibinfo {author} {\bibfnamefont {M.~S.}\ \bibnamefont {Warren}}, \bibinfo
  {author} {\bibfnamefont {G.}~\bibnamefont {Yepes}}, \ and\ \bibinfo {author}
  {\bibfnamefont {S.}~\bibnamefont {Gottl{\"o}ber}},\ }\href {\doibase
  10.1088/0004-637X/724/2/878} {\bibfield  {journal} {\bibinfo  {journal}
  {\apj}\ }\textbf {\bibinfo {volume} {724}},\ \bibinfo {pages} {878} (\bibinfo
  {year} {2010})}\BibitemShut {NoStop}%
\bibitem [{\citenamefont {Bhattacharya}\ \emph {et~al.}(2013)\citenamefont
  {Bhattacharya}, \citenamefont {Habib}, \citenamefont {Heitmann},\ and\
  \citenamefont {Vikhlinin}}]{Bhattacharya_2013}%
  \BibitemOpen
  \bibfield  {author} {\bibinfo {author} {\bibfnamefont {S.}~\bibnamefont
  {Bhattacharya}}, \bibinfo {author} {\bibfnamefont {S.}~\bibnamefont {Habib}},
  \bibinfo {author} {\bibfnamefont {K.}~\bibnamefont {Heitmann}}, \ and\
  \bibinfo {author} {\bibfnamefont {A.}~\bibnamefont {Vikhlinin}},\ }\href
  {\doibase 10.1088/0004-637X/766/1/32} {\bibfield  {journal} {\bibinfo
  {journal} {\apj}\ }\textbf {\bibinfo {volume} {766}},\ \bibinfo {pages} {32}
  (\bibinfo {year} {2013})}\BibitemShut {NoStop}%
\bibitem [{\citenamefont {Aghanim}\ \emph
  {et~al.}(2020{\natexlab{b}})\citenamefont {Aghanim} \emph
  {et~al.}}]{Planckcollab_cosmoparams}%
  \BibitemOpen
  \bibfield  {author} {\bibinfo {author} {\bibfnamefont {N.}~\bibnamefont
  {Aghanim}} \emph {et~al.} (\bibinfo {collaboration} {Planck collaboration}),\
  }\href {\doibase 10.1051/0004-6361/201833910} {\bibfield  {journal} {\bibinfo
   {journal} {\aap}\ }\textbf {\bibinfo {volume} {641}},\ \bibinfo {pages} {A6}
  (\bibinfo {year} {2020}{\natexlab{b}})},\ \bibinfo {note} {[Erratum:
  Astron.Astrophys. 652, C4 (2021)]}\BibitemShut {NoStop}%
\bibitem [{\citenamefont {Narcowich}\ \emph {et~al.}(2006)\citenamefont
  {Narcowich}, \citenamefont {Petrushev},\ and\ \citenamefont
  {Ward}}]{doi:10.1137/040614359}%
  \BibitemOpen
  \bibfield  {author} {\bibinfo {author} {\bibfnamefont {F.~J.}\ \bibnamefont
  {Narcowich}}, \bibinfo {author} {\bibfnamefont {P.}~\bibnamefont
  {Petrushev}}, \ and\ \bibinfo {author} {\bibfnamefont {J.~D.}\ \bibnamefont
  {Ward}},\ }\href {\doibase 10.1137/040614359} {\bibfield  {journal} {\bibinfo
   {journal} {SIAM Journal on Mathematical Analysis}\ }\textbf {\bibinfo
  {volume} {38}},\ \bibinfo {pages} {574} (\bibinfo {year} {2006})}\BibitemShut
  {NoStop}%
\bibitem [{\citenamefont {{Bennett}}\ \emph {et~al.}(1992)\citenamefont
  {{Bennett}} \emph {et~al.}}]{1992ApJ...396L...7B}%
  \BibitemOpen
  \bibfield  {author} {\bibinfo {author} {\bibfnamefont {C.~L.}\ \bibnamefont
  {{Bennett}}} \emph {et~al.},\ }\href {\doibase 10.1086/186505} {\bibfield
  {journal} {\bibinfo  {journal} {\apjl}\ }\textbf {\bibinfo {volume} {396}},\
  \bibinfo {pages} {L7} (\bibinfo {year} {1992})}\BibitemShut {NoStop}%
\bibitem [{\citenamefont {{Bennett}}\ \emph {et~al.}(2003)\citenamefont
  {{Bennett}} \emph {et~al.}}]{2003ApJS..148...97B}%
  \BibitemOpen
  \bibfield  {author} {\bibinfo {author} {\bibfnamefont {C.~L.}\ \bibnamefont
  {{Bennett}}} \emph {et~al.},\ }\href {\doibase 10.1086/377252} {\bibfield
  {journal} {\bibinfo  {journal} {\apjs}\ }\textbf {\bibinfo {volume} {148}},\
  \bibinfo {pages} {97} (\bibinfo {year} {2003})}\BibitemShut {NoStop}%
\bibitem [{\citenamefont {{Page}}\ \emph {et~al.}(2007)\citenamefont {{Page}}
  \emph {et~al.}}]{2007ApJS..170..335P}%
  \BibitemOpen
  \bibfield  {author} {\bibinfo {author} {\bibfnamefont {L.}~\bibnamefont
  {{Page}}} \emph {et~al.},\ }\href {\doibase 10.1086/513699} {\bibfield
  {journal} {\bibinfo  {journal} {\apjs}\ }\textbf {\bibinfo {volume} {170}},\
  \bibinfo {pages} {335} (\bibinfo {year} {2007})}\BibitemShut {NoStop}%
\bibitem [{\citenamefont {{Delabrouille}}\ \emph {et~al.}(2009)\citenamefont
  {{Delabrouille}}, \citenamefont {{Cardoso}}, \citenamefont {{Le Jeune}},
  \citenamefont {{Betoule}}, \citenamefont {{Fay}},\ and\ \citenamefont
  {{Guilloux}}}]{2009A&A...493..835D}%
  \BibitemOpen
  \bibfield  {author} {\bibinfo {author} {\bibfnamefont {J.}~\bibnamefont
  {{Delabrouille}}}, \bibinfo {author} {\bibfnamefont {J.~F.}\ \bibnamefont
  {{Cardoso}}}, \bibinfo {author} {\bibfnamefont {M.}~\bibnamefont {{Le
  Jeune}}}, \bibinfo {author} {\bibfnamefont {M.}~\bibnamefont {{Betoule}}},
  \bibinfo {author} {\bibfnamefont {G.}~\bibnamefont {{Fay}}}, \ and\ \bibinfo
  {author} {\bibfnamefont {F.}~\bibnamefont {{Guilloux}}},\ }\href {\doibase
  10.1051/0004-6361:200810514} {\bibfield  {journal} {\bibinfo  {journal}
  {\aap}\ }\textbf {\bibinfo {volume} {493}},\ \bibinfo {pages} {835} (\bibinfo
  {year} {2009})}\BibitemShut {NoStop}%
\bibitem [{\citenamefont {{Hill}}\ and\ \citenamefont
  {{Spergel}}(2014)}]{2014JCAP...02..030H}%
  \BibitemOpen
  \bibfield  {author} {\bibinfo {author} {\bibfnamefont {J.~C.}\ \bibnamefont
  {{Hill}}}\ and\ \bibinfo {author} {\bibfnamefont {D.~N.}\ \bibnamefont
  {{Spergel}}},\ }\href {\doibase 10.1088/1475-7516/2014/02/030} {\bibfield
  {journal} {\bibinfo  {journal} {\jcap}\ }\textbf {\bibinfo {volume} {2014}},\
  \bibinfo {pages} {030} (\bibinfo {year} {2014})}\BibitemShut {NoStop}%
\bibitem [{\citenamefont {Aghanim}\ \emph {et~al.}(2016)\citenamefont {Aghanim}
  \emph {et~al.}}]{2016A&A...594A..22P}%
  \BibitemOpen
  \bibfield  {author} {\bibinfo {author} {\bibfnamefont {N.}~\bibnamefont
  {Aghanim}} \emph {et~al.} (\bibinfo {collaboration} {Planck collaboration}),\
  }\href {\doibase 10.1051/0004-6361/201525826} {\bibfield  {journal} {\bibinfo
   {journal} {\aap}\ }\textbf {\bibinfo {volume} {594}},\ \bibinfo {pages}
  {A22} (\bibinfo {year} {2016})}\BibitemShut {NoStop}%
\bibitem [{\citenamefont {Madhavacheril}\ \emph {et~al.}(2020)\citenamefont
  {Madhavacheril} \emph {et~al.}}]{2020PhRvD.102b3534M}%
  \BibitemOpen
  \bibfield  {author} {\bibinfo {author} {\bibfnamefont {M.~S.}\ \bibnamefont
  {Madhavacheril}} \emph {et~al.},\ }\href {\doibase
  10.1103/PhysRevD.102.023534} {\bibfield  {journal} {\bibinfo  {journal}
  {\prd}\ }\textbf {\bibinfo {volume} {102}},\ \bibinfo {pages} {023534}
  (\bibinfo {year} {2020})}\BibitemShut {NoStop}%
\bibitem [{\citenamefont {{Rotti}}\ \emph {et~al.}(2022)\citenamefont
  {{Rotti}}, \citenamefont {{Ravenni}},\ and\ \citenamefont
  {{Chluba}}}]{2022MNRAS.515.5847R}%
  \BibitemOpen
  \bibfield  {author} {\bibinfo {author} {\bibfnamefont {A.}~\bibnamefont
  {{Rotti}}}, \bibinfo {author} {\bibfnamefont {A.}~\bibnamefont {{Ravenni}}},
  \ and\ \bibinfo {author} {\bibfnamefont {J.}~\bibnamefont {{Chluba}}},\
  }\href {\doibase 10.1093/mnras/stac2082} {\bibfield  {journal} {\bibinfo
  {journal} {\mnras}\ }\textbf {\bibinfo {volume} {515}},\ \bibinfo {pages}
  {5847} (\bibinfo {year} {2022})}\BibitemShut {NoStop}%
\bibitem [{\citenamefont {{Chandran}}\ \emph {et~al.}(2023)\citenamefont
  {{Chandran}}, \citenamefont {{Remazeilles}},\ and\ \citenamefont
  {{Barreiro}}}]{2023MNRAS.526.5682C}%
  \BibitemOpen
  \bibfield  {author} {\bibinfo {author} {\bibfnamefont {J.}~\bibnamefont
  {{Chandran}}}, \bibinfo {author} {\bibfnamefont {M.}~\bibnamefont
  {{Remazeilles}}}, \ and\ \bibinfo {author} {\bibfnamefont {R.~B.}\
  \bibnamefont {{Barreiro}}},\ }\href {\doibase 10.1093/mnras/stad3156}
  {\bibfield  {journal} {\bibinfo  {journal} {\mnras}\ }\textbf {\bibinfo
  {volume} {526}},\ \bibinfo {pages} {5682} (\bibinfo {year}
  {2023})}\BibitemShut {NoStop}%
\bibitem [{\citenamefont {Coulton}\ \emph
  {et~al.}(2024{\natexlab{b}})\citenamefont {Coulton} \emph
  {et~al.}}]{2023arXiv230701258C}%
  \BibitemOpen
  \bibfield  {author} {\bibinfo {author} {\bibfnamefont {W.~R.}\ \bibnamefont
  {Coulton}} \emph {et~al.} (\bibinfo {collaboration} {ACT collaboration}),\
  }\href {\doibase 10.1103/PhysRevD.109.063530} {\bibfield  {journal} {\bibinfo
   {journal} {\prd}\ }\textbf {\bibinfo {volume} {109}},\ \bibinfo {pages}
  {063530} (\bibinfo {year} {2024}{\natexlab{b}})}\BibitemShut {NoStop}%
\bibitem [{\citenamefont {Ade}\ \emph {et~al.}(2016)\citenamefont {Ade} \emph
  {et~al.}}]{Planck:2015wtm}%
  \BibitemOpen
  \bibfield  {author} {\bibinfo {author} {\bibfnamefont {P.~A.~R.}\
  \bibnamefont {Ade}} \emph {et~al.} (\bibinfo {collaboration} {Planck}),\
  }\href {\doibase 10.1051/0004-6361/201525809} {\bibfield  {journal} {\bibinfo
   {journal} {Astron. Astrophys.}\ }\textbf {\bibinfo {volume} {594}},\
  \bibinfo {pages} {A4} (\bibinfo {year} {2016})},\ \Eprint
  {http://arxiv.org/abs/1502.01584} {arXiv:1502.01584 [astro-ph.CO]}
  \BibitemShut {NoStop}%
\bibitem [{\citenamefont {Adam}\ \emph
  {et~al.}(2016{\natexlab{a}})\citenamefont {Adam} \emph
  {et~al.}}]{Planck:2015aiq}%
  \BibitemOpen
  \bibfield  {author} {\bibinfo {author} {\bibfnamefont {R.}~\bibnamefont
  {Adam}} \emph {et~al.} (\bibinfo {collaboration} {Planck}),\ }\href {\doibase
  10.1051/0004-6361/201525844} {\bibfield  {journal} {\bibinfo  {journal}
  {Astron. Astrophys.}\ }\textbf {\bibinfo {volume} {594}},\ \bibinfo {pages}
  {A7} (\bibinfo {year} {2016}{\natexlab{a}})},\ \Eprint
  {http://arxiv.org/abs/1502.01586} {arXiv:1502.01586 [astro-ph.IM]}
  \BibitemShut {NoStop}%
\bibitem [{\citenamefont {Zonca}\ \emph {et~al.}(2009)\citenamefont {Zonca}
  \emph {et~al.}}]{Zonca:2009vrg}%
  \BibitemOpen
  \bibfield  {author} {\bibinfo {author} {\bibfnamefont {A.}~\bibnamefont
  {Zonca}} \emph {et~al.},\ }\href {\doibase 10.1088/1748-0221/4/12/T12010}
  {\bibfield  {journal} {\bibinfo  {journal} {JINST}\ }\textbf {\bibinfo
  {volume} {4}},\ \bibinfo {pages} {T12010} (\bibinfo {year} {2009})},\ \Eprint
  {http://arxiv.org/abs/1001.4589} {arXiv:1001.4589 [astro-ph.IM]} \BibitemShut
  {NoStop}%
\bibitem [{\citenamefont {Ade}\ \emph {et~al.}(2014)\citenamefont {Ade} \emph
  {et~al.}}]{Planck:2013wmz}%
  \BibitemOpen
  \bibfield  {author} {\bibinfo {author} {\bibfnamefont {P.~A.~R.}\
  \bibnamefont {Ade}} \emph {et~al.} (\bibinfo {collaboration} {Planck}),\
  }\href {\doibase 10.1051/0004-6361/201321531} {\bibfield  {journal} {\bibinfo
   {journal} {Astron. Astrophys.}\ }\textbf {\bibinfo {volume} {571}},\
  \bibinfo {pages} {A9} (\bibinfo {year} {2014})},\ \Eprint
  {http://arxiv.org/abs/1303.5070} {arXiv:1303.5070 [astro-ph.IM]} \BibitemShut
  {NoStop}%
\bibitem [{\citenamefont {Choi}\ \emph {et~al.}(2020)\citenamefont {Choi} \emph
  {et~al.}}]{ACT:2020frw}%
  \BibitemOpen
  \bibfield  {author} {\bibinfo {author} {\bibfnamefont {S.~K.}\ \bibnamefont
  {Choi}} \emph {et~al.} (\bibinfo {collaboration} {ACT}),\ }\href {\doibase
  10.1088/1475-7516/2020/12/045} {\bibfield  {journal} {\bibinfo  {journal}
  {JCAP}\ }\textbf {\bibinfo {volume} {12}},\ \bibinfo {pages} {045} (\bibinfo
  {year} {2020})},\ \Eprint {http://arxiv.org/abs/2007.07289} {arXiv:2007.07289
  [astro-ph.CO]} \BibitemShut {NoStop}%
\bibitem [{\citenamefont {Adam}\ \emph
  {et~al.}(2016{\natexlab{b}})\citenamefont {Adam} \emph
  {et~al.}}]{2016A&A...586A.133P}%
  \BibitemOpen
  \bibfield  {author} {\bibinfo {author} {\bibfnamefont {R.}~\bibnamefont
  {Adam}} \emph {et~al.} (\bibinfo {collaboration} {Planck collaboration}),\
  }\href {\doibase 10.1051/0004-6361/201425034} {\bibfield  {journal} {\bibinfo
   {journal} {\aap}\ }\textbf {\bibinfo {volume} {586}},\ \bibinfo {pages}
  {A133} (\bibinfo {year} {2016}{\natexlab{b}})}\BibitemShut {NoStop}%
\bibitem [{\citenamefont {{Stein}}\ \emph {et~al.}(2020)\citenamefont
  {{Stein}}, \citenamefont {{Alvarez}}, \citenamefont {{Bond}}, \citenamefont
  {{van Engelen}},\ and\ \citenamefont {{Battaglia}}}]{2020JCAP...10..012S}%
  \BibitemOpen
  \bibfield  {author} {\bibinfo {author} {\bibfnamefont {G.}~\bibnamefont
  {{Stein}}}, \bibinfo {author} {\bibfnamefont {M.~A.}\ \bibnamefont
  {{Alvarez}}}, \bibinfo {author} {\bibfnamefont {J.~R.}\ \bibnamefont
  {{Bond}}}, \bibinfo {author} {\bibfnamefont {A.}~\bibnamefont {{van
  Engelen}}}, \ and\ \bibinfo {author} {\bibfnamefont {N.}~\bibnamefont
  {{Battaglia}}},\ }\href {\doibase 10.1088/1475-7516/2020/10/012} {\bibfield
  {journal} {\bibinfo  {journal} {\jcap}\ }\textbf {\bibinfo {volume} {2020}},\
  \bibinfo {pages} {012} (\bibinfo {year} {2020})}\BibitemShut {NoStop}%
\bibitem [{\citenamefont {{Li}}\ \emph {et~al.}(2022)\citenamefont {{Li}},
  \citenamefont {{Puglisi}}, \citenamefont {{Madhavacheril}},\ and\
  \citenamefont {{Alvarez}}}]{2022JCAP...08..029L}%
  \BibitemOpen
  \bibfield  {author} {\bibinfo {author} {\bibfnamefont {Z.}~\bibnamefont
  {{Li}}}, \bibinfo {author} {\bibfnamefont {G.}~\bibnamefont {{Puglisi}}},
  \bibinfo {author} {\bibfnamefont {M.~S.}\ \bibnamefont {{Madhavacheril}}}, \
  and\ \bibinfo {author} {\bibfnamefont {M.~A.}\ \bibnamefont {{Alvarez}}},\
  }\href {\doibase 10.1088/1475-7516/2022/08/029} {\bibfield  {journal}
  {\bibinfo  {journal} {\jcap}\ }\textbf {\bibinfo {volume} {2022}},\ \bibinfo
  {pages} {029} (\bibinfo {year} {2022})}\BibitemShut {NoStop}%
\bibitem [{\citenamefont {{Navarro}}\ \emph {et~al.}(1997)\citenamefont
  {{Navarro}}, \citenamefont {{Frenk}},\ and\ \citenamefont
  {{White}}}]{1997ApJ...490..493N}%
  \BibitemOpen
  \bibfield  {author} {\bibinfo {author} {\bibfnamefont {J.~F.}\ \bibnamefont
  {{Navarro}}}, \bibinfo {author} {\bibfnamefont {C.~S.}\ \bibnamefont
  {{Frenk}}}, \ and\ \bibinfo {author} {\bibfnamefont {S.~D.~M.}\ \bibnamefont
  {{White}}},\ }\href {\doibase 10.1086/304888} {\bibfield  {journal} {\bibinfo
   {journal} {\apj}\ }\textbf {\bibinfo {volume} {490}},\ \bibinfo {pages}
  {493} (\bibinfo {year} {1997})},\ \Eprint
  {http://arxiv.org/abs/astro-ph/9611107} {arXiv:astro-ph/9611107 [astro-ph]}
  \BibitemShut {NoStop}%
\bibitem [{\citenamefont {{Amodeo}}\ \emph {et~al.}(2021)\citenamefont
  {{Amodeo}} \emph {et~al.}}]{2021PhRvD.103f3514A}%
  \BibitemOpen
  \bibfield  {author} {\bibinfo {author} {\bibfnamefont {S.}~\bibnamefont
  {{Amodeo}}} \emph {et~al.},\ }\href {\doibase 10.1103/PhysRevD.103.063514}
  {\bibfield  {journal} {\bibinfo  {journal} {\prd}\ }\textbf {\bibinfo
  {volume} {103}},\ \bibinfo {eid} {063514} (\bibinfo {year} {2021})},\ \Eprint
  {http://arxiv.org/abs/2009.05558} {arXiv:2009.05558 [astro-ph.CO]}
  \BibitemShut {NoStop}%
\bibitem [{\citenamefont {{Schaan}}\ \emph {et~al.}(2021)\citenamefont
  {{Schaan}} \emph {et~al.}}]{2021PhRvD.103f3513S}%
  \BibitemOpen
  \bibfield  {author} {\bibinfo {author} {\bibfnamefont {E.}~\bibnamefont
  {{Schaan}}} \emph {et~al.},\ }\href {\doibase 10.1103/PhysRevD.103.063513}
  {\bibfield  {journal} {\bibinfo  {journal} {\prd}\ }\textbf {\bibinfo
  {volume} {103}},\ \bibinfo {eid} {063513} (\bibinfo {year} {2021})},\ \Eprint
  {http://arxiv.org/abs/2009.05557} {arXiv:2009.05557 [astro-ph.CO]}
  \BibitemShut {NoStop}%
\end{thebibliography}%
\bibliographystyle{apsrev4-1}

\clearpage

\onecolumngrid
\begin{center}
  \textbf{\large Supplemental Material}\\[.2cm]
  \vspace{0.05in}
  {Fiona McCarthy, Dalila P\^\i{}rvu, J.~Colin Hill, Junwu Huang, Matthew~C.~Johnson, and Keir K.~Rogers}
\end{center}

\twocolumngrid

\setcounter{equation}{0}
\setcounter{figure}{0}
\setcounter{table}{0}
\setcounter{section}{0}
\setcounter{page}{1}
\makeatletter
\renewcommand{\theequation}{S\arabic{equation}}
\renewcommand{\thefigure}{S\arabic{figure}}
\renewcommand{\theHfigure}{S\arabic{figure}}%
\renewcommand{\thetable}{S\arabic{table}}

\onecolumngrid

This Supplemental Material contains supporting material for the main \emph{Letter}, including details about the signal modeling and needlet ILC mapmaking, as well as discussions about the impact of deprojection choices, frequency coverage, sky area, and masks.

\section{Signal Modeling}\label{sec:halomodel_unwISE}

In this work we use the same assumptions to model the DP-induced spectral distortion anisotropies as in Ref.~\cite{Pirvu:2023lch}. We extend the analysis of this work to include the cross-correlation between the dark screening optical depth $\tau^\darksc$ and the halo occupation distributions (HODs)~\cite{2000MNRAS.318.1144P} representing the \textit{unWISE} Blue and Green galaxy catalogs. A similar calculation is performed in~\cite{Mondino:2024rif} for the case of axion-induced dark screening.

The optical depth for photon-dark photon conversion is the sum over all resonance events that take place along each line of sight $\nhat$, as defined in Eq.~\eqref{eq:totalprobHEP}. Within the halo model, this quantity can be factorized into contributions from each halo $i$, which can be described by the product of an isotropic and an anisotropic term:
\begin{equation}
    \tau^{\darksc}(\nhat,\nu) \equiv \sum_i P^{i}_{\gamma \rightarrow \darkph} = \sum_i P(z_i, m_i,\nu) u(\hat{n}_i - \hat{n} | z_i, m_i),
\end{equation}
where we have defined
\begin{equation}
    P(z, m,\nu) =\left.\frac{2 \pi \varepsilon^2 m_{\darkph}^4}{2\pi \nu(z)} \frac{\dd \rho^{\rm gas}}{\dd r}\right|_{r_{\rm res}} ^{-1} \Theta(r_{\rm res }-r_{\rm vir}), \quad \quad
    u(\theta| z, m) = \left[1-\frac{(\chi(z) \theta/ r_{\rm res})^2}{(1+z)^2}\right]^{-1/2}.
\end{equation}
Note that we assume that these quantities are functions of halo mass $m$ and redshift $z$ only, and $\chi(z)$ is comoving distance to $z$. Above, $r_{\rm res}(z,m)$ is the halo-centric radius at which the resonance condition is met and the virial radius $r_{\rm vir}(z,m)$ is taken as the halo boundary, beyond which no conversion takes place. The angle between the halo center and the photon arrival direction is defined as $\theta \simeq |\hat{n}_i - \hat{n}| \leq r_{\rm res}(1+z)/\chi(z)$, such that $\theta \ll 1$. The $\Theta$ function counts the number of resonance crossings per halo; it equals $2$ when $r_{\rm res} < r_{\rm vir}$, since the photon crosses the resonance once as it enters the halo and a second time exiting the halo, and is set to 1 when the photon grazes the resonance surface at $r_{\rm res} = r_{\rm vir}$. Finally, $\rho^{\rm gas}(r,z,m)$ is the radial profile of the ionized gas density within a spherically symmetric halo. We adopt the ``AGN feedback'' profile from Ref.~\cite{Battaglia:2016xbi}, which is obtained from hydrodynamical cosmological simulations, and is defined in detail in Ref.~\cite{Pirvu:2023lch}.

Taking the ensemble average over the halos distributed along the line of sight and performing a harmonic transform, we arrive at the expression for the DP optical depth:
\begin{equation}
    \tau_{\ell}^\darksc(z,m,\nu) = \sqrt{\frac{4 \pi}{2 \ell+1}} P(z,m,\nu) u_{\ell}(z,m), \quad \quad
    u_{\ell}(z, m) \equiv \sqrt{\frac{2\ell+1}{4\pi}}\int \dd^2 \hat{n}\, u(\theta|z,m) \mathcal{P}_{\ell}(\cos\theta),
\end{equation}
where $\mathcal{P}_{\ell}$ is the Legendre polynomial of degree $\ell$. The DP-induced fluctuation in the CMB temperature is
\begin{equation}
    \Tdsc_{\ell}(z,m,\nu) = \frac{1-e^{-x}}{x} T_{\rm CMB} \tau_{\ell}^\darksc(z,m,\nu),
\end{equation}
where $x\equiv 2\pi \nu /T_{\rm CMB}$ and $T_{\rm CMB} \simeq 2.726$ K is the CMB monopole temperature.

The cross-power spectrum between a patchy dark screening map and a galaxy number density map is formally obtained by taking the product of the multipole-space kernels of the dark screening-induced temperature fluctuations $\Tdsc_{\ell}(z,m)$ and a galaxy overdensity field $u_{\ell}^g(z,m)$, weighted by the halo mass function $n(\chi, m)$, defined as the isotropic average halo number density per comoving volume element and halo mass. Within the halo model framework, the cross-correlation is defined as a sum of a 1-halo term, which carries the small-scale dependence on the halo density profile's projected angular shape, and a 2-halo term, proportional to the clustering of halos:
\begin{equation}
\begin{aligned}
    C_\ell^{ g \Tdsc} &= C_\ell^{1-{\rm halo}} + C_\ell^{2-{\rm halo}}, \\
    C_\ell^{1-{\rm halo}} &= \int \dd z \, \frac{\chi(z)^2}{H(z)} \int \dd m \, n(z,m) \Tdsc_{\ell}(z,m) u_\ell^g (z,m), \\
    C_\ell^{2-{\rm halo}} &= \int \dd z \, \frac{\chi(z)^2}{H(z)} \left( \prod_{i=1,2} \int \dd m_i \, n(z,m_i) \, b(z,m_i) \right)  \Tdsc_{\ell} (z,m_1) u_\ell^g (z,m_2) P^{\rm lin}\left(\frac{\ell+\frac{1}{2}}{\chi(z)}, z\right),
\end{aligned}
\end{equation}
where {$H(z)$ is the Hubble rate at $z$}, $P^{\rm lin}$ is the linear matter power spectrum evaluated at redshift $z$ and comoving wavenumber $k = \left( \ell + \frac{1}{2}\right) / \chi(z)$, and $b(z,m)$ is the linear halo bias.

To model how the observed galaxies populate the underlying dark matter halo distribution, we use the HOD described in Ref.~\cite{2022PhRvD.106l3517K}. The galaxy multipole space kernel $u_{\ell}^g (z,m)$ is given by:
\begin{equation}\label{eq:ug_distrib}
    u_{\ell}^g(z, m) = W(z) \bar{n}_g^{-1} \left[ N_c(m) + N_s(m) u_{\ell}^{\rm NFW}(z,m) \right],
\end{equation}
where
\begin{equation}
    \bar{n}_g(z) = \int \dd  m \, n(z,m) \left[ N_c(m) + N_s(m) \right], \quad \quad
    W(z) =  \frac{H(z)}{\chi(z)^2} \frac{\dd N_g}{\dd z},
\end{equation}
and the functions $N_c(m)$ and $N_s(m)$ represent the expectation values for the number of central and satellite galaxies respectively. These functions are parametrized as a function of halo mass $m$ as:
\begin{equation}
    N_c(m) = \frac{1}{2} + \frac{1}{2}\operatorname{erf}\left(\frac{\log m-\log m_{\rm min }}{\sigma_{\log m}}\right), \quad \quad
    N_s(m) = N_c(m) \left( \frac{m}{m_*}\right)^{\alpha_s}.
\end{equation}
Eq.~\eqref{eq:ug_distrib} above defines central galaxies as lying at the center of each halo, while satellites are distributed according to an NFW profile, where $u_{\ell}^{\rm NFW} (z,m)$ is the normalized harmonic transform of the NFW density profile~\cite{1996ApJ462563N}, truncated at $r = \lambda \, r_{\rm 200c}$. Finally, the quantity $\dd N_g / \dd z$ is the redshift distribution of the empirical galaxy sample, normalized to unity. Overall, this HOD model is defined by five free parameters: $\left\{\alpha_{s}, \sigma_{\log m}, m_{\rm min}, m_\star, \lambda \right\}$. The HOD values used in this work are the best-fit set of parameters obtained for the \textit{unWISE} Blue and Green samples in Ref.~\cite{2022PhRvD.106l3517K}.

For the numerical modeling of the template signal we use a modified version of the code \texttt{hmvec}.\footnote{\url{https://github.com/simonsobs/hmvec}} We assume that within the virial radius, the halo density is 178 times the critical density of the universe,~\ie $r_{\rm vir} \equiv r_{178c}$. We use the halo mass function of Ref.~\cite{tinker2008} that fixes the halo bias function~\cite{Tinker_2010}, as well as the concentration-mass relation from Ref.~\cite{Bhattacharya_2013}. To evaluate the integrals, we use $100$ equal redshift bins in the range $z \in \left[0.005, 4\right]$ and $100$ logarithmically-spaced halo mass bins $m \in  10^{11} - 10^{17} \, {\rm M_{\odot}}$. We assume a flat $\Lambda$CDM cosmology, with parameters from the best-fit \emph{Planck} $2018$ model~\cite{Planckcollab_cosmoparams}: $\Omega_{\rm cdm} = 0.1193$, $\Omega_{\rm b} = 0.02242$, $H_0 = 67.66\ {\rm km/s/Mpc}$, $\ln(10^{10} A_{s}) = 3.047$, $n_{s} = 0.9665$.

\section{NILC mapmaking}

In CMB thermodynamic temperature units, the frequency dependence of the DP-induced spectral distortion is
\begin{align}\label{eq.full_DP_SED}
    	\frac{\Delta T(\nu)}{T_{\rm CMB}}
     &=\sum_{t_{\rm res}} \frac{ \varepsilon^2 \, m_{\darkph}^2}{2 \nu (t_{\rm res})} \left| \frac{\dd}{\dd t} \ln m_{\gamma}^2(\vec{x}(t) )\right|^{-1}_{t=t_{\rm res}}\,\left(\frac{1-e^{-2\pi \nu/T_{\rm CMB}}}{2\pi \nu/T_{\rm CMB}}\right)\\   
&\propto \frac{1}{x}
     \left(\frac{1-e^{-x}}{x}\right) \,\label{eq.full2},
\end{align}
where $\varepsilon$ is the kinetic mixing parameter (see Eq.~\eqref{eq:darkphoton}), $T_{\rm CMB} \simeq 2.726$~K is the CMB monopole temperature, and $x \equiv 2\pi \nu/T_{\rm CMB}$ is the redshift-independent dimensionless photon frequency.  We normalize this distortion at a reference frequency of 353~GHz, and thus we denote our NILC reconstruction of this distortion as $\Tdsc_{353 \, {\rm GHz}}$.  The ILC estimate is a linear combination of frequency maps that preserves the spectral energy distribution (SED) of a signal of interest --- in our case, that in Eq.~\eqref{eq.full2} --- while minimizing the variance of the estimated map, given the frequency-frequency covariance matrix, which is estimated directly from the data. For NILC, the covariance matrix is estimated on a needlet frame~\cite{doi:10.1137/040614359}, allowing for simultaneous harmonic (angular)-space and pixel-space localization.  ILC has a long history of applications to CMB data (\eg~\cite{1992ApJ...396L...7B,2003ApJS..148...97B,2007ApJS..170..335P,2009A&A...493..835D,2014JCAP...02..030H,2016A&A...594A..22P,2020PhRvD.102b3534M,2022MNRAS.515.5847R,2023MNRAS.526.5682C,2024PhRvD.109b3528M,2023arXiv230701258C}) to isolate components with well-understood SEDs such as the blackbody CMB temperature anisotropy and the Compton-$y$ distortion induced by the thermal SZ (tSZ) effect~\cite{1970Ap&SS...7....3S}.  We use a constrained ILC~\cite{2009ApJ...694..222C,2011MNRAS.410.2481R} to deproject the SEDs of other sky components that could bias our measurement --- in particular, the tSZ SED; a modified blackbody SED with spectral index $\beta^{\rm CIB}=1.6$ and temperature $T^{\rm{CIB}}=20$~K; and the first derivative of this modified blackbody with respect to $\beta$.  The latter two constraints effectively remove the cosmic infrared background (CIB) signal, following the moment expansion approach of Ref.~\cite{2017MNRAS.472.1195C}.  Lastly, on large scales (the first five of the harmonic needlet scales defined below) we deproject a blackbody SED with temperature $T=2.726$ K~\cite{1996ApJ...473..576F,2009ApJ...707..916F} to remove the blackbody CMB anisotropies, which are correlated on large scales with the galaxy overdensity field due to the late-Universe integrated Sachs-Wolfe (ISW) effect~\cite{1967ApJ...147...73S}.  

\begin{figure}
\includegraphics[width=0.5\columnwidth]{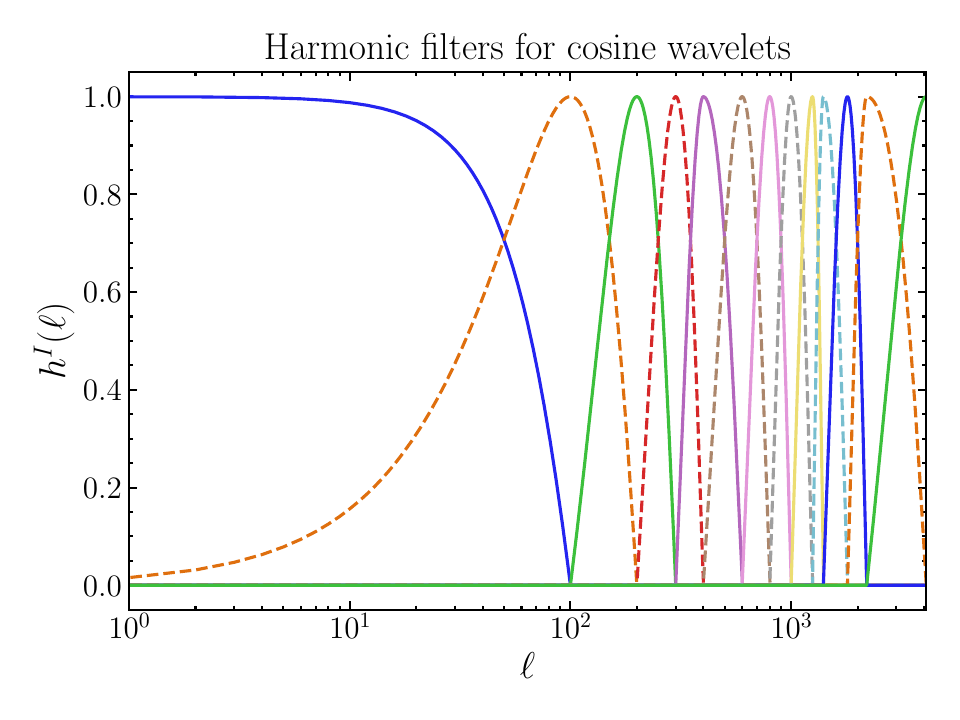}
\caption{The harmonic filters used in the construction of the NILC dark screening map.}\label{fig:needlet_scales}
\end{figure}

We use cosine needlets in our NILC, which have the following harmonic filters:
\begin{equation}
h^I(\ell) = \begin{cases}
\cos\left(\frac{\pi}{2}\frac{\ell^I_{\rm peak}-\ell}{\ell^I_{\rm peak}-\ell^{I-1}_{\rm peak}}\right)&\ell^{I-1}_{\rm peak}\le\ell<\ell^{I}_{\rm peak}\\
\cos\left(\frac{\pi}{2}\frac{\ell-\ell^I_{\rm peak}}{\ell^{I+1}_{\rm peak}-\ell^I_{\rm peak}}\right)&\ell^{I}_{\rm peak}\le\ell<\ell^{I+1}_{\rm peak}\\
0 & {\rm otherwise}.
\end{cases}
\end{equation}
We use 13 harmonic needlet scales with the $\ell_{\rm peak}$ values given by \{0, 100, 200, 300, 400, 600, 800, 1000, 1250, 1400, 1800, 2200, 4097\} (following Ref.~\cite{2023arXiv230701258C}). The harmonic needlet filters are shown in Fig.~\ref{fig:needlet_scales}.

The calculation of the covariance matrix from the data leads to the well-known ``ILC bias''~\cite{2009A&A...493..835D}, where chance fluctuations in the noise can be used to artificially minimize the variance of the inferred signal. This bias can be controlled by measuring the covariance matrix from regions which contain a large enough number of modes. In particular, by using Gaussian real-space kernels with standard deviation given by~\cite{2024PhRvD.109b3528M}
\begin{equation}
\sigma_{{\rm real},I}^2 = 2  \left( \frac{|1+N_{{\rm deproj}}-N_{{\rm freq}}|}{b^{{\rm tol}}N^I_{{\rm modes}}} \right),
\end{equation}
where $N_{{\rm deproj}}$ is the number of components deprojected and $N_{{\rm freq}}$ is the number of frequency channels used in the ILC, we maintain a fractional ILC bias less than $b^{{\rm tol}}$.
Our real-space filters are Gaussian, with FWHMs chosen at each scale to ensure that the fractional ILC bias is 0.01 for an undeprojected map with full frequency coverage.  We drop frequency channels when their beam functions become smaller than a certain threshold depending on the needlet filters; the frequency channels included, as well as the real-space filter sizes, at each needlet scale $I$ are given in Table~\ref{tab:realspace_filters}.  {Note that we treat the \emph{Planck} beams~\cite{Planck:2015wtm,Planck:2015aiq} as isotropic Gaussians, as described in Ref.~\cite{2024PhRvD.109b3528M}.}

\begin{figure*}
\includegraphics[width=0.49\columnwidth]{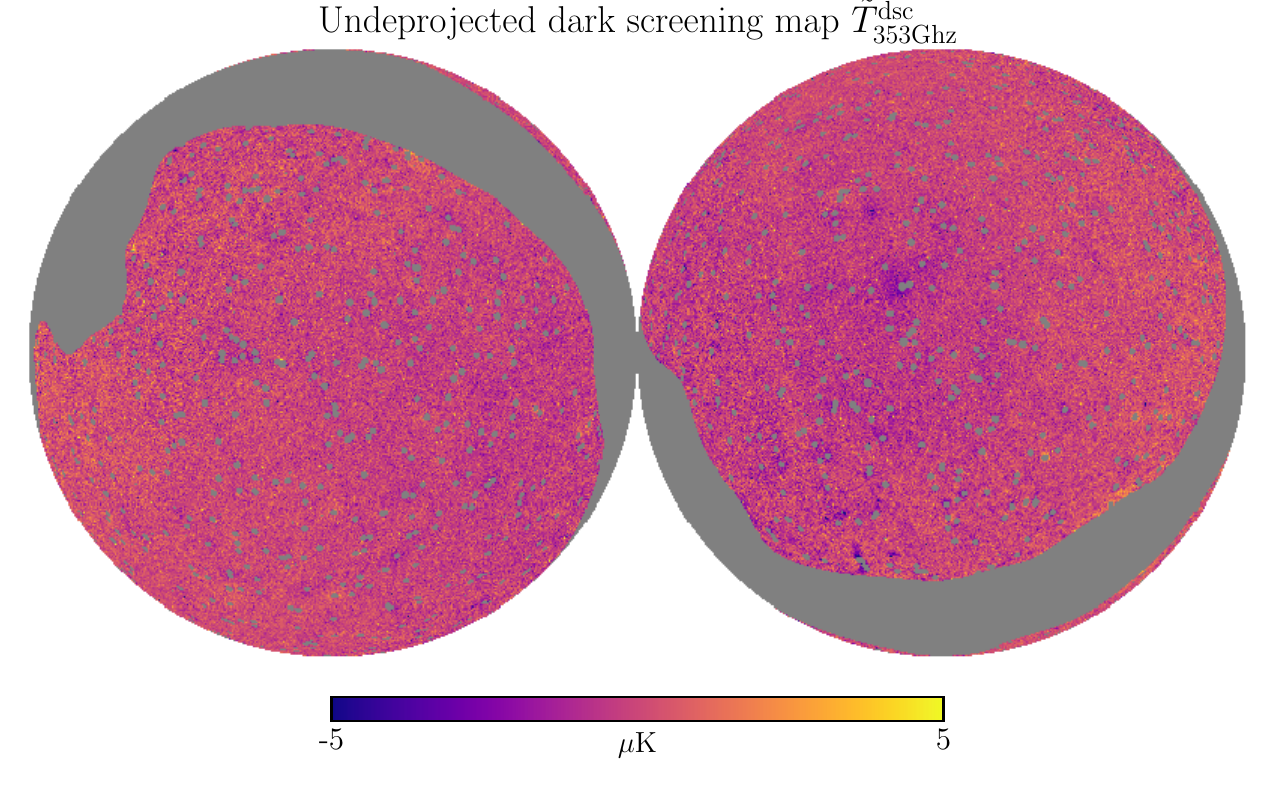}
\includegraphics[width=0.49\columnwidth]{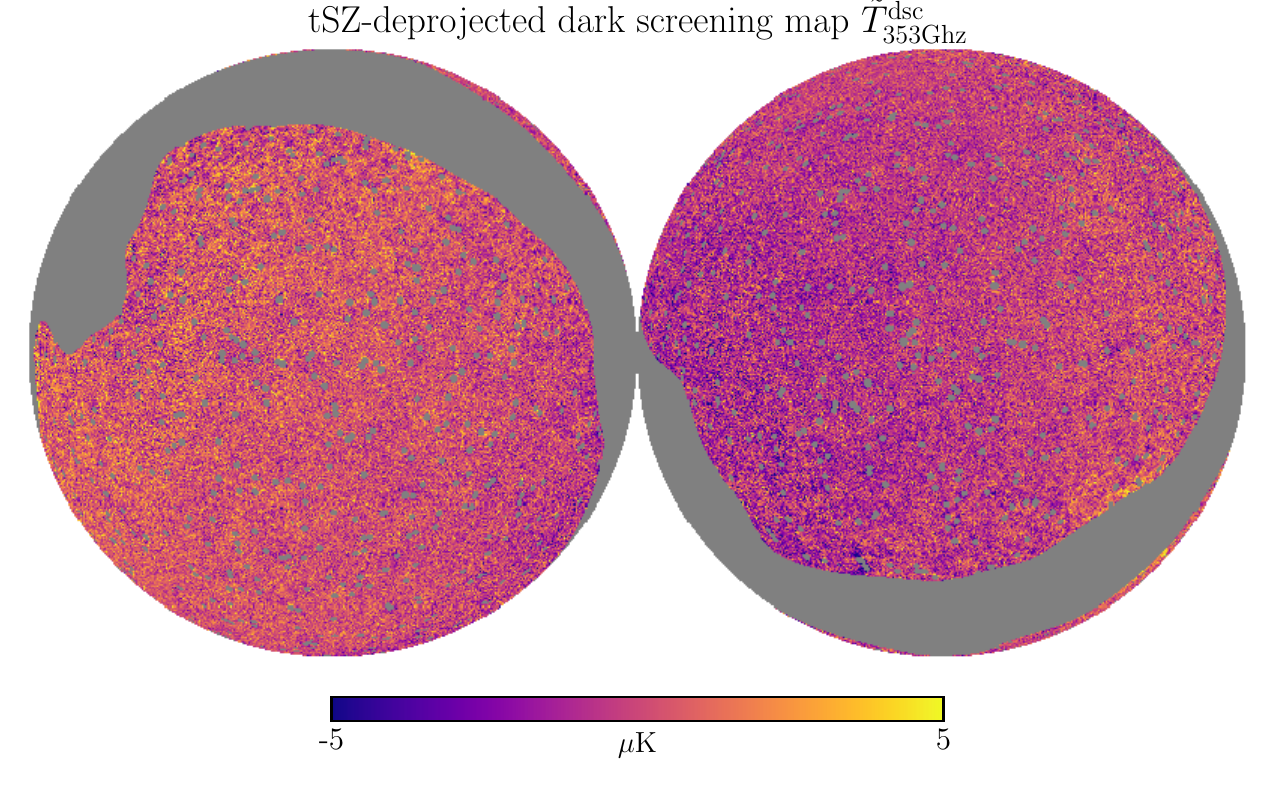}
\includegraphics[width=0.9\columnwidth]{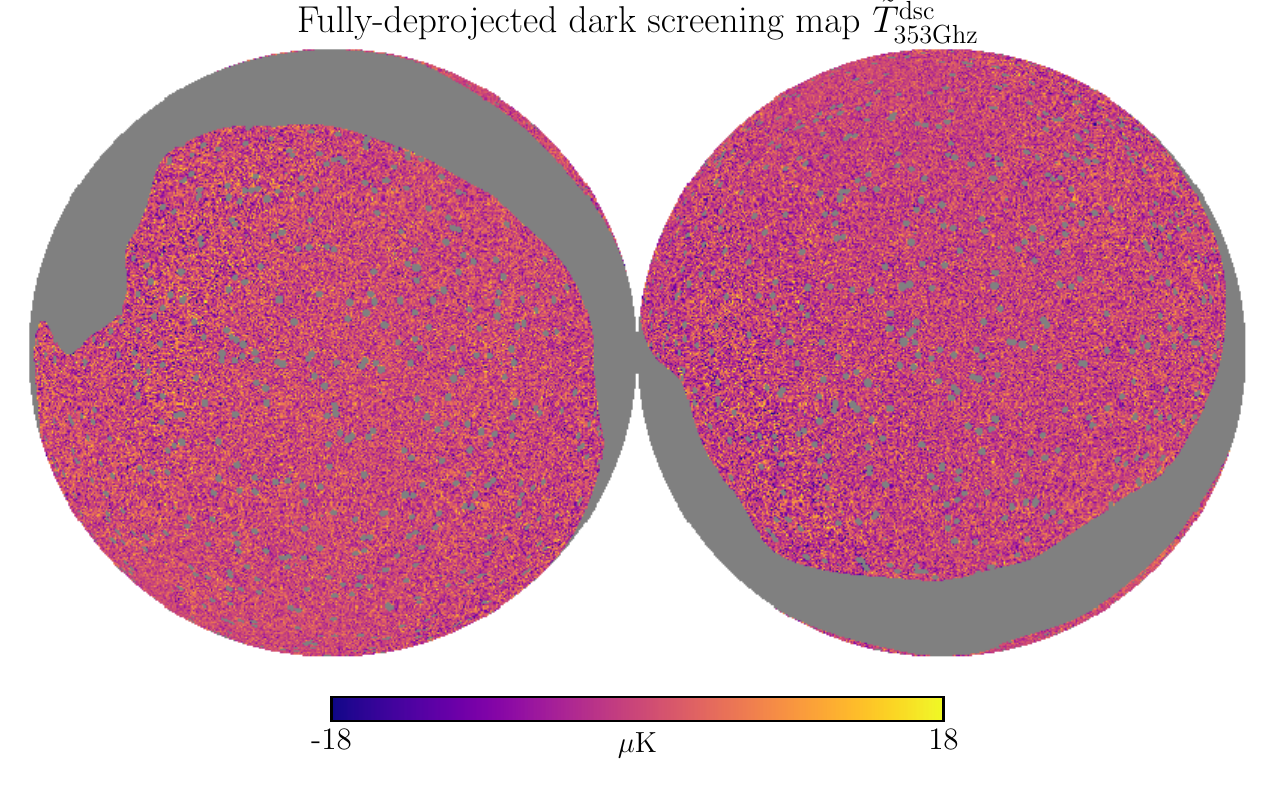}

\caption{NILC dark screening maps, with the undeprojected map on the top left, the tSZ-deprojected map on the top right, and the fully-deprojected map (tSZ + CIB + $\beta^{\rm CIB}$ + large-scale CMB) on the bottom. We show the maps in orthographic projection and equatorial coordinates. The southern Galactic hemisphere is on the left, and the northern Galactic hemisphere is on the right of each figure. Notable features are the removal of very dark negative features in the undeprojected map upon deprojection of the tSZ contamination; these features coincide with the locations of known, bright tSZ sources (\eg the Coma cluster, which is visible by eye in the undeprojected map in the northern hemisphere).  There is also a substantial increase in variance in the fully-deprojected map compared to the undeprojected and tSZ-deprojected maps, due to the known bias-variance tradeoff in constrained NILC constructions.  }\label{fig:darkscreeningmaps}
\end{figure*}

We construct all of our NILC dark screening maps at a reference frequency of 353 GHz. The auto-power spectra of our maps are shown in Fig.~\ref{fig:autopower}. The maps themselves are shown in Fig.~\ref{fig:darkscreeningmaps}.

\begin{table*}
\begin{tabular}{|c|c|c|c|}
\hline
Harmonic scale $I$ & $\ell_{\rm peak}$&Frequencies included & Real-space filter FWHM (degrees)\\\hline
    0 & 0 & {30,44,70,100,143,217,353,545} GHz & 91.8\\\hline
    1 & 100 &{30,44,70,100,143,217,353,545} GHz & 35.6\\\hline
    2 & 200 &{30,44,70,100,143,217,353,545} GHz & 25.2\\\hline
    3 & 300 & {30,44,70,100,143,217,353,545} GHz& 20.6\\\hline
    4 & 400 & {30,44,70,100,143,217,353,545} GHz& 14.1\\\hline
    5 & 600 & {30,44,70,100,143,217,353,545} GHz& 10.3\\\hline
    6 & 800 & {44,70,100,143,217,353,545} GHz& 8.26\\\hline
    7 & 1000 & {70,100,143,217,353,545} GHz& 6.31\\\hline
    8 & 1250 & {70,100,143,217,353,545} GHz& 6.11\\\hline
    9 & 1400 & {70,100,143,217,353,545} GHz& 4.74\\\hline
    10 & 1800 & {70,100,143,217,353,545} GHz & 3.55\\\hline
    11 & 2200 & {143,217,353,545} GHz& 1.34\\\hline
    12 & 4097 &{143,217,353,545} GHz & 1.28\\\hline

\end{tabular}
\caption{The frequencies included at each needlet scale, and the size (FWHM) of the Gaussian real-space filters.}\label{tab:realspace_filters}
\end{table*}

\begin{figure}
    \centering
    \includegraphics[width=0.5\textwidth]{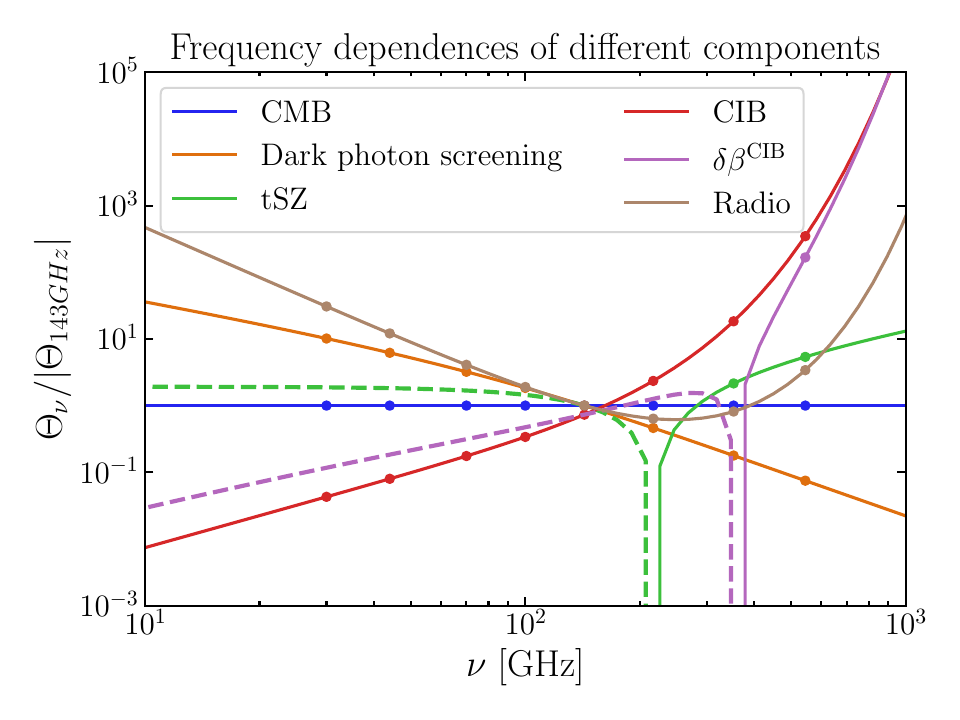}
    \caption{The relative frequency scalings of various components of interest in CMB thermodynamic temperature units. All are normalized to their absolute value at 143 GHz, and negative SEDs are shown with dashed lines (while the dark photon signal is negative, we define the SED in Eq.~\eqref{eq.full_DP_SED} with a positive sign and look for a negative cross-correlation in the data, which is why we show it as positive here). We choose ($T^{\rm CIB}=20 \,\, \mathrm{K}$, $\beta^{\rm CIB}=1.4$) for the CIB modified blackbody SED here, and the radio SED is modeled as a power-law with index $\beta=-0.5$ in specific intensity (as in, e.g.,~\cite{ACT:2020frw}). The points indicate the \emph{Planck} frequency channels used in this work.}
    \label{fig:freqscalings}
\end{figure}

The frequency scalings of various components of interest for the problem are shown in Fig.~\ref{fig:freqscalings}.  {To compute the signal due to each component in a given \emph{Planck} frequency map, we integrate these SEDs over the \emph{Planck} passbands~\cite{Zonca:2009vrg,Planck:2013wmz}, as described in Ref.~\cite{2024PhRvD.109b3528M}.}

\section{Impact of deprojection choices and CIB model}~\label{app:deprojections}

We measure $C_\ell^{g\Tdsc}$ using a $\Tdsc$ map that has the tSZ spectral distortion, the CIB, and the first moment of the CIB with respect to $\beta$ deprojected, along with the primary CMB blackbody on large scales (in particular, the first five harmonic needlet scales, so $\ell \lesssim 400$). In this section, we show the impact of these deprojections and demonstrate the insensitivity of our measurements to the assumed frequency dependence of the CIB SED.

\begin{figure*}
\includegraphics[width=0.49\columnwidth]{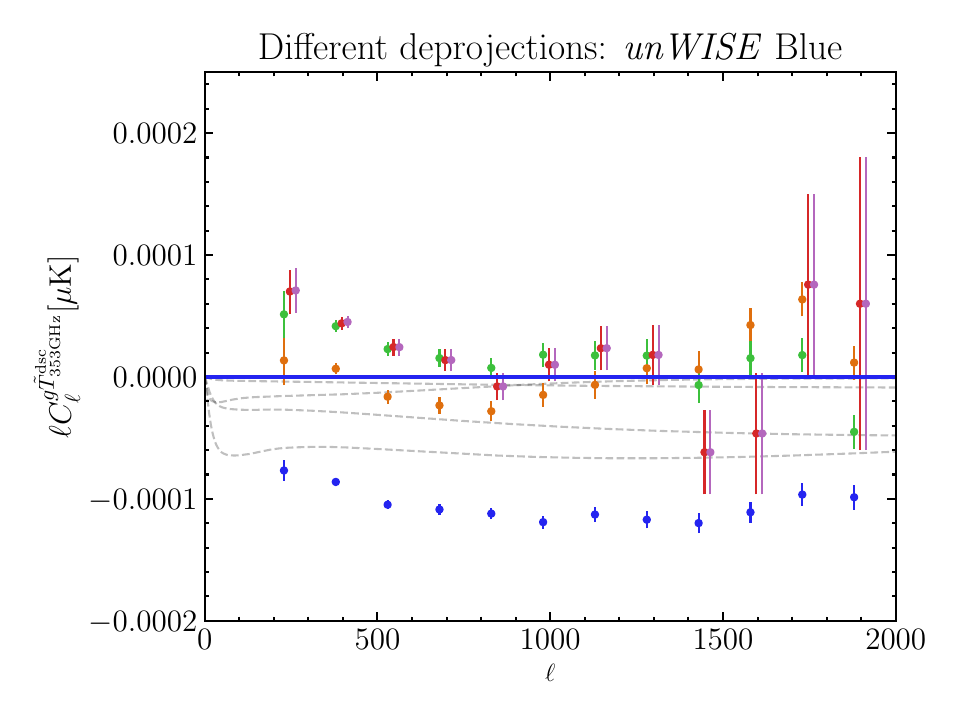}
\includegraphics[width=0.49\columnwidth]{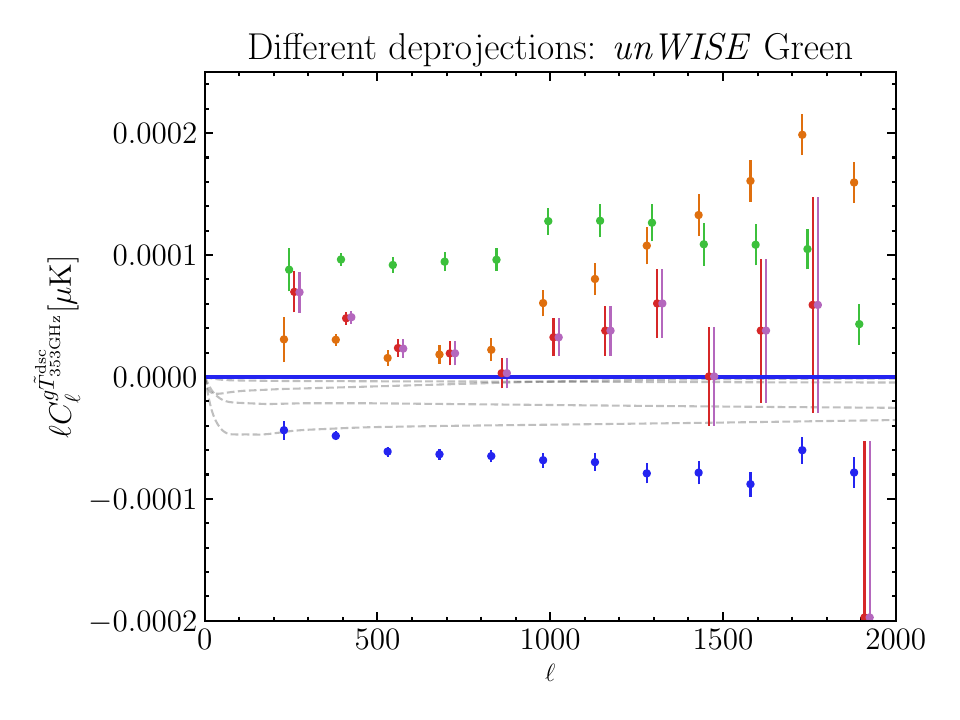}
\includegraphics[width=0.7\columnwidth]{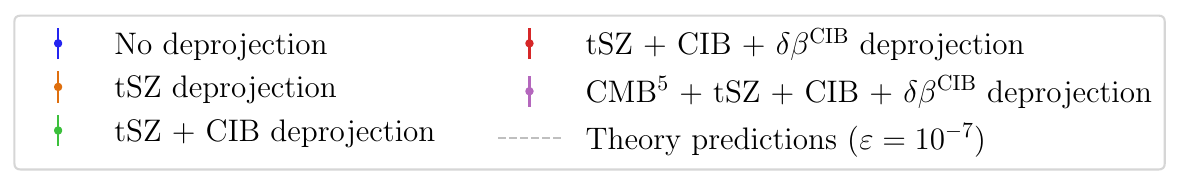}
\caption{The impact of deprojections on the cross-correlation measurements. Deprojection of tSZ contamination, and thus removing the $C_\ell^{gy}$ bias to the measurement, has a significant effect, as does the CIB deprojection and the $\delta\beta$ deprojection in green and red, respectively. Note that moving from a tSZ+CIB+$\delta\beta$-deprojected map to a CMB$^5$tSZ+CIB+$\delta\beta^{\rm{CIB}}$-deprojected  map (which additionally removes the CMB component on large scales) makes a negligible difference. Note that, for visual clarity, the points are offset slightly along the $x$ axis. }\label{fig:deprojections}
\end{figure*}

In Fig.~\ref{fig:deprojections} we show the impact of the deprojection choices in the NILC map on the final measurement. A striking feature of this plot is the necessity for tSZ deprojection. If an unconstrained (pure minimum-variance) ILC is used, the residual tSZ foreground is significant enough to bias the measurement very high.

\begin{figure*}
\includegraphics[width=0.49\columnwidth]{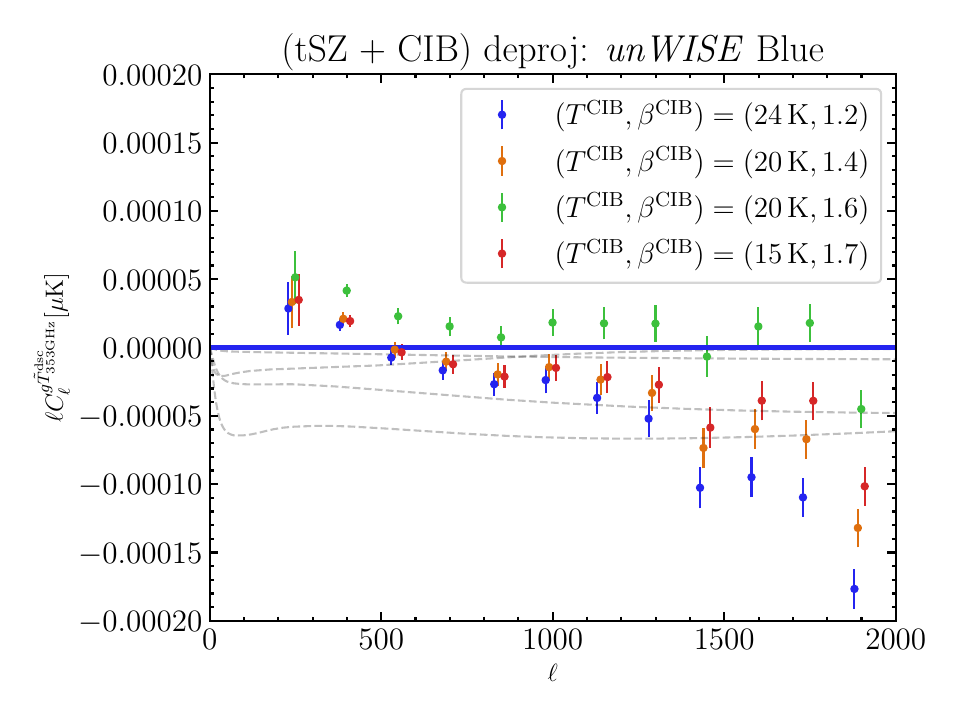}
\includegraphics[width=0.49\columnwidth]{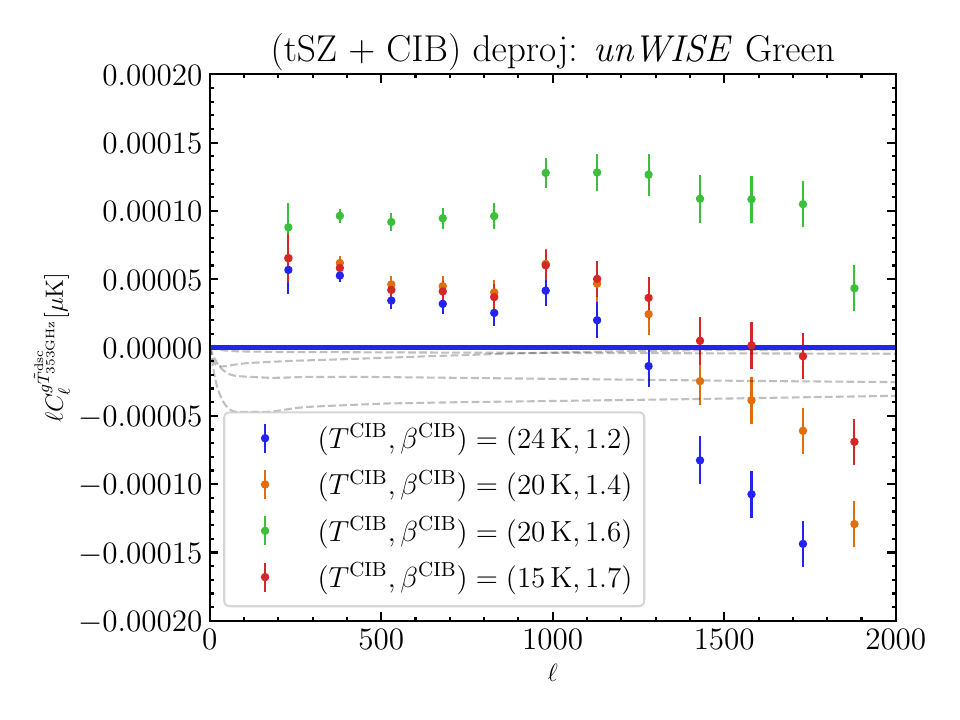}
\includegraphics[width=0.49\columnwidth]{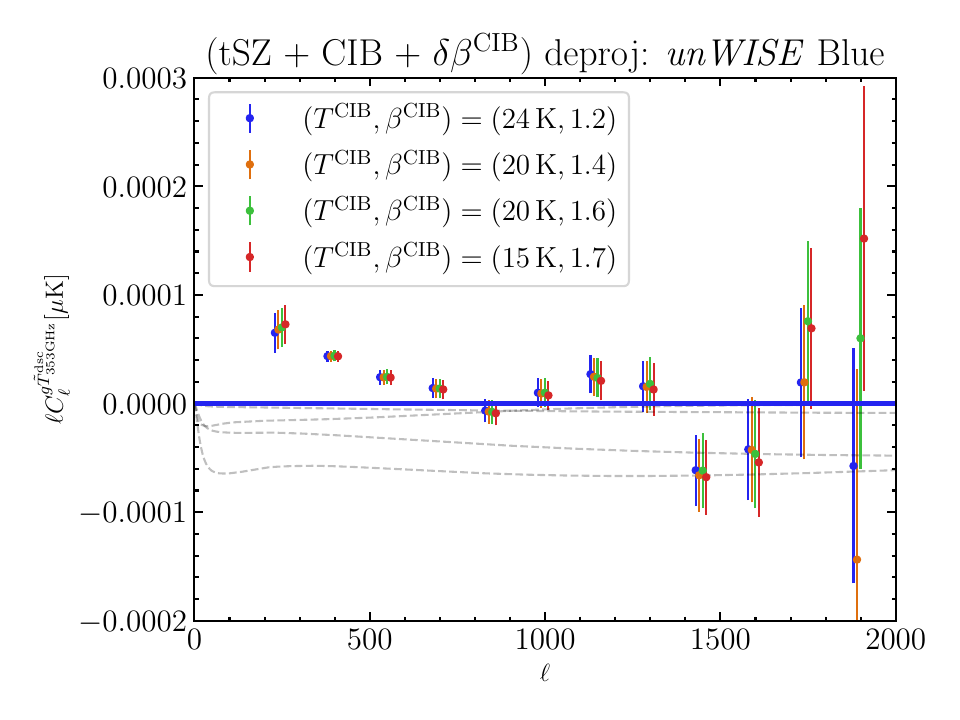}
\includegraphics[width=0.49\columnwidth]{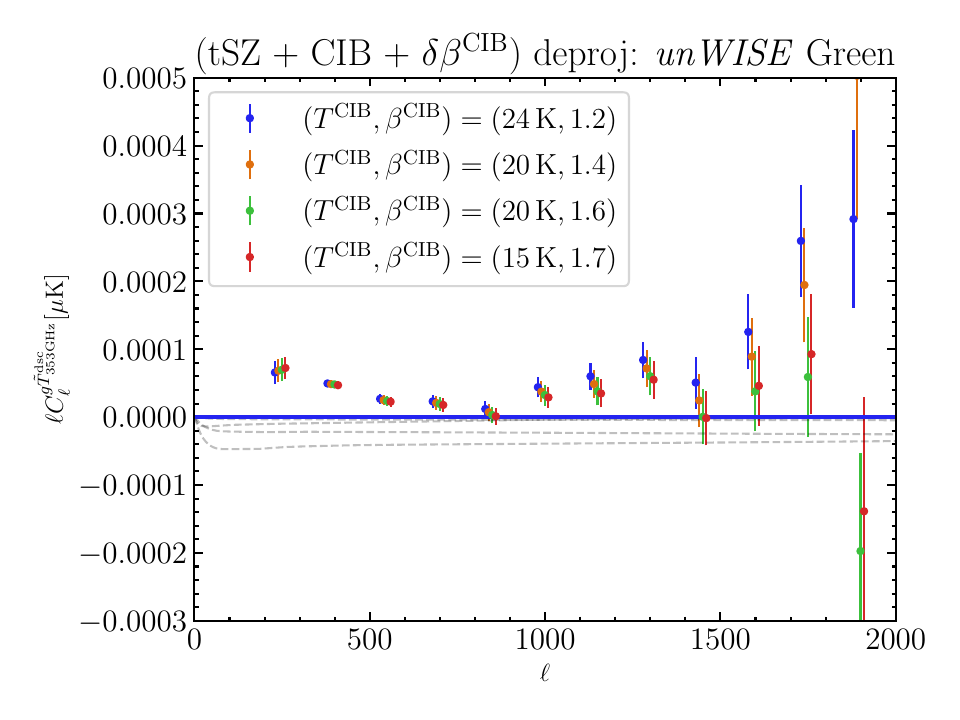}

\caption{The changes in the cross-correlation measurements when we change the CIB SED  used to deproject tSZ+CIB (top) contamination, compared to when we deproject tSZ+CIB+$\delta\beta^{\rm CIB}$ contamination (bottom). In the former case, the measurement is sensitive to the choice of CIB model; in the latter case, the measurement is insensitive (within the error bars of the measurement).}\label{fig:CIBSED}
\end{figure*}

In Fig.~\ref{fig:CIBSED} we demonstrate the stability of the measurement with respect to the choices of the parameters used to deproject the CIB. In particular, the CIB SED  $\Theta^{\rm CIB}$ is modeled as a modified blackbody at temperature $T^{\rm CIB}$ and with spectral index $\beta^{\rm CIB}$:
\begin{equation}
\Theta^{\rm CIB}(\nu) \propto B(\nu, T^{\rm CIB}) \nu^{\beta^{\rm CIB}},
\end{equation}
where $B(\nu, T)$ is the Planck function. The parameters $T^{\rm CIB}$ and $\beta^{\rm CIB}$ are uncertain and indeed different for each object that comprises the CIB. On average, however, the appropriate $(T^{\rm CIB}$,  $\beta^{\rm CIB})$ may depend on the mean redshift of the objects one is interested in. In particular, Ref.~\cite{2024PhRvD.109b3528M} found that the CIB monopole (as predicted by the CIB model constrained by the \textit{Planck} CIB auto-power spectrum in Ref.~\cite{2016A&A...586A.133P}) is best fit by $\left(T^{\rm CIB}, \beta^{\rm CIB}\right)\simeq(11.87\,  {\rm K}, 1.75)$. However, the CIB monopole is sourced at higher redshift than the anisotropies we are interested in, and so the appropriate temperature is likely higher than this (the CIB temperature increases with redshift, but does so slower than $(1+z)$, the factor that describes the redshift of the photons, so the effective temperature decreases with redshift). 

In any case, it is clear from Fig.~\ref{fig:CIBSED} that the deprojection with respect to  CIB$+\delta\beta^{\rm CIB}$ is insensitive to the choice of CIB SED, within a reasonable range, indicating that we have removed the systematics due to residual CIB contamination in the cross-correlation measurement.  Without the additional moment deprojection (referred to as $\delta\beta^{\rm CIB}$ deprojection), the measurement is strongly dependent on the choice of parameters.

\section{Sky area and masks}

\begin{figure*}
\includegraphics[width=0.49\columnwidth]{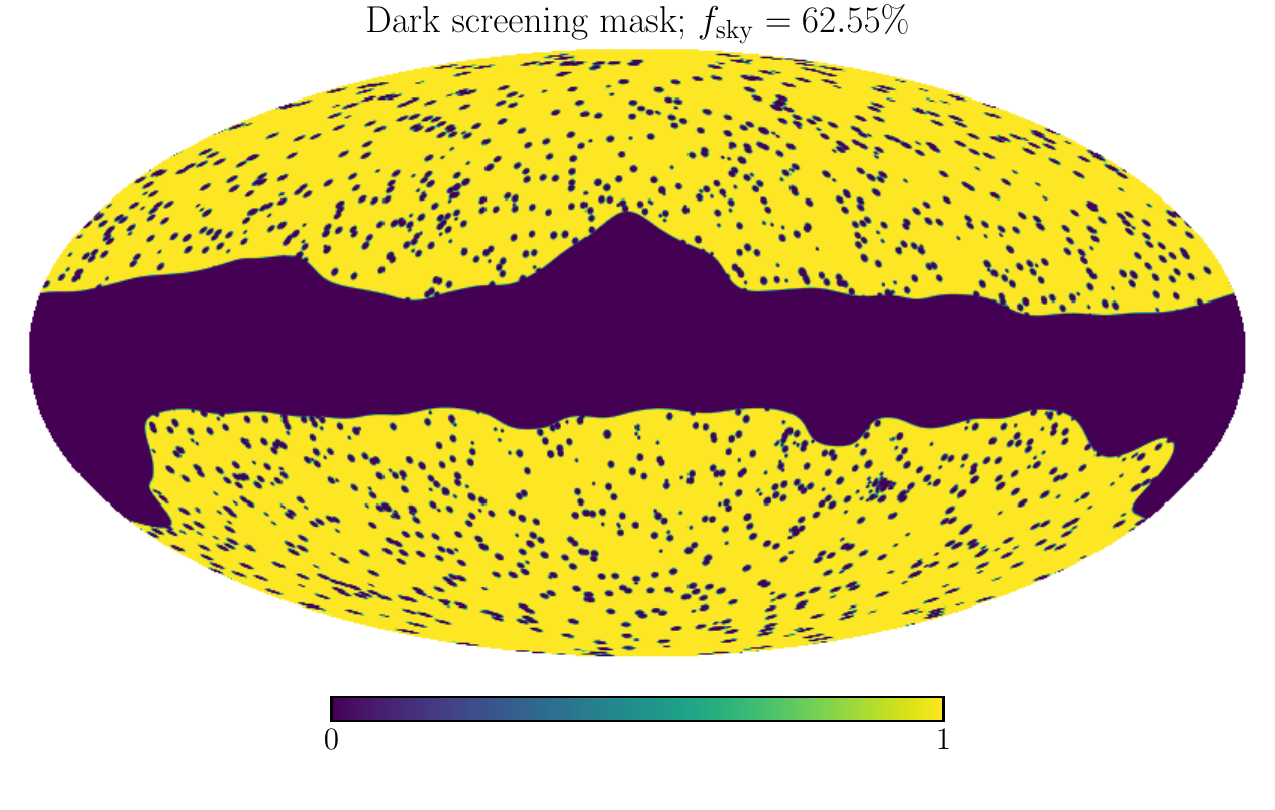}
\includegraphics[width=0.49\columnwidth]{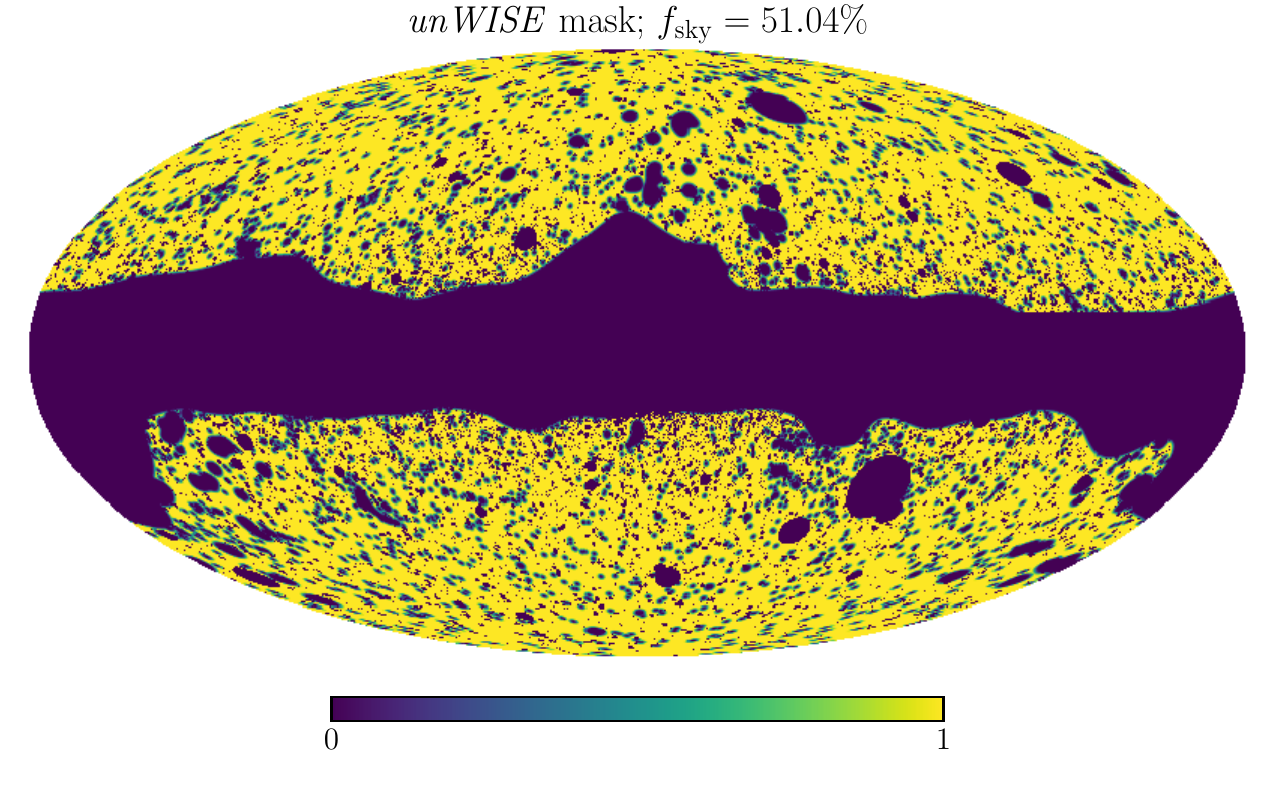}

\caption{The masks used in the nominal measurement, for the dark screening map (\textit{left}) and the \textit{unWISE} galaxies (\textit{right}). The dark screening mask leaves $f_{\rm sky}=56.50\%$ of the sky unmasked, and the \textit{unWISE} mask leaves $f_{\rm sky}=51.04\%$ of the sky unmasked. We show the masks in Mollweide projection and Galactic coordinates. All areas masked in the dark screening mask are masked in the \emph{unWISE} mask, such that the union of these masks has $f_{\mathrm{sky}}=51.04\%$.}\label{fig:masks}
\end{figure*}

Our  measurement is made with the dark screening map masked by the union of the preprocessing mask and the \textit{Planck} 70\% Galactic plane mask (the mask that leaves 70\% of the sky \textit{unmasked}), apodized with a scale of 30 arcminutes using the ``C2'' apodization routine of \texttt{namaster}. The \textit{unWISE} galaxies are masked by the \textit{unWISE} mask, which is the union of a point source mask and the \textit{Planck} CMB lensing mask. Following Ref.~\cite{2021PhRvD.104d3518K}, we apodize only the \textit{Planck} CMB lensing contribution to this mask, with an apodization scale of 60 arcminutes, again using the ``C2'' routine of \texttt{namaster}. These masks are shown in Fig.~\ref{fig:masks}, and the dark screening maps themselves (for various deprojection choices) are shown in Fig.~\ref{fig:darkscreeningmaps}.

\begin{figure*}
\includegraphics[width=0.49\columnwidth]{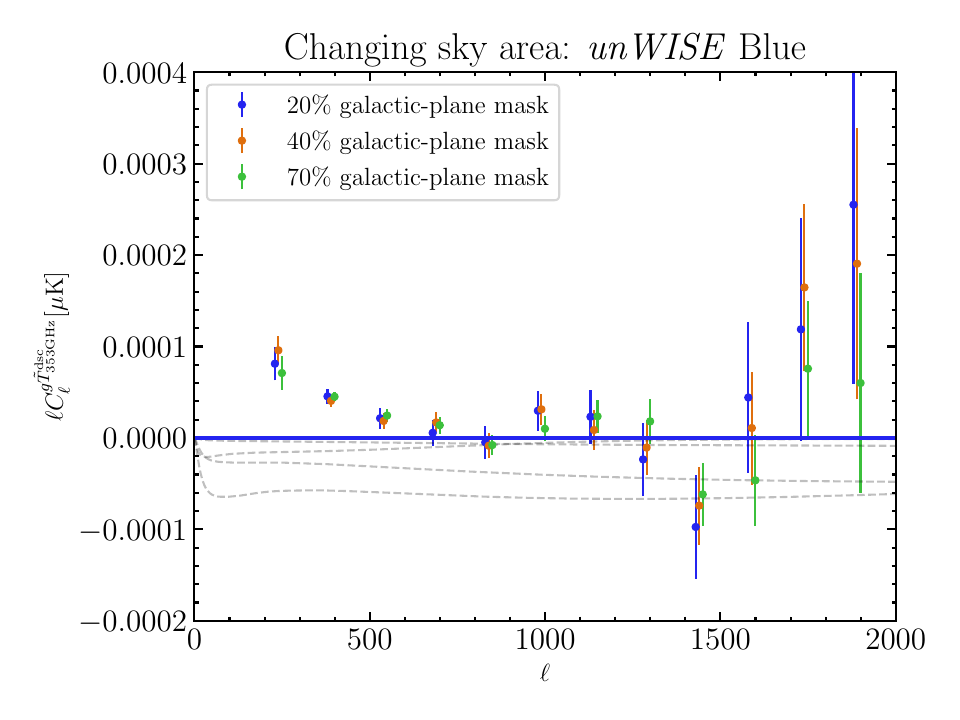}
\includegraphics[width=0.49\columnwidth]{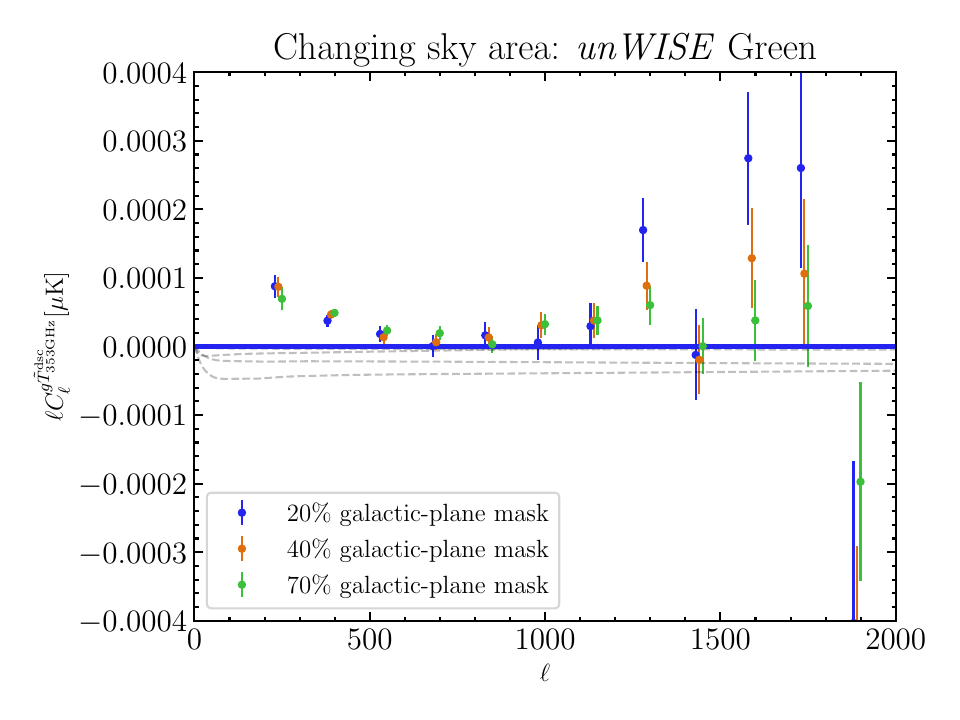}
\caption{Cross-correlation measurements performed with different amounts of the Galactic plane masked. The number in the legend refers to the fraction of sky left unmasked. Overall, the measurements are broadly consistent, with no indication of a residual Galactic contamination bias. Thus, in our nominal analysis, we use the measurement on the largest area of sky (the 70\% Galactic plane mask), which yields the smallest error bars. }\label{fig:compare_skyareas}
\end{figure*}

In Fig.~\ref{fig:compare_skyareas} we show the cross-correlation measurements as we vary the amount of the Galactic plane that we mask, to test for Galactic contamination in our data. Generally, there is no evidence of such contamination, as the measurements are all broadly consistent.

\section{Radio bias subtraction}

\begin{figure}
    \includegraphics[width=0.49\textwidth]{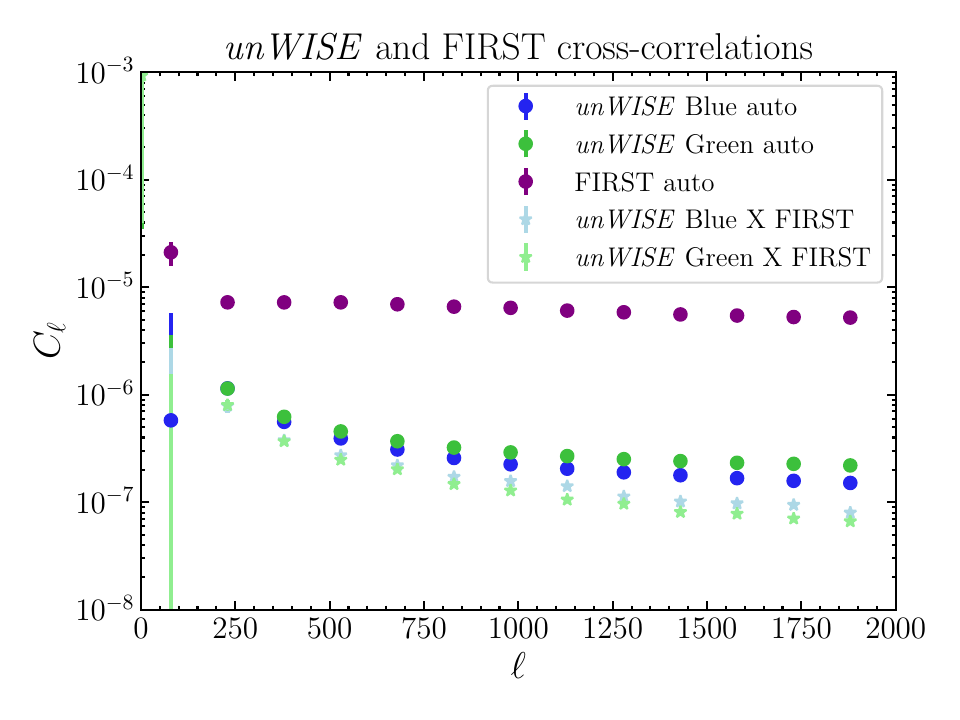}
    \includegraphics[width=0.49\textwidth]{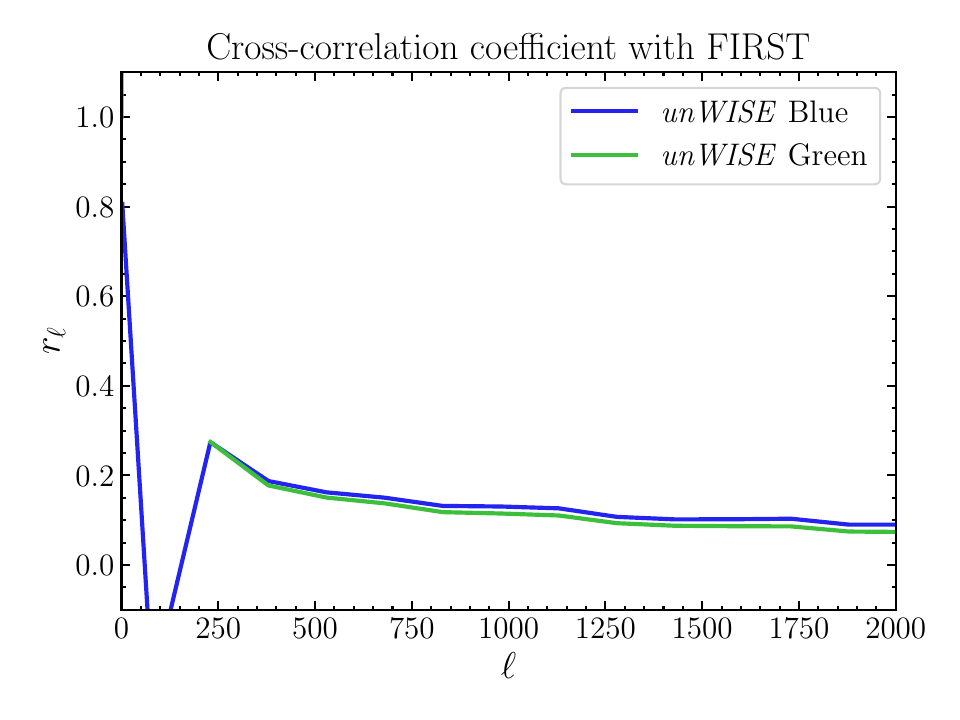}
\caption{Auto- and cross-power spectra of the FIRST and \textit{unWISE} samples (left), with the correlation coefficient $r_\ell$ between the \textit{unWISE} samples and FIRST explicitly calculated on the right.}
    \label{fig:FIRST_unwise}
\end{figure}

As we have not explicitly deprojected foreground contamination due to radio sources in the NILC dark screening map, the possibility of a residual extragalactic radio contribution that is correlated with the \textit{unWISE} galaxies remains.\footnote{Note that the NILC procedure will naturally clean a significant amount of radio contamination via its variance-minimization objective.} To assess the level of bias from this contribution, we directly measure the cross-correlation of the  \textit{unWISE} galaxy samples with a separate radio galaxy catalog from the FIRST survey~\cite{1997ApJ...475..479W}. We measure the cross-correlation on the area of sky left unmasked by the union of the \textit{unWISE} mask and the mask removing the sky area not covered by FIRST. The \textit{unWISE} Green and Blue samples have a $10-20\%$ correlation with the FIRST catalog, as shown in Fig.~\ref{fig:FIRST_unwise}. The correlation coefficient $r_\ell$ between fields $X$ and $Y$ is estimated as
\begin{align}
r_\ell = \frac{\hat C^{XY}}{\sqrt{\hat C^{XX}\hat C^{YY}}}.
\end{align}

\begin{figure}
    \includegraphics[width=0.49\textwidth]{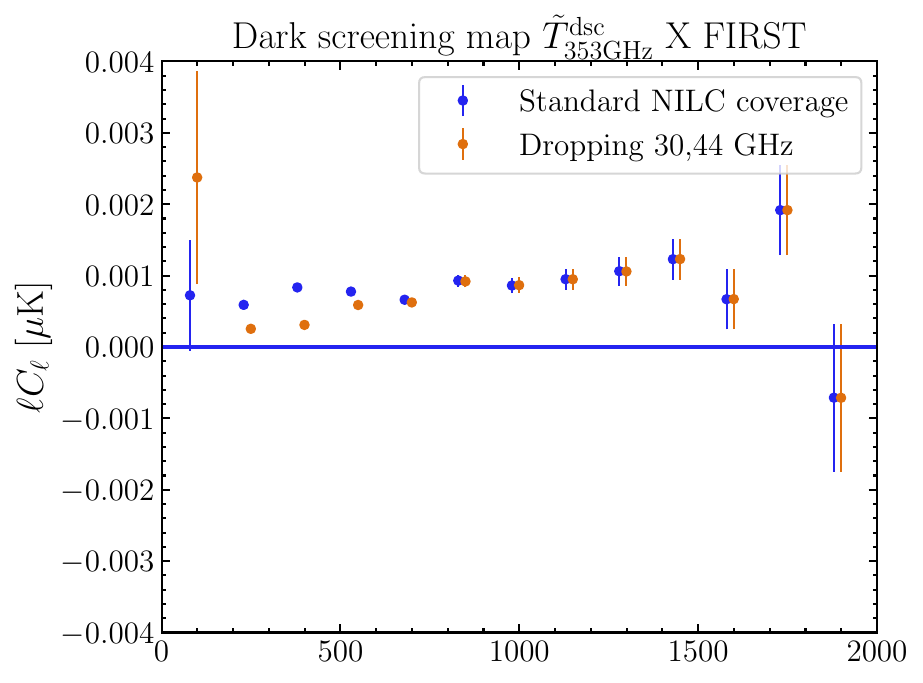}
    \includegraphics[width=0.49\textwidth]{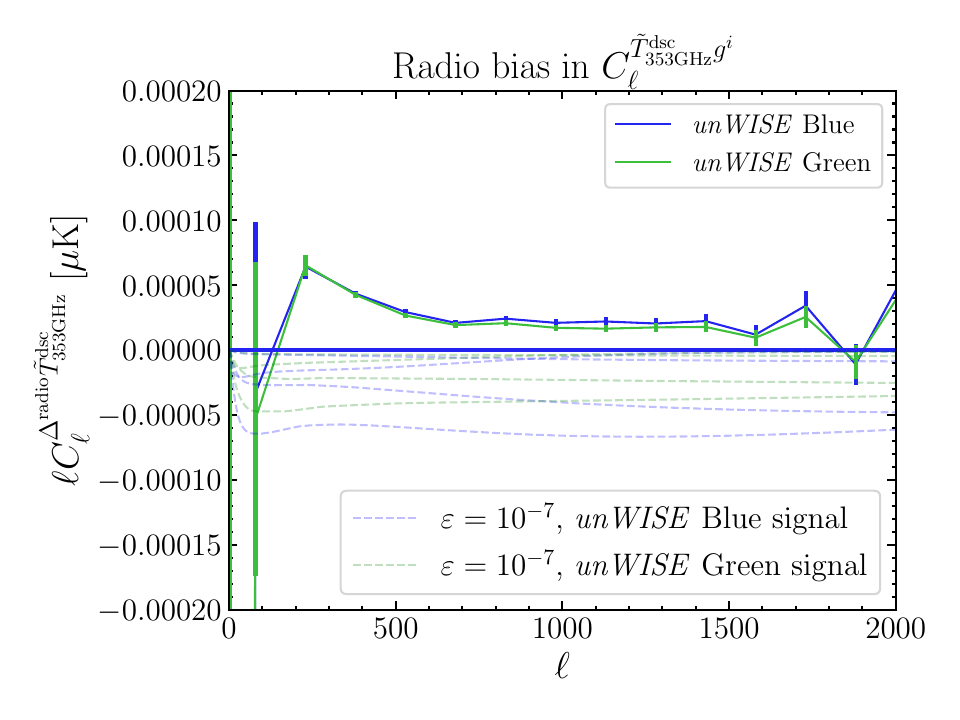}
    \caption{The measured cross-correlation of our dark screening map with the FIRST catalog (\textit{left}), which we then convert into a prediction for the radio bias in the \textit{unWISE} cross-correlation measurement on the \textit{right}.}
    \label{fig:dpFIRST}
\end{figure}

The cross-correlation between our fully-deprojected dark screening map and the FIRST catalog is shown in Fig.~\ref{fig:dpFIRST}. There is a clear non-zero positive correlation, a sign of residual positive radio emission in the dark screening map. Additionally, the measurement depends on the frequency coverage of the map (see below), a sign that we are detecting something with a different SED to the DP signal.

Note that  radio sources are brightest at low frequencies, and would be expected to  leak into the measurement as a positive (emission) signal, due to the relatively similar shape of the synchrotron SED and the DP distortion SED that our NILC preserves.   Indeed, we verify with simulations that this is the case --- in particular, we perform our NILC procedure on the \textit{Planck} mm-wave sky simulations described in Ref.~\cite{2024PhRvD.109b3529M}, which include extragalactic signals from the Websky simulations~\cite{2020JCAP...10..012S}, including a radio component that is correlated with LSS~\cite{2022JCAP...08..029L}.  In the absence of a proper \textit{unWISE} mock galaxy catalog built from the Websky matter density field, we correlate our NILC dark screening map (constructed from the Websky frequency maps) with the Websky halo catalog, reweighted to have the same redshift distribution as the \textit{unWISE} Blue galaxy catalog (as well as the Websky CMB lensing convergence map, which is sourced at higher redshift than the \emph{unWISE} galaxies), and verify that the radio bias is positive. 

Under the assumption that this measurement is dominated by the radio residuals in the dark screening map correlating with the FIRST catalog, we can use it to create a model of the residual radio -- \textit{unWISE} correlation. Specifically, writing the dark screening map as
\begin{align}
    \tilde T^{\mathrm{dsc}} = \tilde T^{\rm{dsc, \,\mathrm{true}}} + \Delta^{\mathrm{radio}} + N \,,
\end{align}
where $N$ are the remaining uncorrelated foregrounds and noise, $ \tilde T^{\mathrm{dsc},\, \mathrm{true}}$ is the true dark screening signal, and $\Delta^{\mathrm{radio}}$ is the radio residual, we have 
\begin{align}
\hat C_\ell^{\tilde T^{\mathrm{dsc}} \rm{FIRST}} = C_\ell^{\tilde T^{\mathrm{dsc}\,\mathrm{true}}\rm{ FIRST}}+ C_\ell^{\Delta^{\mathrm{radio}}\rm{ FIRST}}\approx C_\ell^{\Delta^{\mathrm{radio}}\rm{ FIRST}}.
\end{align}

In the \textit{unWISE} -- DP cross-correlation, we have
\begin{align}
\hat C_\ell^{\tilde T^{\mathrm{dsc}} g^i} = C_\ell^{\tilde T^{\mathrm{dsc}\,\mathrm{true}}g^i}+ C_\ell^{\Delta^{\mathrm{radio}}g^i};
\end{align}
we can thus use the FIRST measurements to estimate
\begin{align}
C_\ell^{\Delta^{\mathrm{radio}}g^i} = \frac{ \hat C_\ell^{\rm{FIRST\,}g^i}}{\hat C_\ell^{\rm{FIRST\,FIRST}}}\hat C_\ell^{\tilde T^{\mathrm{dsc}} \rm{FIRST}}.
\end{align}
We show these bias estimates explicitly in the right panel of Fig.~\ref{fig:dpFIRST}. They have {a similar} scale dependence to the expected dark screening signal, with a magnitude corresponding to $\varepsilon\sim5\times10^{-8}$ for the masses that our data are most sensitive to (but with opposite sign, as they are due to emission rather than screening). 

\section{Frequency coverage}~\label{app:drop30}

\begin{figure*}
\includegraphics[width=0.49\columnwidth]{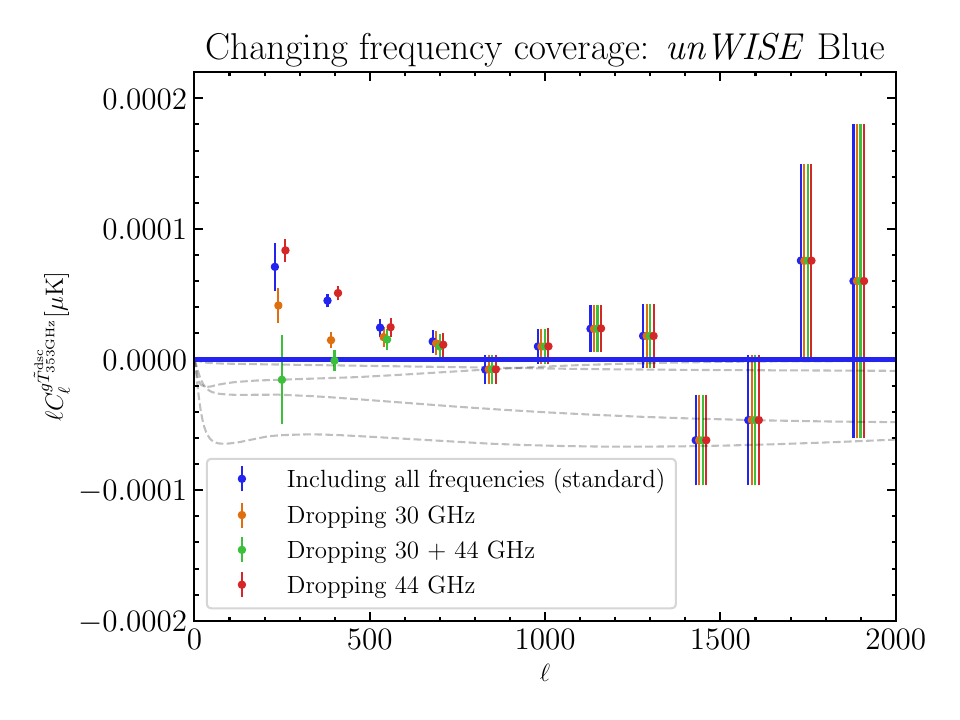}
\includegraphics[width=0.49\columnwidth]{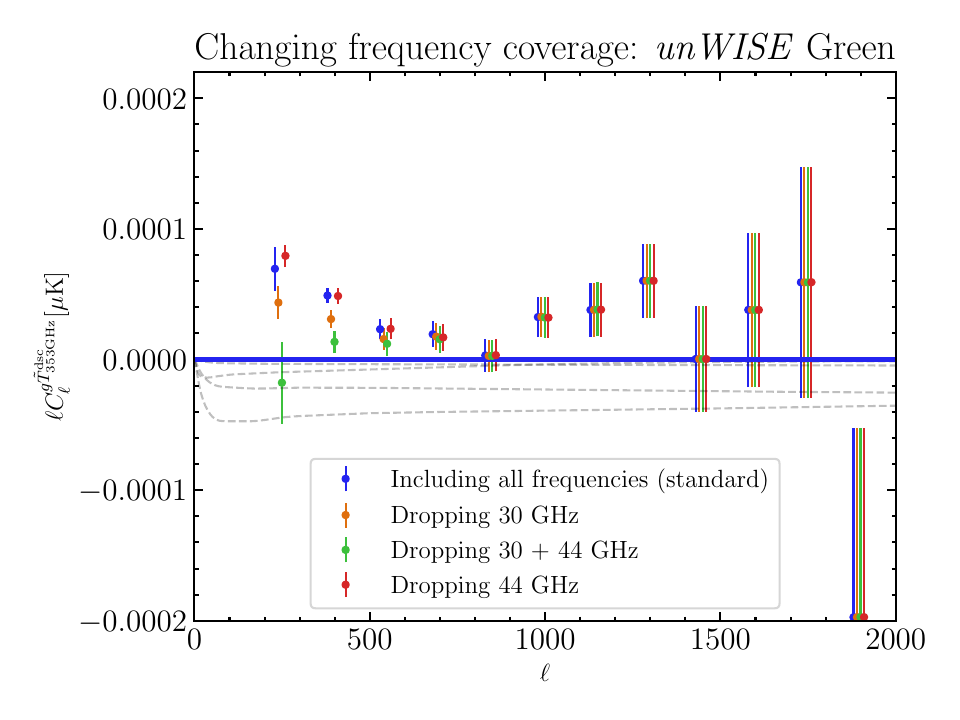}
\includegraphics[width=0.49\columnwidth]{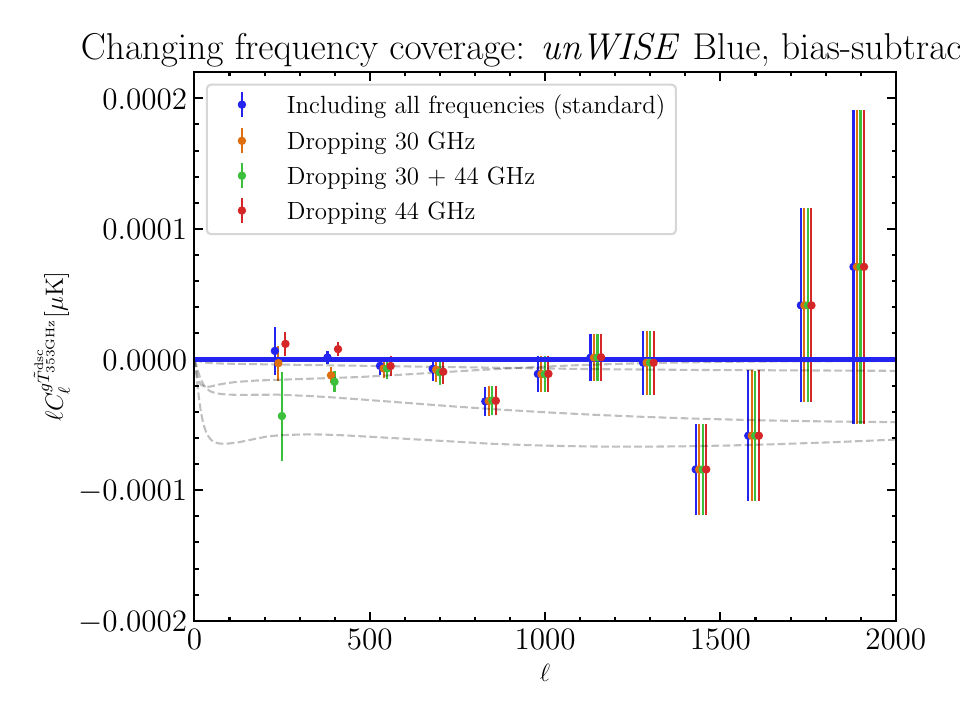}
\includegraphics[width=0.49\columnwidth]{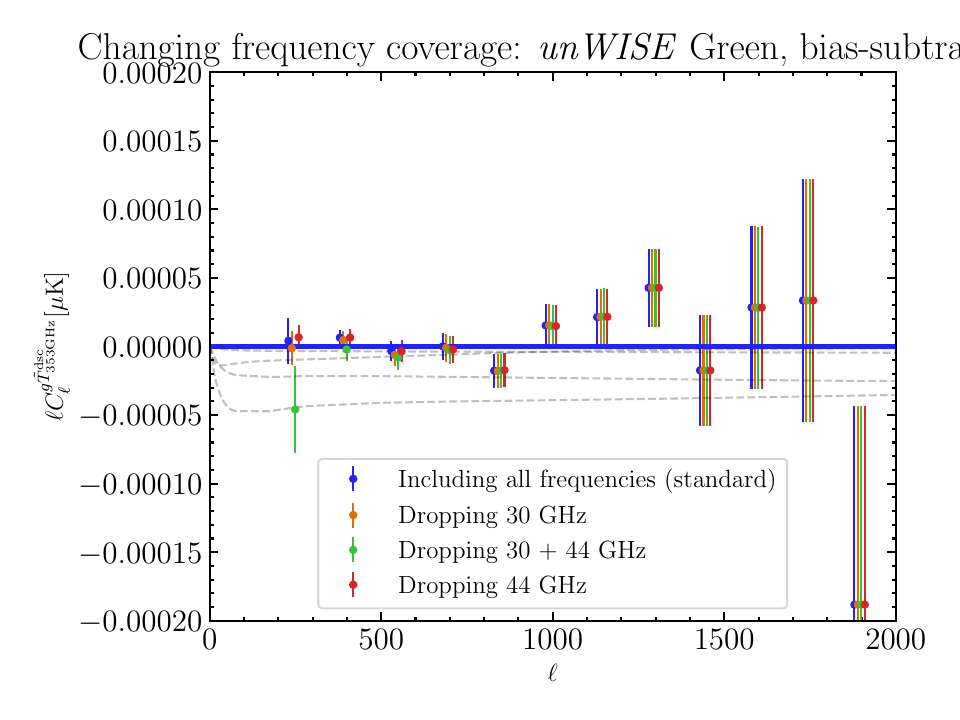}

\caption{The change in the cross-correlation measurements when we change the frequency coverage used in the NILC map construction, both for the raw data (\textit{top}) and the bias-subtracted data (\textit{bottom}).  Note that the low-frequency \textit{Planck} maps have larger beams than the high-frequency maps, and so do not contribute on small scales (high $\ell$), which is why the points at $\ell \gtrsim 1000$ are not sensitive to these tests.}\label{fig:drop_frequencies}
\end{figure*}

In our nominal results, we include the \{30, 44, 70, 100, 143, 217, 353, 545\} GHz \textit{Planck} intensity maps. In the top of Fig.~\ref{fig:drop_frequencies}, we show the cross-correlation measurements when the lower-frequency channels, 30 and 44 GHz, are excluded from the construction of the NILC dark screening map.

When all frequency channels are included, before radio-bias subtraction a positive (emission) signal is evident at low $\ell$ ($\ell\lesssim500$). This residual bias is removed when we drop the lowest-frequency channels (30 and 44 GHz) from the measurement, and is generally stable to the additional removal of the 70 GHz channel (not shown in the plots). 

This instability to the choice of frequency coverage is evidence for a frequency-dependent residual foreground in the NILC map, which we attribute to radio emission (see above). However, after performing the radio-bias subtraction described in the previous section (with the radio bias recalibrated for each choice of frequency coverage), we note that this procedure brings the data into agreement for all choices of frequency coverage, as shown in the bottom of Fig.~\ref{fig:drop_frequencies}.

\section{Recovery of an artificial signal}

To ensure that our pipeline is unbiased, we inject an artificial signal into the data and demonstrate that we can recover it.  We create the artificial signal by first filtering the \textit{unWISE} map to create a $\tilde T^{\mathrm{dsc}}$ map with the appropriate correlation:
\begin{align}
\tilde T^{\mathrm{dsc}} = \frac{C_\ell^{g\tilde T^{\mathrm{dsc}}}}{\hat C_\ell^{gg}}
\end{align}
where $C_\ell^{g\tilde T^{\mathrm{dsc}}}$ is our signal prediction and $\hat C_\ell^{gg}$ is the measured \textit{unWISE} Blue auto-power spectrum, which is effectively noiseless over the range of scales considered here due to its extremely low shot noise (\ie high number density). We choose a signal corresponding to $\varepsilon=1.06\times10^{-7}$ and $m_{A^\prime}=5.2\times10^{-13} \,\, \mathrm{eV}$. We add this to the \textit{Planck} single-frequency maps with the appropriate DP-distortion frequency dependence and then perform NILC component separation. 

We measure the cross-correlation of this signal-injected NILC map with the \textit{unWISE} Blue sample, and we re-calibrate the radio bias \textit{from this map} using the FIRST data as described above. This is very important, as we aim to explicitly verify that the radio bias subtraction step does not artificially subtract the true signal. 

\begin{figure}
\includegraphics[width=0.5\textwidth]{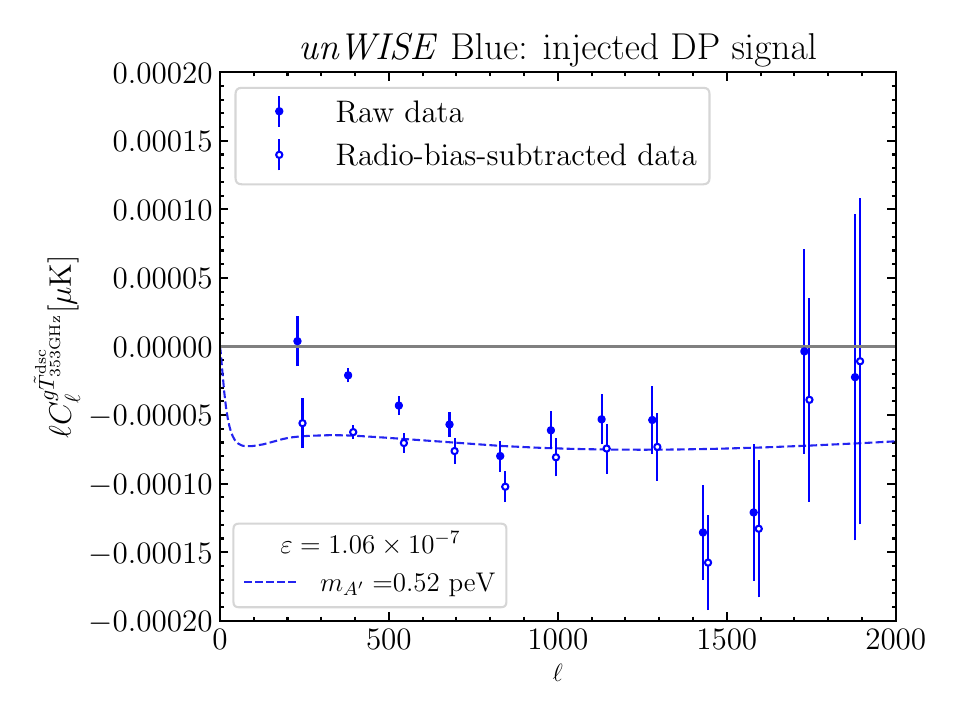}
\caption{Recovery of an injected DP distortion signal.}\label{fig:injection}
\end{figure}

The recovery of the input signal is demonstrated in Fig.~\ref{fig:injection}, thereby validating our pipeline, including the radio bias subtraction. {The $\Delta \chi^2$ between the data and the model curve is 15.58 for 12 points, which corresponds to a PTE of 0.21, indicating good agreement.}

\section{Changes to the galaxy and electron models}~\label{app:vary_halomodel}

In our nominal analysis, we hold fixed the astrophysical quantities appearing in the halo model used to compute $C_\ell^{g \Tdsc}$, as described in SM~\ref{sec:halomodel_unwISE}. In reality, there is some uncertainty associated with both the galaxy HOD and electron density profile models. Properly, one should marginalize over these uncertainties,~\eg by performing a full Markov Chain Monte Carlo (MCMC) analysis with the HOD and electron profile parameters allowed to vary. However, such an operation would be computationally expensive and, as we show below, unlikely to significantly change our derived constraints on $\varepsilon$. In this section, we explore the dependence of our DP constraints on such changes to the underlying astrophysical modeling assumptions, to justify our decision to ignore the associated marginalization.

\subsection{Uncertainties in the galaxy HOD model}

\begin{figure}
    \includegraphics[width=0.6\columnwidth]{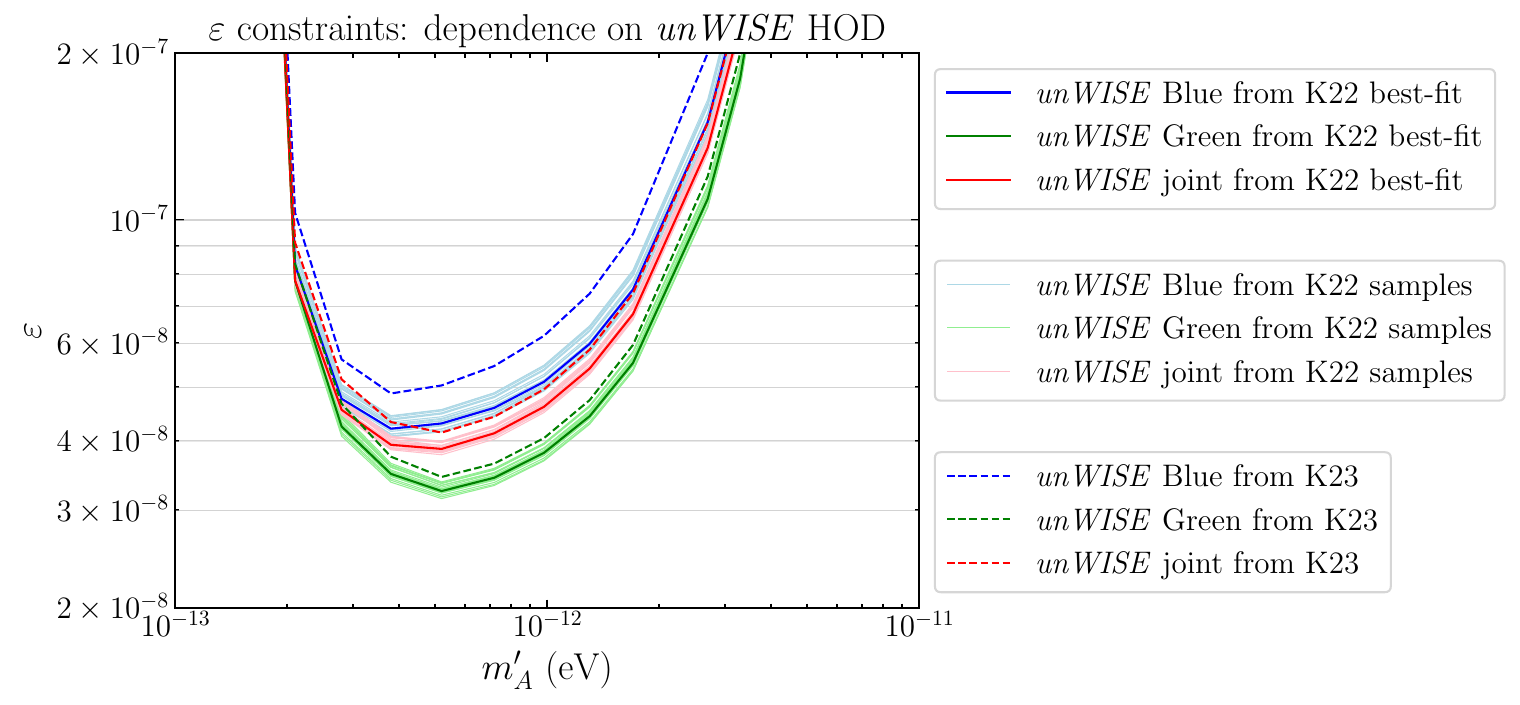}
    \includegraphics[width=0.39\columnwidth]{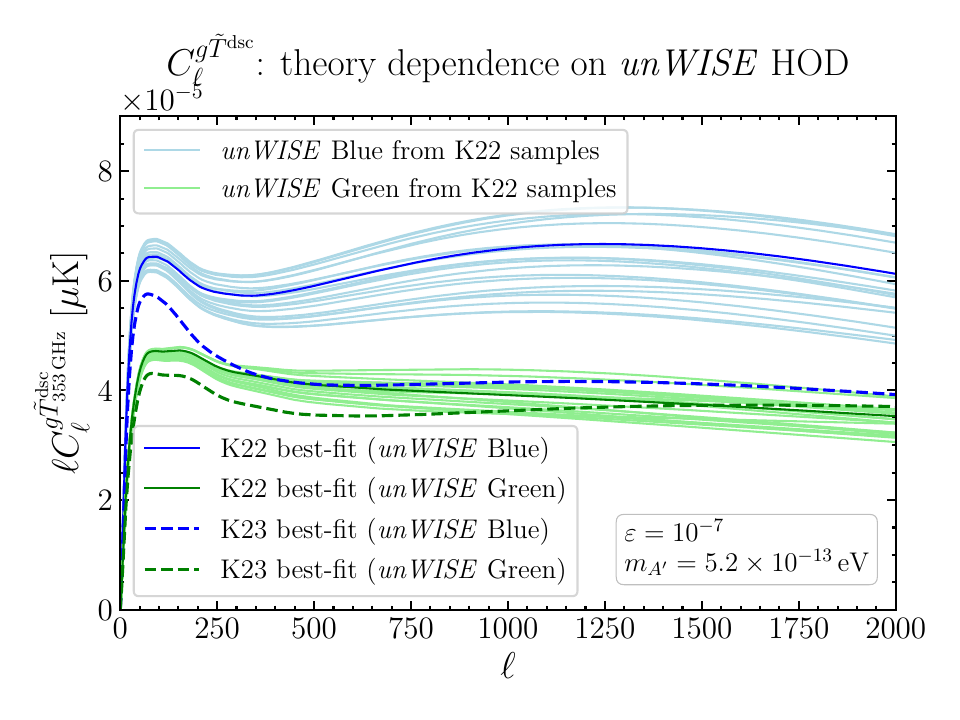}
    \caption{\textit{Left:} The change in our constraints on $\varepsilon$ when we change the parameters used in the HOD model that describes the \textit{unWISE} galaxies, both by directly drawing samples from the MCMC chains from Ref.~\cite{2022PhRvD.106l3517K} (K22) to constrain the model (light solid lines) and by switching to the best-fit model of Ref.~\cite{2023PhRvD.108l3501K} (K23) (dashed lines). Our nominal constraints, from the best-fit model of K22, are in dark solid lines. \textit{Right:} The explicit changes in the model prediction for $C_\ell^{g \tilde T^{\rm dsc}}$ under the same modeling assumption changes, for $\varepsilon=10^{-7}$ and $m_{A^\prime}=5.2\times10^{-13} \,\, \mathrm{eV}$.}\label{fig:HOD_switching}
\end{figure}

The \emph{unWISE} HOD parameters we use were obtained in Ref.~\cite{2022PhRvD.106l3517K} from fits to galaxy clustering and galaxy -- CMB lensing cross-correlation data. To explore the uncertainty involved in these fits, we draw 20 samples from the MCMC chains used to approximate the posterior distributions in Ref.~\cite{2022PhRvD.106l3517K}; these can be thought of as a prior on the galaxy HOD model.  We then re-run our analysis pipeline to obtain constraints on $\varepsilon$ for each set of galaxy HOD parameters.  We show in Fig.~\ref{fig:HOD_switching} the distribution of the constraints we obtain when we vary the HOD model in this way.  In general, the results obtained with the best-fit HOD model were representative; the constraints degrade at most from $\varepsilon<2.08\times10^{-8}$ at their strongest to $\varepsilon<2.17\times10^{-8}$, a degradation of only 5\%. 

A larger, but not significant, degradation in the constraints is found when we switch from the best-fit model of Ref.~\cite{2022PhRvD.106l3517K} (K22) to that of Ref.~\cite{2023PhRvD.108l3501K} (K23) (see their Appendix B). In K23, the HOD parameters were fit to data at higher $\ell$ than in K22 (where only the large scales, $\ell\lesssim1000$, were used). In particular, due to the change in the best-fit value of the $\sigma_{\mathrm{log}\, m}^{\mathrm{Blue}}$ parameter from 0.687 to 0.02, the prediction of $C_\ell^{g\tilde T^{\mathrm{dsc}}}$ is much lower in the K23 model.  These changes in the predictions for $C_\ell^{g\tilde T^{\mathrm{dsc}}}$ are explicitly shown in the right panel of Fig.~\ref{fig:HOD_switching}.

\subsection{Uncertainties in the electron model}

\begin{figure}
\includegraphics[width=0.5\textwidth]{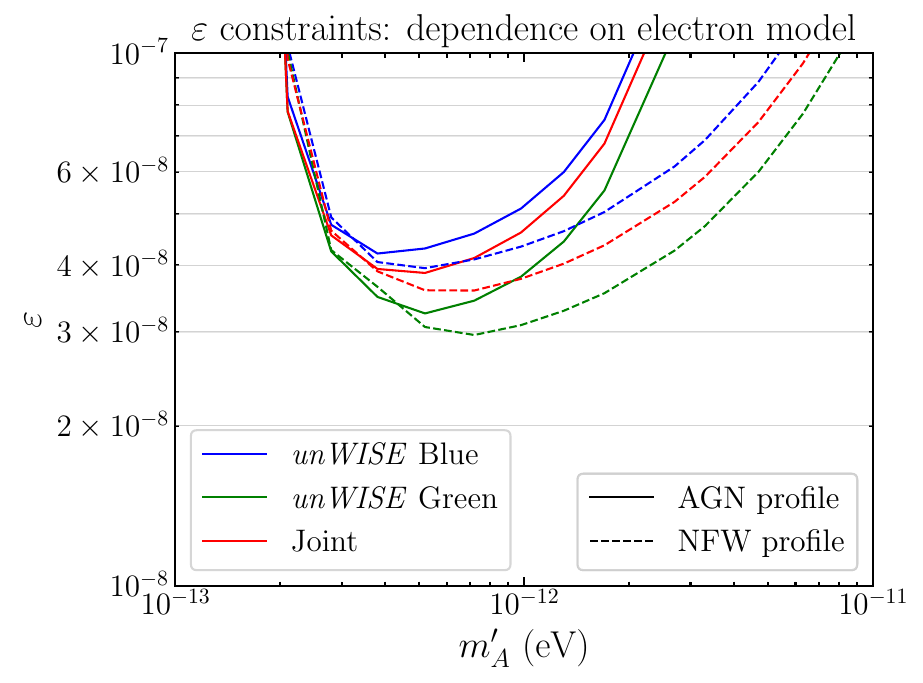}
\caption{The change in our constraints on $\varepsilon$ when we assume that the electrons follow the dark matter (NFW) profile (dashed lines) compared to our nominal  model from Ref.~\cite{Battaglia:2016xbi} (solid lines), in which the electron profile has a much shallower slope at large halo-centric radii due to the effects of AGN feedback. We see that the feedback reduces our sensitivity to $\varepsilon$ at the high end of the mass range, because the associated high electron densities are significantly lessened by the feedback-induced smoothing of the electron distribution.  }\label{fig:changing_electrons}
\end{figure}

There are also uncertainties in the electron density profile model used in the computation of $C_\ell^{g\tilde T^{\mathrm{dsc}}}$. In our nominal analysis, we use the spherically-symmetric generalized Navarro-Frenk-White (gNFW) profiles fit in Ref.~\cite{Battaglia:2016xbi} to hydrodynamic simulations incorporating feedback from active galactic nuclei (AGNs). Responding to the feedback, the electrons are distributed more smoothly than the dark matter, which follows the standard NFW profile~\cite{1997ApJ...490..493N}. To indicate how our constraints on $\varepsilon$ could change due to different assumptions about the electron distribution, we calculate $C_\ell^{g\tilde T^{\mathrm{dsc}}}$ assuming that the electrons exactly trace the dark matter; this is an extreme assumption that is empirically ruled out~\cite{2021PhRvD.103f3514A,2024arXiv240113033C}. The resulting constraints are shown in Fig.~\ref{fig:changing_electrons}.

We see that the inclusion of AGN feedback (and the other non-gravitational physics included in the simulations of Ref.~\cite{Battaglia:2016xbi}) reduces our sensitivity to $\varepsilon$ at the high end of the relevant DP mass range. Thus, if AGN feedback is weaker than that in the simulations on which these profiles are based, our constraints are conservative.   Updated DP analyses using new electron density profile modeling constraints, such as from kinematic SZ measurements~\cite{2021PhRvD.103f3514A,2021PhRvD.103f3513S,2021PhRvD.104d3518K} would be interesting.

\end{document}